\documentclass{mavrl}

\usepackage[T1]{fontenc}
\usepackage{subcaption}
\usepackage{booktabs}
\usepackage{multirow}
\usepackage{listings}
\usepackage{xcolor}
\usepackage{array}
\usepackage{longtable}
\definecolor{lstkeyword}{RGB}{0,51,179}
\definecolor{lstbuiltin}{RGB}{128,0,128}
\definecolor{lststring}{RGB}{6,125,23}
\definecolor{lstcomment}{RGB}{110,110,110}

\title{MatCalc: A Workflow Harness for Materials Property Calculation and Benchmarking}

\author{Runze Liu}
\affiliation[UCSD]{Aiiso Yufeng Li Family Department of Chemical and Nano Engineering, University of California San Diego, La Jolla, California, United States}
\email{rul048@ucsd.edu}

\author{Atul C. Thakur}
\affiliation[UCSD]{Aiiso Yufeng Li Family Department of Chemical and Nano Engineering, University of California San Diego, La Jolla, California, United States}

\author{Mgcini Keith Phuthi}
\affiliation[UCSD]{Aiiso Yufeng Li Family Department of Chemical and Nano Engineering, University of California San Diego, La Jolla, California, United States}

\author{Sojung Koo}
\affiliation[UCSD]{Aiiso Yufeng Li Family Department of Chemical and Nano Engineering, University of California San Diego, La Jolla, California, United States}

\author{Ji Qi}
\affiliation[UCSD]{Aiiso Yufeng Li Family Department of Chemical and Nano Engineering, University of California San Diego, La Jolla, California, United States}

\author{Tsz Wai Ko}
\affiliation[UCSD]{Aiiso Yufeng Li Family Department of Chemical and Nano Engineering, University of California San Diego, La Jolla, California, United States}

\author{Andrew S. Rosen}
\affiliation[Princeton]{Department of Chemical and Biological Engineering, Princeton University, Princeton, New Jersey, United States}

\author{Shyue Ping Ong}
\affiliation[NUS]{Department of Materials Science and Engineering,
National University of Singapore, Singapore}
\alsoaffiliation[UCSD]{Aiiso Yufeng Li Family Department of Chemical and Nano Engineering, University of California San Diego, La Jolla, California, United States}
\email{shyuep@nus.edu.sg}


\begin{document}

\maketitle

\begin{abstract}
Atomistic simulations driven by a parameterization of the potential energy surface (PES) provide access to a wide range of materials properties. However, computing many properties requires multi-step workflows whose parameters, such as relaxation criteria, strain magnitudes, supercell sizes, and thermostat settings, materially affect the results and are rarely applied consistently across studies. Here we present MatCalc (Materials Calculator), an open-source Python framework that provides a single, consistent interface for computing materials properties from any PES parameterization. MatCalc currently supports relaxed structures, formation and cohesive energies, equations of state, elastic constants, harmonic phonons, quasi-harmonic thermal expansion, lattice thermal conductivity, migration barriers, molecular dynamics observables, and surface, grain boundary, interfacial, and adsorption energies. Its core abstraction, \texttt{PropCalc}, accepts a Python Materials Genomics (pymatgen) \texttt{Structure} or Atomic Simulation Environment (ASE) \texttt{Atoms} object, delegates PES evaluations to any ASE \texttt{Calculator}, and runs the underlying relaxations and molecular dynamics with either ASE or the Large-scale Atomic/Molecular Massively Parallel Simulator (LAMMPS) as the simulation backend, so identical workflow code runs with density functional theory (DFT), classical interatomic potentials, machine learning interatomic potentials (MLIPs), or foundation potentials (FPs). MatCalc also ships MatCalc-Bench, a curated suite of property-level reference datasets and standardized evaluation protocols spanning structural, thermodynamic, mechanical, vibrational, kinetic, transport, surface, and interfacial properties, with Perdew--Burke--Ernzerhof (PBE) and r$^2$SCAN DFT reference data and experimental values. To our knowledge, MatCalc-Bench is the first public FP benchmark that evaluates models against r$^2$SCAN and against experiment across multiple properties. Benchmarking ten state-of-the-art FPs on 17 properties, we find that the best FPs approach the accuracy of element-specific MLIPs and that no single FP leads on more than a few properties. The training data, through their configurational coverage, DFT settings, and exchange-correlation functional, largely determine property-level accuracy: against experiment, FPs trained on r$^2$SCAN data reduce the median errors of eight bulk properties to within 7\%, compared with up to 22\% for PBE-trained FPs. MatCalc lowers the barrier for non-specialists to apply state-of-the-art MLIPs, while giving domain experts and AI agents a reproducible, calculator-agnostic harness for materials simulation and benchmarking.
\end{abstract}


\section{Introduction}

Many properties of materials can be derived from the potential energy surface (PES), i.e., the energy as a function of atomic positions and lattice vectors, together with its derivatives, the forces and stresses. For instance, obtaining the equilibrium structure of a crystal is an exercise in navigating the PES to find a local minimum. Other properties, such as elastic constants and phonon dispersions, are obtained from derivatives of the PES about this equilibrium. Yet other properties, such as diffusivities, can be obtained from time-correlation functions computed from molecular dynamics (MD) simulations, in which the PES supplies the forces used to integrate the equations of motion.

The PES has traditionally been obtained either from accurate but expensive \textit{ab initio} methods, such as density functional theory (DFT),\cite{hohenbergInhomogeneousElectronGas1964,kohnSelfConsistentEquationsIncluding1965} or from inexpensive but less accurate classical interatomic potentials. The advent and rapid maturation of machine learning interatomic potentials (MLIPs)\cite{behler2021,deringer2021,unke2021,ko2023recent} has broken this trade-off by offering linear-scaling evaluations of the PES at near-\textit{ab initio} accuracy.\cite{zuoPerformanceCostAssessment2020} More recently, foundation potentials (FPs), also known as universal MLIPs,\cite{chenUniversalGraphDeep2022,dengCHGNetPretrainedUniversal2023,Batatia2025MACEMP,Wood2025UMA} trained on large, chemically diverse PES datasets such as MPtrj\cite{dengCHGNetPretrainedUniversal2023}, Alexandria\cite{schmidt2024alexandria}, OMat24\cite{barroso-luqueOpenMaterials20242024}, MatPES\cite{kaplan2025matpes}, and OMol25\cite{levine2025}, extend MLIPs across the periodic table and are increasingly used as general-purpose, low-cost surrogates for DFT.\cite{ko2026foundational}

\begin{figure}
    \centering
    \includegraphics[width=1\linewidth]{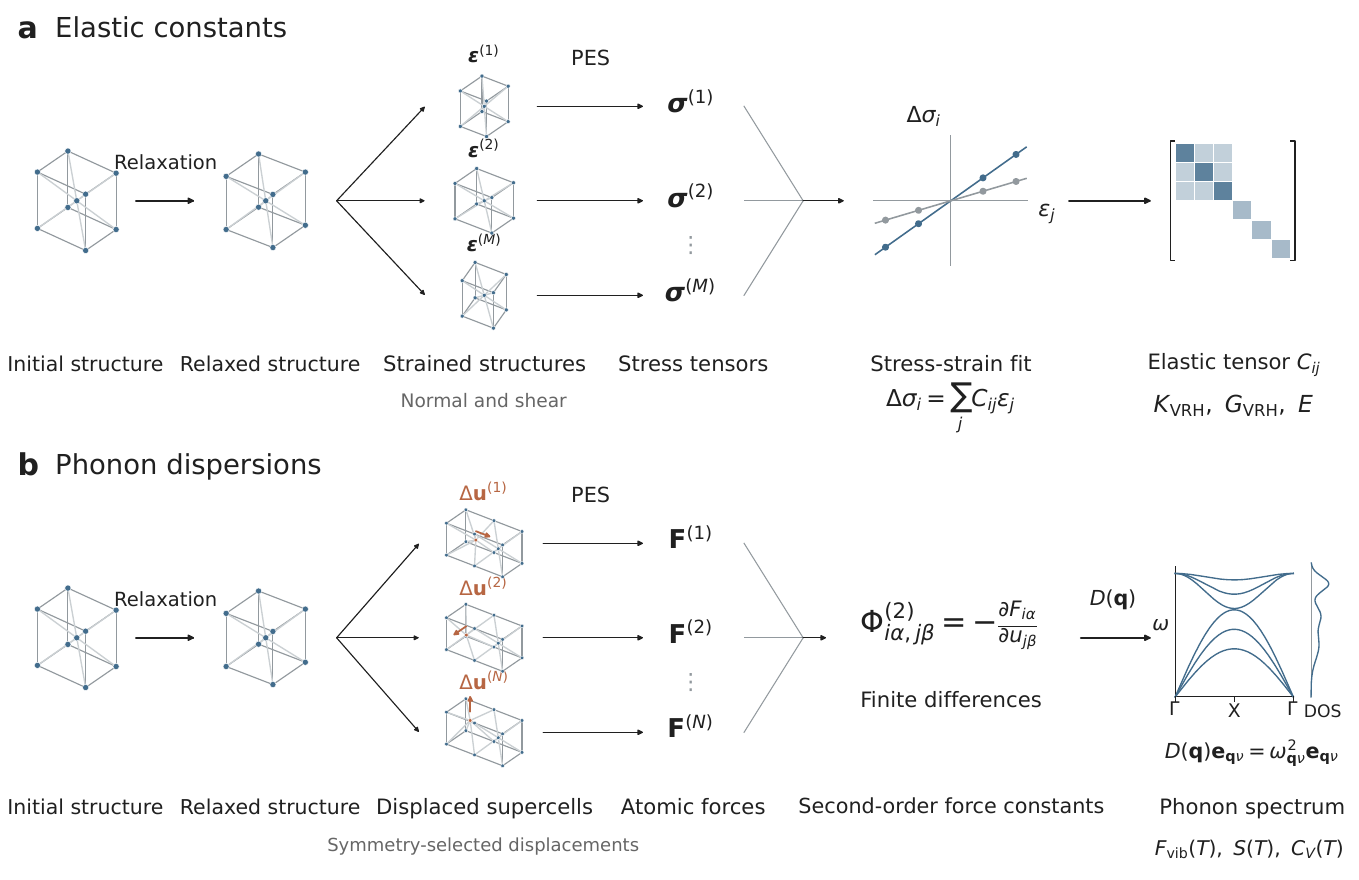}
    \caption{\textbf{Representative multi-step property workflows.} (a)~Elastic constants: the structure is relaxed, $M$ normal and shear strains $\varepsilon^{(m)}$ are applied, the stress tensor $\sigma^{(m)}$ of each strained structure is evaluated, and a linear fit $\Delta\sigma_i=\sum_j C_{ij}\varepsilon_j$ yields the elastic tensor $C_{ij}$, the Voigt--Reuss--Hill bulk and shear moduli $K_\mathrm{VRH}$ and $G_\mathrm{VRH}$, and the Young's modulus $E$. (b)~Phonon dispersions: the structure is relaxed, the atomic forces $\mathbf{F}^{(n)}$ are evaluated for $N$ symmetry-selected atomic displacements $\Delta\mathbf{u}^{(n)}$ of a supercell,  and finite differences yield the second-order force constants $\Phi^{(2)}_{i\alpha,j\beta}$. Diagonalizing the dynamical matrix $D(\mathbf{q})$ gives the phonon dispersion and density of states, and the harmonic vibrational free energy $F_\mathrm{vib}(T)$, entropy $S(T)$, and heat capacity $C_V(T)$. In both workflows, the PES enters only through the relaxation and the stress or force evaluations; all other steps are independent of the PES provider but parameters, such as strain magnitudes, displacement amplitudes, and supercell sizes, can materially affect the result.}
    \label{fig:workflows}
\end{figure}

Regardless of the choice of PES, a central bottleneck remains: computing a materials property requires orchestrating a sequence of PES evaluations and post-processing analyses (Fig.~\ref{fig:workflows}). Obtaining the elastic tensor, for example, requires relaxing the structure, evaluating the stresses of a set of normal- and shear-strained structures, and fitting the stress--strain response (Fig.~\ref{fig:workflows}a). Obtaining the phonon dispersion, and from it the dynamical stability, vibrational entropy, and harmonic free energy, requires relaxing the structure, evaluating the forces on supercells with a symmetry-determined set of atomic displacements, and constructing and diagonalizing the dynamical matrix (Fig.~\ref{fig:workflows}b). Obtaining the lattice thermal conductivity additionally requires third-order force constants.\cite{togoDistributionsPhononLifetimes2015} This orchestration creates two problems. First, each workflow typically chains several packages, such as Python Materials Genomics (pymatgen),\cite{ongPythonMaterialsGenomics2013} the Atomic Simulation Environment (ASE),\cite{larsenAtomicSimulationEnvironment2017} and phonopy,\cite{togoFirstPrinciplesPhonon2015} that must be adapted for each new property or PES. Second, the computed property depends on workflow parameters, such as relaxation criteria, strain magnitudes, displacement amplitudes, and supercell sizes, that must be carefully calibrated and are rarely applied consistently across studies, making results from different studies difficult to compare. Agentic artificial intelligence (AI) built on large language models (LLMs) can reduce the human effort of setting up such calculations, but agents also make errors and produce inconsistent results across repeated prompts.\cite{parkExtensibleAgenticFramework2026,branAugmentingLargeLanguage2024,dawsonLARAValidationDrivenAgentic2026} There is therefore a need for a domain-expert-validated harness that ensures standardized, reproducible application of the relevant materials physics across repeated evaluations.

Evaluation consistency is especially important for materials property benchmarks used to compare different MLIPs. At present, the most widely used FP benchmark is Matbench-Discovery.\cite{riebesellMatbenchDiscoveryEvaluation2023} While Matbench-Discovery has served the community well as an early benchmark for FP development, its narrow focus on energetics and thermodynamic stability, only supplemented recently by lattice thermal conductivity metrics,\cite{Pota2024kSRME} has become limiting as FP performance on these metrics saturates. High performance on stability classification does not guarantee high performance across the broad spectrum of properties of interest to materials designers and scientists.\cite{Deng2025Softening,loewUniversalMachineLearning2025} Property-specific benchmarks have since emerged for phonons,\cite{loewUniversalMachineLearning2025} lattice thermal conductivity,\cite{Pota2024kSRME} PES softening,\cite{Deng2025Softening} and physical awareness and stability under extreme conditions,\cite{Chiang2025MLIPArena} but each adopts its own workflow, calculation parameters, and reference data, so that results are difficult to compare across properties and studies. All of these benchmarks, as well as multi-property platforms such as CHIPS-FF and LAMBench, compare FPs mainly with DFT reference data computed with PBE or another single functional. FPs trained on r$^2$SCAN data are now available,\cite{kaplan2025matpes,kunerMPALOE2025} but their evaluation against r$^2$SCAN references has been limited to equilibrium structures, energies, and forces,\cite{kaplan2025matpes,kunerMPALOE2025} and comparisons of FPs with experiment have been limited to individual properties, material classes, or models. No public benchmark yet tests whether the higher accuracy of r$^2$SCAN over PBE carries over to the properties predicted by FPs.

In this work, we introduce MatCalc (Materials Calculator), an open-source Python library that addresses both problems by providing a single, consistent interface for computing materials properties from any PES. MatCalc is built around three API classes: \texttt{PESCalculator}, which provides unified access to custom MLIPs and FPs as ASE \texttt{Calculator} objects; \texttt{PropCalc}, which implements property calculation workflows such as those in Fig.~\ref{fig:workflows} and executes their relaxations and MD simulations with either an ASE or a Large-scale Atomic/Molecular Massively Parallel Simulator (LAMMPS)\cite{Thompson2022LAMMPS} backend; and \texttt{Benchmark}, which evaluates a \texttt{PropCalc} against a curated reference dataset. Because a single architecture is now often trained on several foundation datasets, with substantially different results, MatCalc also defines a unified naming scheme for pre-trained FPs, \texttt{<Architecture>-\allowbreak<Dataset>\allowbreak[-<Functional>]-\allowbreak<Version>\allowbreak[-<Variant>]} (e.g., TensorNet-MatPES-PBE-2025.2, MACE-MPA-0-medium), which places the training dataset and reference functional on an equal footing with the architecture.

In the following sections, we first describe the MatCalc API and its usage. We then use MatCalc to compare FPs with five element-specific MLIPs on six elemental crystals,\cite{zuoPerformanceCostAssessment2020} and introduce MatCalc-Bench, a suite that covers 17 properties and comprises about 35,000 DFT reference calculations from 12 published data sources and this work, including r$^2$SCAN reference values for a subset of properties, and more than 2,000 experimental values. We find that the errors of FPs trained on OMat24 and MatPES lie within the range of the element-specific MLIPs for lattice parameters, elastic constants, and defect energetics, with a clear element-specific advantage only in equation-of-state fidelity. The training data, through their coverage of off-equilibrium configurations, their DFT settings, and their exchange-correlation functional, largely determine property-level accuracy, and no FP leads on more than a few properties. FPs trained on MPtrj, Alexandria, and OMat24, which mix PBE and PBE+$U$ energies following the Materials Project scheme, have errors of 3--6~eV in the adsorption energies of O-containing adsorbates on eight transition metals and of 2--5~J~m$^{-2}$ in metal/oxide interfacial energies, whereas FPs trained on the uniformly PBE MatPES data do not. Training on reference data computed with the r$^2$SCAN meta-generalized gradient approximation (meta-GGA) functional\cite{furnessAccurateNumericallyEfficient2020} rather than the Perdew--Burke--Ernzerhof (PBE) GGA functional\cite{perdewGeneralizedGradientApproximation1996} reduces the errors relative to experiment: for each of the three MatPES-trained architectures, the mean absolute percentage error (MAPE) of the elemental lattice parameters falls (from 0.58--1.07\% to 0.33--0.59\%), and the median overestimation of the volumetric thermal expansion coefficient falls from $+19\%$ to $+6.9\%$. Finally, failure rates, such as the fractions of migration paths that fail to converge and of MD trajectories that miss the diffusion observed in \textit{ab initio} MD, vary across FPs and are a dimension of model performance that is invisible to accuracy metrics computed only on successful calculations. MatCalc is extensible to new models through ASE compatibility and to new properties through the \texttt{PropCalc} interface, and we maintain a public leaderboard at \url{https://bench.matcalc.ai}.

\section{Results}

\subsection{API design}

\begin{figure}[htbp]
\centering
\includegraphics[width=\textwidth]{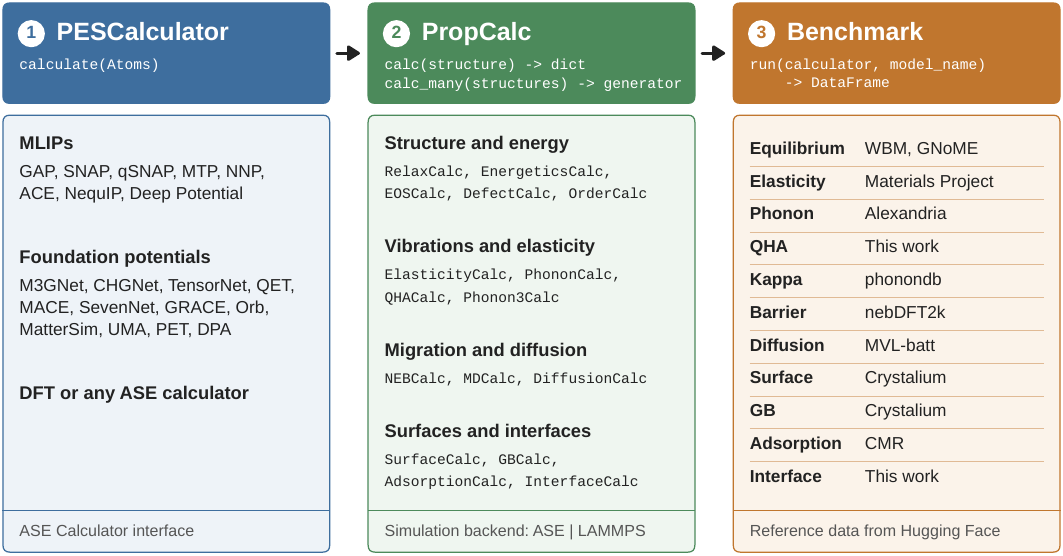}
\caption{\textbf{Overview of the MatCalc API design.} (1)~\texttt{PESCalculator} provides unified access to the PES: foundation potentials are instantiated from canonical \texttt{<Architecture>-\allowbreak<Dataset>\allowbreak[-<Functional>]-\allowbreak<Version>\allowbreak[-<Variant>]} names; local-environment MLIPs are supported through the \texttt{maml} package's LAMMPS-backed evaluators or their native ASE calculators; and DFT and other codes are supported through the ASE \texttt{Calculator} interface. (2)~\texttt{PropCalc} subclasses implement property workflows behind a uniform \texttt{calc(structure) $\rightarrow$ dict} contract, with the underlying potential as the sole model-specific constructor argument, and execute their relaxations and MD simulations with either the ASE or the LAMMPS simulation backend; \texttt{calc\_many} parallelizes evaluation over structure collections. (3)~Each \texttt{Benchmark} binds a property calculator to a curated DFT reference dataset, retrieved from Hugging Face and cached automatically, and its \texttt{run} method returns a \texttt{pandas} DataFrame of reference values and model predictions with built-in checkpointing.}
\label{fig:matcalc_architecture}
\end{figure}

The MatCalc API comprises three primary classes (Fig.~\ref{fig:matcalc_architecture}):
\begin{itemize}
    \item \texttt{PESCalculator} is a subclass of the ASE \texttt{Calculator} class that provides PES evaluations. It also implements factory methods for initializing popular PES providers. Local-environment MLIPs, such as the Gaussian Approximation Potential (GAP)\cite{Bartok2010GAP}, Spectral Neighbor Analysis Potential (SNAP)\cite{Thompson2015SNAP}, Moment Tensor Potential (MTP)\cite{Shapeev2016MTP}, and Behler--Parrinello Neural Network Potential (NNP)\cite{Behler2007NNP}, are supported via the LAMMPS-backed evaluators in the \texttt{maml} package\cite{chenMamlMaterialsMachine2021,zuoPerformanceCostAssessment2020}; Atomic Cluster Expansion (ACE)\cite{Drautz2019ACE} potentials are supported through \texttt{pyace},\cite{lysogorskiyPerformantImplementationAtomic2021} and NequIP\cite{batznerEquivariantGraphNeural2022} and DeePMD-kit\cite{zengDeePMDkitV2Software2023} models through their native ASE calculators. FPs are loaded through \texttt{load\_fp}, which dispatches to the implementation package of each supported architecture, such as M3GNet\cite{chenUniversalGraphDeep2022}, CHGNet\cite{dengCHGNetPretrainedUniversal2023}, TensorNet\cite{Simeon2023TensorNet}, QET\cite{ko2026}, MACE\cite{Batatia2022MACE}, GRACE\cite{bochkarevGraphAtomicCluster2024a}, SevenNet\cite{park2024}, Orb\cite{neumann2024}, PET\cite{pozdnyakovSmoothExactRotational2023}, and UMA\cite{Wood2025UMA} (Supplementary Table~\ref{si:tab-supported}). FPs registered in MatCalc are loaded by canonical names of the form \texttt{<Architecture>-\allowbreak<Dataset>\allowbreak[-<Functional>]-\allowbreak<Version>\allowbreak[-<Variant>]} (e.g., TensorNet-MatPES-r2SCAN-2025.2-m) where the upstream release provides these fields. Unregistered models from a supported package can be loaded through a provider override of \texttt{load\_fp}, and any other ASE \texttt{Calculator}, including interfaces to DFT codes, can be used directly. New models can be added to MatCalc by the community through additional factory methods or registry entries.
    \item \texttt{PropCalc} is the central abstract base class for workflows that convert PES predictions on an input structure (pymatgen \texttt{Structure}, ASE \texttt{Atoms}, or the output dictionary of an upstream \texttt{PropCalc}) into a property prediction. The output is always a Python dictionary that contains the computed properties with metadata such as units. The \texttt{PropCalc} subclasses are listed in Table~\ref{tab:propcalc_summary}. A \texttt{calc\_many} method parallelizes property calculations over a collection of structures using the \texttt{joblib} package. Analogous to the \texttt{Pipeline} class in scikit-learn,\cite{Pedregosa2011sklearn} the \texttt{ChainedCalc} subclass composes multiple \texttt{PropCalc} instances into a sequential pipeline in which each stage receives the accumulated results of the preceding stages. A common use case is performing a structure relaxation followed by multiple property evaluations (e.g., energetics, phonons, and elasticity) at the relaxed geometry without repeating the relaxation step.
    \item Finally, \texttt{Benchmark} is an abstract base class for evaluating a \texttt{PESCalculator} against reference data through a \texttt{PropCalc}. Each \texttt{Benchmark} subclass binds a property calculator to a curated DFT reference dataset. For example, \texttt{ElasticityBenchmark} runs \texttt{ElasticityCalc} on 12,122 Materials Project structures and compares the results with the PBE elastic moduli, and its \texttt{run} method returns a \texttt{pandas}\cite{reback2020pandas} DataFrame with reference values and model predictions side by side. Reference datasets are hosted in the \texttt{materialyze/matcalc-bench} dataset repository on Hugging Face and are downloaded and cached automatically on first use. The \texttt{run} method saves a checkpoint at a user-specified interval (by default, every 1,000 structures), so that an interrupted run resumes from the last checkpoint rather than from the beginning. Running a full comparison across several FPs at once is handled by \texttt{BenchmarkSuite}, which takes a dictionary of named \texttt{PESCalculator} objects together with a list of benchmarks and returns one DataFrame per benchmark, in which the predictions of all models are joined to the reference values.

\end{itemize}

Below the \texttt{PropCalc} interface, the relaxations and MD simulations of every workflow are executed by one of two interchangeable simulation backends. The default ASE backend runs relaxations with ASE optimizers and cell filters and MD with ASE integrators, and therefore works with any ASE \texttt{Calculator}. The LAMMPS backend runs single-point evaluations, relaxations, and MD in an in-process LAMMPS instance, so that the entire optimization or MD loop executes in compiled code rather than in a Python loop that exchanges positions and forces with the PES model at every step. Under this backend, MatGL TensorNet and M3GNet potentials\cite{Ko2025MatGL} are exported to TorchScript and evaluated natively through MatGL's \texttt{pair\_style matgl} (or its Kokkos-accelerated variant, \texttt{matgl/kk}), \texttt{maml} potentials (GAP, MTP, NNP, and SNAP) through their LAMMPS pair styles, and any other LAMMPS force field through the \texttt{LAMMPSCalculator} class. Because the backend is resolved below the \texttt{PropCalc} interface, the same workflow code and parameters apply to both, and the MatCalc test suite runs every \texttt{PropCalc} under both backends to verify their parity.

\subsection{Usage}

To illustrate how the \texttt{PropCalc} abstraction of Fig.~\ref{fig:matcalc_architecture} translates into practice, consider the task of computing the Voigt--Reuss--Hill elastic moduli of diamond-cubic Si from an FP. Loading the potential and instantiating \texttt{ElasticityCalc} reduces the task to a few lines:

\begin{lstlisting}
import matcalc as mtc
from pymatgen.ext.matproj import MPRester

with MPRester() as mpr:
    si = mpr.get_structure_by_material_id("mp-149")

calc = mtc.load_fp("TensorNet-MatPES-r2SCAN-2025.2")
props = mtc.ElasticityCalc(calc, relax_structure=True, units_GPa=True).calc(si)
print(f"K_VRH = {props['bulk_modulus_vrh']:.1f} GPa")
print(f"G_VRH = {props['shear_modulus_vrh']:.1f} GPa")
\end{lstlisting}

\texttt{ElasticityCalc} relaxes the input structure, generates normal and shear deformations at multiple strain amplitudes (by default, normal strains of $\pm 0.005$ and $\pm 0.01$ and shear strains of $\pm 0.03$ and $\pm 0.06$), and evaluates the stress of each deformed configuration. Linear fits to the stress--strain responses give the elastic tensor, from which the Voigt--Reuss--Hill bulk and shear moduli and the Young's modulus are calculated. The returned dictionary contains these properties and their unit annotations under \texttt{\_units}; with \texttt{units\_GPa=True}, moduli are reported in GPa rather than in the native pymatgen unit of eV~\AA$^{-3}$. The same property workflow applies to FPs, custom MLIPs, or a DFT calculator wrapped in an ASE \texttt{Calculator}, which can be loaded as follows:

\begin{lstlisting}
import matcalc as mtc

# FPs: alias or unregistered model
calc = mtc.load_fp("uma")
calc = mtc.load_fp("grace", model="GRACE-2L-SMAX-OMAT-large-fp64")

# Custom MLIPs: MatGL, NequIP or MTP
calc = mtc.PESCalculator.load_matgl("path/to/matgl_model")
calc = mtc.PESCalculator.load_nequip("path/to/nequip_model")
calc = mtc.PESCalculator.load_mtp("path/to/mtp_model", elements=["Si"])

# Any other ASE calculator: DFT, Tersoff, eSEN
from ase.calculators.vasp import Vasp
calc = Vasp(xc="PBE", encut=520, kpts=(8, 8, 8))

from ase.calculators.tersoff import Tersoff
calc = Tersoff.from_lammps("Si.tersoff")

from fairchem.core import OCPCalculator
calc = OCPCalculator(model_name="eSEN-30M-OAM", local_cache=".")
\end{lstlisting}

Individual calculators compose into multi-stage pipelines through \texttt{ChainedCalc}. Each stage receives the accumulated output dictionary from the preceding stages, allowing multiple properties to be evaluated from a shared relaxed structure. Setting \texttt{relax\_structure=False} in the downstream calculators ensures that the relaxation is performed only once:

\begin{lstlisting}
pipeline = mtc.ChainedCalc([
    mtc.RelaxCalc(calc, fmax=0.01, relax_cell=True),
    mtc.ElasticityCalc(calc, relax_structure=False),
    mtc.PhononCalc(calc, relax_structure=False),
])
results = pipeline.calc(si)
\end{lstlisting}

The same workflows generalize directly to collections of structures via \texttt{calc\_many}, which dispatches the underlying \texttt{calc} method over an arbitrary iterable with optional \texttt{joblib} parallelization and returns a generator, so that results can be consumed as they complete. With \texttt{allow\_errors=True}, a failed calculation yields \texttt{None} rather than aborting the batch. This is the default for all \texttt{Benchmark} runs:

\begin{lstlisting}
# structures: any iterable of pymatgen Structure or ASE Atoms objects
eos = mtc.EOSCalc(calc, relax_structure=True)
results = list(eos.calc_many(structures, n_jobs=16))
\end{lstlisting}

On a single NVIDIA A100-SXM4-40GB GPU, this workflow completes approximately 5,000 equation-of-state calculations per hour with TensorNet-MatPES-PBE-2025.2 for a typical distribution of unit-cell sizes (5--30 atoms) using the above settings with the ASE backend.

Benchmarking follows the same pattern. The example below evaluates two PBE-trained FPs, consistent with the PBE reference data, on random 100-material subsets of the elasticity and phonon benchmarks:

\begin{lstlisting}
import matcalc as mtc
from matcalc import benchmark as bm

suite = bm.BenchmarkSuite([
    bm.ElasticityBenchmark(n_samples=100),
    bm.PhononBenchmark(n_samples=100),
])
elastic_df, phonon_df = suite.run({
    "TensorNet-MatPES-PBE": mtc.load_fp("TensorNet-MatPES-PBE-2025.2"),
    "MACE-MatPES-PBE": mtc.load_fp("MACE-OMat24-ft-MatPES-PBE-0"),
})
\end{lstlisting}

All benchmark results in the following section were obtained with MatCalc and its ASE backend, using the settings given in Methods.

\subsection{Benchmarks}

We now use MatCalc to benchmark FPs in the prediction of materials properties. The main panel comprises seven PBE-trained FPs: MACE-MPtrj, MACE-MPA, MACE-OMat24, GRACE-OMat24, MACE-MatPES-PBE, TensorNet-MatPES-PBE, and QET-MatPES-PBE.\cite{Batatia2025MACEMP,Simeon2023TensorNet,Lysogorskiy2026GRACE,kaplan2025matpes,ko2026} Their names in the model registry are given in Methods. The panel spans four training datasets (MPtrj\cite{dengCHGNetPretrainedUniversal2023}, MPA\cite{dengCHGNetPretrainedUniversal2023,schmidt2024alexandria,barroso-luqueOpenMaterials20242024}, OMat24\cite{barroso-luqueOpenMaterials20242024}, and MatPES-PBE\cite{kaplan2025matpes}) and four architectures (MACE\cite{Batatia2022MACE}, TensorNet\cite{Simeon2023TensorNet}, QET\cite{ko2026}, and GRACE\cite{bochkarevGraphAtomicCluster2024a}). QET adds a linear-scaling charge equilibration block to TensorNet, which enables it to capture non-local charge transfer. The panel is supplemented by three r$^2$SCAN-trained variants, MACE-MatPES-r2SCAN, TensorNet-MatPES-r2SCAN, and QET-MatPES-r2SCAN, for a subset of benchmarks. Evaluating MACE trained on different datasets isolates the effect of training data, while the groups MACE-OMat24/GRACE-OMat24, MACE-MatPES-PBE/TensorNet-MatPES-PBE/QET-MatPES-PBE, and MACE-MatPES-r2SCAN/TensorNet-MatPES-r2SCAN/QET-MatPES-r2SCAN share training data and isolate the effects of architecture and training procedure. Because the MACE-MatPES FPs were fine-tuned from MACE-OMat24, their differences from the TensorNet and QET FPs also reflect the OMat24 pretraining. The panel is not intended to be exhaustive; other FPs, such as UMA, Orb, SevenNet, MatterSim, and PET, can be evaluated with the same code. Comparisons against DFT are meaningful only when the reference data and the FP training data use the same exchange-correlation functional and, preferably, the same computational parameters; we therefore compare PBE-trained FPs with PBE references, and r$^2$SCAN-trained FPs with r$^2$SCAN references. Strictly, MPtrj, Alexandria, and OMat24 are not pure PBE datasets: following the Materials Project, they apply an empirical Hubbard $U$ correction to V, Cr, Mn, Fe, Co, Ni, Mo, and W in oxides and fluorides,\cite{jainFormationEnthalpiesMixing2011} whereas MatPES-PBE applies no $U$. We nevertheless refer to all FPs trained on these datasets as PBE-trained. Where the reference data allow, each FP is compared with DFT values computed at the settings of its training data, including $U$; the adsorption and interface references are PBE without $U$, and the effect of $U$ on the adsorption energies is assessed separately (Supplementary Section~\ref{si:ads-pbeu}). All ten FPs are additionally compared with experimental references where available.

The benchmarking is conducted in two parts. We first compare the FPs with element-specific MLIPs for six elemental systems to assess the accuracy gap between universal and system-specific models. We then evaluate the FPs across a broad range of materials and properties, which we will collectively call MatCalc-Bench. MatCalc-Bench is intended to serve as a more comprehensive evaluation of FP performance across properties of relevance to materials design, from the PES minimum and its curvature to saddle points, from ground-state to finite-temperature behavior, and from bulk to low-coordination and heterogeneous environments.

\subsubsection{Elemental crystals}

The benchmark of \citet{zuoPerformanceCostAssessment2020} covers six elemental crystals (fcc Ni and Cu, bcc Li and Mo, and diamond-cubic Si and Ge) and five MLIP architectures (GAP, MTP, NNP, SNAP, and quadratic SNAP (qSNAP)). These element-specific MLIPs provide a stringent baseline as each is fitted to a single element and to configurations of the types probed below, whereas an FP describes the entire periodic table with one set of parameters. All models were evaluated with an identical \texttt{ChainedCalc} pipeline, in which \texttt{RelaxCalc} first relaxes the structure and \texttt{ElasticityCalc}, \texttt{DefectCalc}, \texttt{NEBCalc}, and \texttt{EOSCalc} then operate on the relaxed result (Methods). Errors were computed against the DFT-PBE values.\cite{zuoPerformanceCostAssessment2020}

Figure~\ref{fig:elemental_scorecard} provides a summary of the average errors, and per-element errors are given in Supplementary Figs.~\ref{si:elem-c11-c12}--\ref{si:elem-ea-deos} and Table~\ref{si:tab-elemental-lattice}. Among the element-specific MLIPs, MTP generally yields the lowest errors, and we therefore use it as the best-case reference for the FPs.

MACE-MPtrj is by far the worst-performing model, especially for the elastic constants ($C_{ij}$), migration energies ($E_m$) and activation energies ($E_a$). Most of its errors are underestimates: MACE-MPtrj underestimates $E_v$ and $E_m$ for all six elements ($E_v$ by 1.4 and 1.5~eV for Si and Ge) and $C_{11}$ and $C_{44}$ for five of the six ($C_{11}$ of Mo by 128~GPa). We attribute this systematic underestimation to the composition of MPtrj, which consists mostly of near-equilibrium structures from DFT relaxations and therefore constrains the PES mainly near its minima. This attribution is consistent with the systematic PES softening reported for FPs trained on Materials Project relaxation trajectories, which underpredict the curvature of the PES and the energies of configurations away from equilibrium.\cite{Deng2025Softening,kaplan2025matpes} For the same reason, MACE-MPtrj performs well in the equation-of-state deviation $\Delta_\mathrm{EOS}$, which probes only the near-equilibrium PES.

The four FPs trained on OMat24 and MatPES (MACE-OMat24, GRACE-OMat24, MACE-MatPES-PBE, and TensorNet-MatPES-PBE) are competitive with the element-specific MLIPs. MACE-OMat24 and MACE-MatPES-PBE have larger errors in the lattice parameter $a$ (MAPEs of 0.48\% and 0.59\%, against 0.25\% for MTP), whereas GRACE-OMat24 and TensorNet-MatPES-PBE (0.14\% and 0.27\%) match MTP (Supplementary Table~\ref{si:tab-elemental-lattice}). The errors of these four FPs in $C_{ij}$, $E_v$, $E_m$ and $E_a$ are generally higher than those of the best performing MTPs, but still within the range of errors among all the element-specific MLIPs. QET-MatPES-PBE, trained on the same data as TensorNet-MatPES-PBE, has the lowest $C_{44}$ error of all models (4.3~GPa) and a low $E_m$ error (61~meV), but a large $E_v$ error (645~meV). All five FPs perform poorly on $\Delta_\mathrm{EOS}$, suggesting that the broader configurational coverage of OMat24 and MatPES comes at some cost to the fidelity of the near-equilibrium PES. This advantage of the element-specific MLIPs, reflecting their dense near-equilibrium strain coverage, is confined to the $\pm6\%$ volume window over which $\Delta_\mathrm{EOS}$ is defined.\cite{lejaeghere2014} Under larger compression, the element-specific MLIPs tend to deviate more from DFT than the FPs (Supplementary Fig.~\ref{si:eos-full}), consistent with training sets that contain strains of at most $\pm10\%$ along single deformation modes.\cite{zuoPerformanceCostAssessment2020}

\begin{figure}[htbp]
\centering
\includegraphics[width=0.8\textwidth]{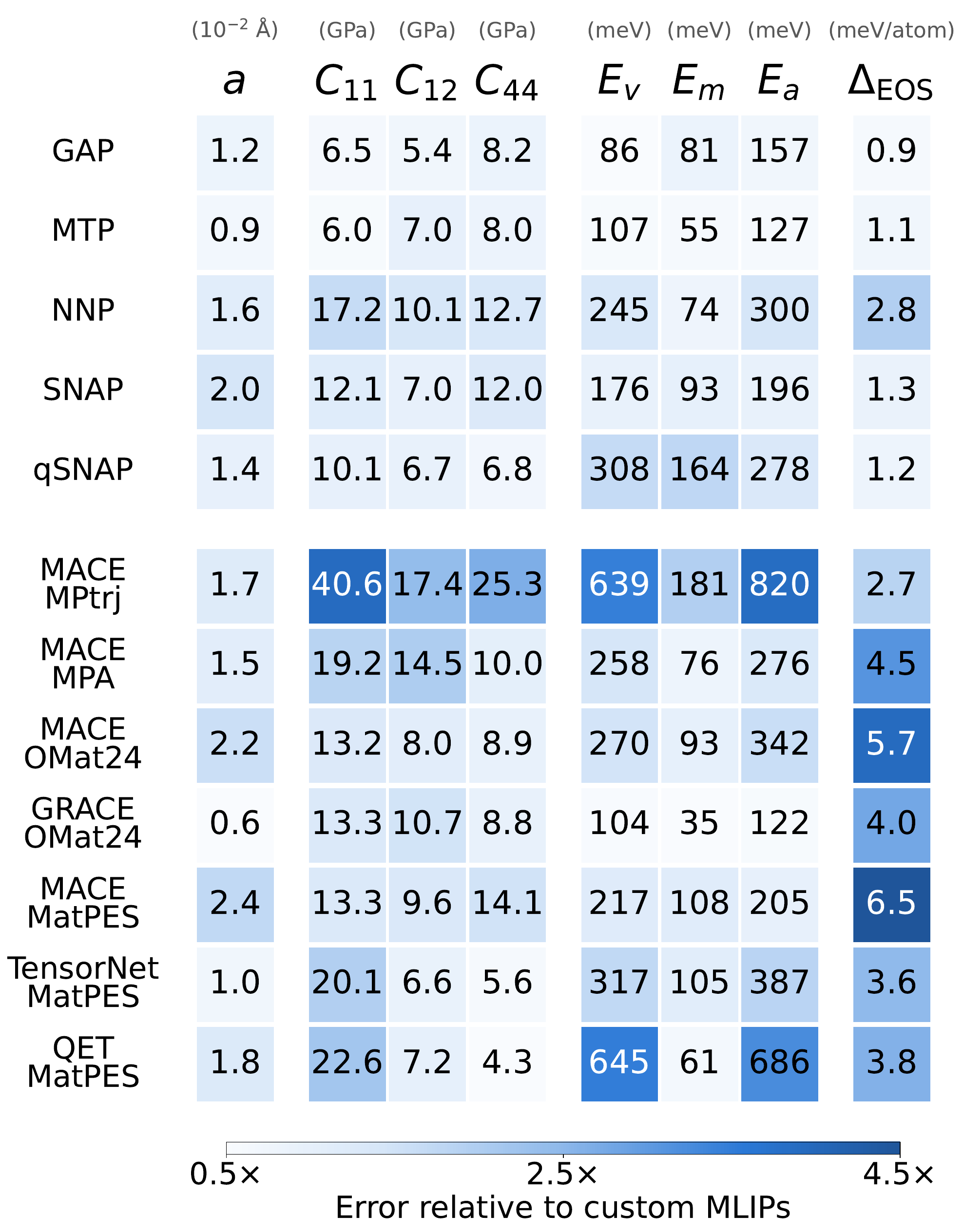}
\caption{\textbf{Accuracy of foundation potentials (FPs) relative to element-specific MLIPs for six elemental crystals.} Each cell gives the mean absolute error of a model over Ni and Cu (fcc), Li and Mo (bcc), and Si and Ge (diamond) for the lattice parameter $a$, the elastic constants $C_{11}$, $C_{12}$, and $C_{44}$, the vacancy formation energy $E_v$, the vacancy migration barrier $E_m$, the activation energy $E_a = E_v + E_m$, and the equation-of-state deviation $\Delta_\mathrm{EOS}$. Colors show the error divided by the mean error of the five element-specific MLIPs for the same property. All FPs shown are PBE-trained; MatPES refers to MatPES-PBE. References are the DFT-PBE values of \citet{zuoPerformanceCostAssessment2020}. Per-element errors are given in Supplementary Figs.~\ref{si:elem-c11-c12}--\ref{si:elem-ea-deos}.}
\label{fig:elemental_scorecard}
\end{figure}

Agreement with DFT does not necessarily imply agreement with experiment. Relative to experimental 0~K lattice parameters (room-temperature values corrected for thermal expansion\cite{lejaeghere2014}), the PBE-trained models overestimate $a$ by a median of $+0.8\%$, whereas the three r$^2$SCAN-trained FPs deviate by only $-0.2\%$, $+0.1\%$, and $-0.2\%$ (Supplementary Table~\ref{si:tab-elemental-lattice}). The overestimation by the PBE-trained models mirrors the well-known underbinding of the PBE functional,\cite{haasCalculationLatticeConstant2009} and training on r$^2$SCAN data reduces it.

\subsubsection{MatCalc-Bench}

The preceding benchmark shows that FPs approach the accuracy of element-specific MLIPs for simple metallic and covalent elemental crystals. However, FPs are intended for the full compositional and structural diversity of inorganic materials, which requires evaluation across thousands of materials and many property classes. MatCalc-Bench evaluates 17 properties, comprising about 35,000 DFT reference calculations from 12 published data sources and this work, and more than 2,000 experimental values. These properties are arranged along two axes. The first is how far a property probes the PES beyond its minimum: the minimum itself (relaxed structures and formation energies), its curvature (elastic moduli and heat capacities), the volume dependence of the curvature (thermal expansion coefficients and isothermal bulk moduli), its third derivatives (lattice thermal conductivities), and configurations far from the minimum, at saddle points (migration barriers) and along finite-temperature trajectories (ionic conductivities). The second is the atomic environment, from the bulk structures that make up most of the training data to atoms with reduced or distorted coordination at surfaces and grain boundaries, and to atoms where chemically different components meet, as for adsorbates on surfaces and at interfaces between different materials. Table~\ref{tab:overview} summarizes the errors of the PBE-trained FPs, and the results are presented first along the first axis for bulk materials and then along the second. For most properties, the lowest MAE is not significantly different from the second lowest (Supplementary Table~\ref{si:table1-ci}), so we compare groups of FPs rather than rank individual models.

\begin{table}[htbp]
\centering
\caption{\textbf{Performance of PBE-trained foundation potentials across MatCalc-Bench.} Values are mean absolute errors (MAEs) against PBE reference data, except for $\bar{d}_\mathrm{fp}$, the mean CrystalNN fingerprint distance between FP-relaxed and DFT-relaxed structures. $\bar{d}_\mathrm{fp}$ and $E_\mathrm{f}^{0}$ are averaged over five runs with different random perturbations to reduce sampling noise. Thermal properties ($C_V$, $\alpha_V$, $K_T$, and $\kappa$) are at 300\,K. The thermal and ionic conductivities are compared on a logarithmic scale ($\kappa$ in W\,m$^{-1}$\,K$^{-1}$, $\sigma$ in mS\,cm$^{-1}$). $E_\mathrm{m}$ and $\log_{10}\sigma$ are computed over the calculations of each model that did not fail. Bold: lowest value in each row; for most properties, it is not significantly different from the second lowest. The $\gamma_\mathrm{surf}$ MAE of MACE-MPA is dominated by negative surface energies of Y (see text). Reference data and protocols are given in Supplementary Table~\ref{si:bench-tasks} and Methods, the elements covered by each benchmark in Supplementary Fig.~\ref{si:element-coverage}, and the 95\% bootstrap confidence intervals in Supplementary Table~\ref{si:table1-ci}.}
\label{tab:overview}
\footnotesize
\setlength{\tabcolsep}{3pt}
\renewcommand{\arraystretch}{1.1}
\resizebox{\textwidth}{!}{%
\begin{tabular}{c ccccccc}
\toprule
\multirow{2}{*}{\textbf{Property}} & \multicolumn{1}{c}{MPtrj} & \multicolumn{1}{c}{MPA} & \multicolumn{2}{c}{OMat24} & \multicolumn{3}{c}{MatPES-PBE} \\
\cmidrule(lr){2-2}\cmidrule(lr){3-3}\cmidrule(lr){4-5}\cmidrule(lr){6-8}
 & MACE & MACE & MACE & GRACE & MACE & TensorNet & QET \\
\midrule
\multicolumn{8}{l}{\textit{Relaxation and elasticity}} \\ \addlinespace[2pt]
$\bar{d}_\mathrm{fp}$ & 0.39 & 0.26 & 0.24 & \textbf{0.22} & 0.27 & 0.31 & 0.32 \\
$E_\mathrm{f}^{0}$ (meV\,atom$^{-1}$) & 78 & 34 & 36 & \textbf{31} & 53 & 56 & 78 \\
$K_\mathrm{VRH}$ (GPa) & \textbf{16} & 19 & 17 & \textbf{16} & 18 & 20 & 18 \\
$G_\mathrm{VRH}$ (GPa) & 24 & 17 & 17 & 19 & \textbf{14} & 16 & 16 \\
\addlinespace[4pt] \multicolumn{8}{l}{\textit{Vibration and thermodynamics}} \\ \addlinespace[2pt]
$C_V$ (J\,mol-atom$^{-1}$\,K$^{-1}$) & 0.64 & 0.36 & 0.23 & \textbf{0.18} & 0.60 & 0.67 & 0.52 \\
$\alpha_V$ (ppm\,K$^{-1}$) & 18 & \textbf{11} & 13 & 13 & \textbf{11} & 12 & 15 \\
$K_T$ (GPa) & 24 & 25 & 44 & 19 & 26 & 19 & \textbf{18} \\
$\log_{10}\kappa$ & 0.20 & 0.09 & 0.05 & \textbf{0.03} & 0.14 & 0.19 & 0.16 \\
\addlinespace[4pt] \multicolumn{8}{l}{\textit{Migration and diffusion}} \\ \addlinespace[2pt]
$E_\mathrm{m}$ (eV) & 0.16 & 0.16 & 0.16 & 0.16 & 0.22 & 0.18 & \textbf{0.15} \\
$\log_{10}\sigma$ & 0.37 & 0.19 & \textbf{0.17} & 0.19 & 0.20 & 0.20 & 0.22 \\
\addlinespace[4pt] \multicolumn{8}{l}{\textit{Surfaces and grain boundaries}} \\ \addlinespace[2pt]
$\gamma_\mathrm{surf}$ (J\,m$^{-2}$) & 0.23 & 52.66 & \textbf{0.12} & \textbf{0.12} & 0.14 & 0.20 & 0.25 \\
$\gamma_\mathrm{GB}$ (J\,m$^{-2}$) & 0.24 & \textbf{0.06} & 0.07 & 0.09 & 0.07 & 0.07 & 0.09 \\
\addlinespace[4pt] \multicolumn{8}{l}{\textit{Adsorption and interfaces}} \\ \addlinespace[2pt]
$E_\mathrm{ads}$ (eV) & 0.86 & 0.93 & 0.99 & 1.10 & 0.26 & \textbf{0.24} & 0.36 \\
$\gamma_\mathrm{int}$ (J\,m$^{-2}$) & 2.25 & 3.40 & 4.70 & 4.89 & 0.24 & \textbf{0.20} & 0.22 \\
$W_\mathrm{adh}$ (J\,m$^{-2}$) & 2.06 & 3.16 & 4.47 & 4.71 & \textbf{0.17} & 0.19 & 0.29 \\
\bottomrule
\end{tabular}}
\end{table}

\paragraph{Relaxation and elastic properties.}

As shown in Table~\ref{tab:overview}, the OMat24-trained FPs recover the PBE-relaxed structures most reliably, with mean fingerprint distances $\bar{d}_\mathrm{fp}$ of 0.22 (GRACE-OMat24) and 0.24 (MACE-OMat24), trailed by MatPES and MPA models (0.26--0.32), whereas MACE-MPtrj shows the largest deviation (0.39). Formation energies $E_\mathrm{f}^{0}$ follow a similar order, with MAEs of 31--36\,meV\,atom$^{-1}$ for the MPA and OMat24 models, 53 and 56\,meV\,atom$^{-1}$ for MACE-MatPES-PBE and TensorNet-MatPES-PBE, and 78\,meV\,atom$^{-1}$ for MACE-MPtrj and QET-MatPES-PBE (see Supplementary Fig.~\ref{si:wbm-ef-ptable} for the errors by element). These comparisons depend on the consistency between the reference and the training data. Two references are available for this benchmark: the original WBM data at MP settings,\cite{wangPredictingStableCrystalline2021} and a re-relaxation with the MatPES settings,\cite{kaplan2025matpes} which use a higher plane-wave cutoff (680 instead of 520\,eV), a denser $k$-point mesh, updated PAW pseudopotentials, and no Hubbard $U$ correction. In Table~\ref{tab:overview}, the MatPES-trained FPs are compared with the recalculated reference and all other FPs with the original one. Against the tighter reference, the $E_\mathrm{f}^{0}$ MAEs of the MPA and OMat24 models rise from 30--35 to 48--56\,meV\,atom$^{-1}$ and that of MACE-MPtrj from 77 to 84\,meV\,atom$^{-1}$ on the structures common to both references, whereas $\bar{d}_\mathrm{fp}$ changes by at most 0.02 (Supplementary Table~\ref{si:wbm-references}). Against the same reference, the MPA and OMat24 FPs (48--56\,meV\,atom$^{-1}$) and MACE-MatPES-PBE and TensorNet-MatPES-PBE (53 and 56\,meV\,atom$^{-1}$) are therefore indistinguishable, and the larger $E_\mathrm{f}^{0}$ MAEs of the MatPES-trained FPs in Table~\ref{tab:overview} reflect the difference in reference settings rather than in model accuracy.

Among the r$^2$SCAN-trained FPs, MACE-MatPES-r2SCAN recovers the r$^2$SCAN-relaxed structures most reliably ($\bar{d}_\mathrm{fp}$ of 0.16, against 0.21 for TensorNet-MatPES-r2SCAN and 0.23 for QET-MatPES-r2SCAN). MACE-MatPES-r2SCAN and TensorNet-MatPES-r2SCAN give similar $E_\mathrm{f}^{0}$ MAEs (44 and 45\,meV\,atom$^{-1}$), and QET-MatPES-r2SCAN a slightly higher one (62\,meV\,atom$^{-1}$).

For elastic moduli, GRACE-OMat24 and MACE-MPtrj achieve the lowest $K_\mathrm{VRH}$ MAE (16\,GPa), although all seven FPs lie within 16--20\,GPa, and MACE-MatPES-PBE the lowest $G_\mathrm{VRH}$ MAE (14\,GPa). MACE-MPtrj has the highest $G_\mathrm{VRH}$ MAE (24\,GPa), compared with 14--19\,GPa for the other FPs. It underestimates $G_\mathrm{VRH}$ for most materials (Supplementary Fig.~\ref{si:elastic-parity}), which suggests that the relaxation trajectories in MPtrj undersample shear-deformed configurations.

\paragraph{Vibrational and thermodynamic properties.}

GRACE-OMat24 achieves the lowest MAEs for both the harmonic constant-volume heat capacity $C_V$ and the anharmonic lattice thermal conductivity $\kappa$ at 300\,K (0.18\,J\,mol-atom$^{-1}$\,K$^{-1}$ and 0.03 for $\log_{10}\kappa$), followed by MACE-OMat24 on $C_V$ (0.23\,J\,mol-atom$^{-1}$\,K$^{-1}$) and on $\log_{10}\kappa$ (0.05), whereas MACE-MPtrj shows the largest $\log_{10}\kappa$ error (0.20). Among the MACE models, MACE-OMat24 also lowers the $C_V$ MAE of MACE-MPtrj and MACE-MPA (0.64 and 0.36\,J\,mol-atom$^{-1}$\,K$^{-1}$) and their $\log_{10}\kappa$ MAEs (0.20 and 0.09); because these models share the same architecture, the improvements can be attributed to the training data. The MatPES-trained FPs have higher MAEs than the OMat24-trained FPs for both properties (0.52--0.67\,J\,mol-atom$^{-1}$\,K$^{-1}$ for $C_V$ and 0.14--0.19 for $\log_{10}\kappa$). A possible reason is that the finite-displacement calculations of these properties probe small displacements around equilibrium, which are densely sampled by the relaxation trajectories of rattled structures in OMat24 but less so by the molecular dynamics snapshots in MatPES. The r$^2$SCAN-trained FPs reproduce their r$^2$SCAN reference values with errors similar to those of the corresponding PBE-trained FPs against PBE (MAEs of 0.13--0.43 against 0.24--0.40\,J\,mol-atom$^{-1}$\,K$^{-1}$ for $C_V$ and of 0.12--0.18 against 0.15--0.17 for $\log_{10}\kappa$; Supplementary Section~\ref{si:cv-r2scan}).

The quasi-harmonic volumetric thermal expansion coefficient $\alpha_V$ and isothermal bulk modulus $K_T$ show a different trend: MACE-MatPES-PBE and MACE-MPA achieve the lowest $\alpha_V$ MAE (11\,ppm\,K$^{-1}$) and QET-MatPES-PBE the lowest $K_T$ MAE (18\,GPa), followed closely by TensorNet-MatPES-PBE and GRACE-OMat24 (19\,GPa), whereas MACE-MPtrj shows the largest $\alpha_V$ error (18\,ppm\,K$^{-1}$). With the same architecture, MACE-MatPES-PBE reduces the $\alpha_V$ and $K_T$ MAEs of MACE-OMat24 (13\,ppm\,K$^{-1}$ and 44\,GPa), possibly because the MatPES snapshots, taken from constant-pressure molecular dynamics at 300\,K, sample the thermally displaced and thermally expanded configurations at the temperature at which $\alpha_V$ and $K_T$ are evaluated.

\paragraph{Migration and diffusion properties.}

Most of the PBE-trained FPs reach similar accuracy on the Li-ion migration barrier $E_\mathrm{m}$ and the ionic conductivity $\log_{10}\sigma$, with MAEs of 0.15--0.18\,eV and 0.17--0.22, respectively (Table~\ref{tab:overview}). These MAEs are computed only over the NEB calculations that converged and the MD simulations in which diffusion was measured, and therefore do not show how often a calculation fails.

\begin{figure}[htbp]
\centering
\includegraphics[width=\textwidth]{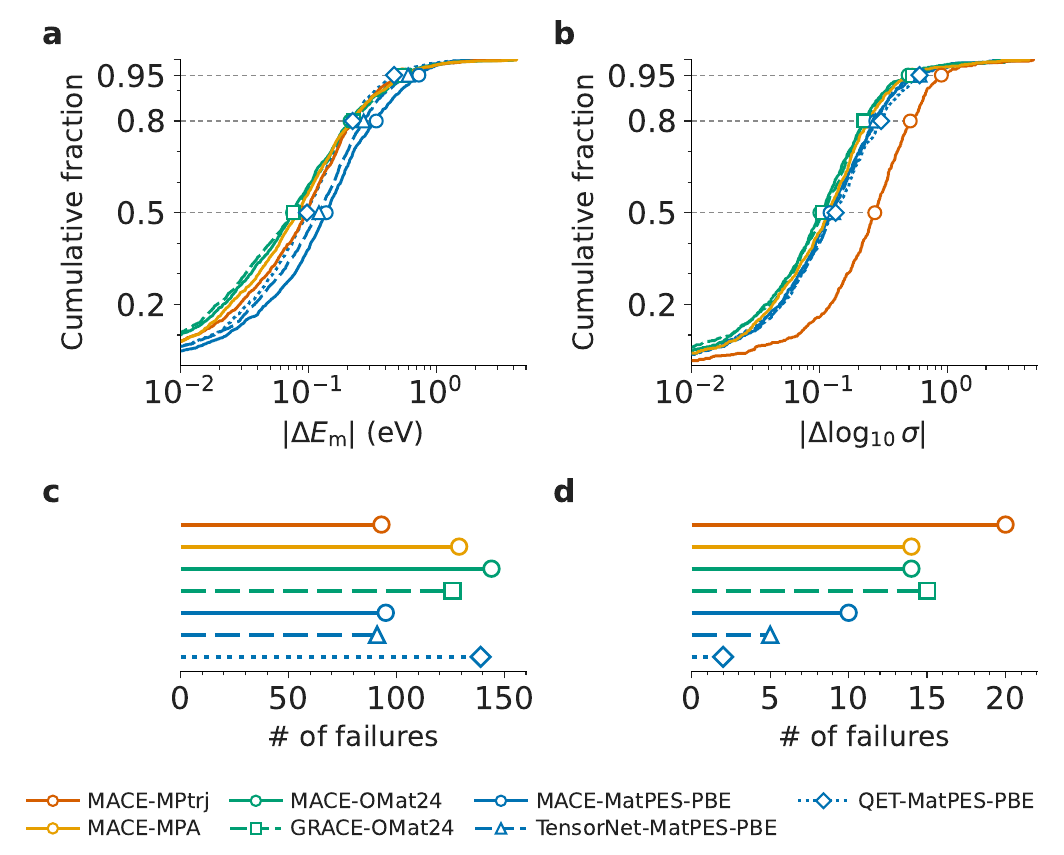}
\caption{\textbf{Migration barriers and ionic conductivities from PBE-trained foundation potentials.} (a, b)~Cumulative fraction of samples as a function of the absolute error in (a)~the Li-ion migration barrier $E_\mathrm{m}$ and (b)~the ionic conductivity $\log_{10}\sigma$. For each FP, the fraction excludes the failures counted in (c) and (d). Horizontal dashed lines are drawn at fractions of 0.5, 0.8, and 0.95, with open symbols where each curve crosses them. (c, d)~Number of failures of each FP. In (c), a failure is a migration path on which the largest force component over all images and both endpoints remained at or above 0.1~eV~\AA$^{-1}$ after 100 optimization steps, following the criterion of \citet{dembitskiy2025benchmarking}. In (d), a failure is an MD simulation for which the fitted diffusivity was at the numerical floor of the fit, $1.667\times10^{-17}$~cm$^2$~s$^{-1}$. Color identifies the training dataset; line style and symbol shape identify the architecture.}
\label{fig:migration-diffusion}
\end{figure}

Figure~\ref{fig:migration-diffusion} shows the distribution of the absolute errors (a, b) together with the number of failures (c, d). The error distributions of the FPs lie close together for $E_\mathrm{m}$ (a); for $\log_{10}\sigma$ (b), those of six FPs nearly coincide and that of MACE-MPtrj is shifted to larger errors over the whole range. Accuracy and reliability, however, do not necessarily coincide. For $E_\mathrm{m}$, QET-MatPES-PBE has the lowest MAE (0.15\,eV) but leaves 139 of the 1,681 migration paths unconverged, whereas MACE-MatPES-PBE has the highest MAE (0.22\,eV) but only 95. For $\log_{10}\sigma$, MACE-OMat24 has the lowest MAE (0.17) but 14 failures among the 748 simulations, whereas QET-MatPES-PBE has the highest MAE of the other six FPs (0.22) but only 2. MACE-MPtrj is the only FP that is both the least accurate and the least reliable for $\log_{10}\sigma$, with the largest error (0.37) and the most failures (20). It overestimates $\sigma$ in 78\% of the simulations, consistent with PES softening\cite{Deng2025Softening}. Because the MAE of each FP is computed over a different subset of paths and simulations, FPs that fail on the most difficult cases are favored. On the paths and simulations on which an FP fails, the other FPs have MAEs that are 30--50\% higher for $E_\mathrm{m}$ and three to seven times higher for $\log_{10}\sigma$ than on those on which it succeeds (Supplementary Table~\ref{si:migration-failed}). On the paths and simulations on which all seven FPs succeed, the MAEs are lower by about 0.01\,eV and 0.02, respectively (Supplementary Table~\ref{si:migration-common}).

\paragraph{Surface and grain boundary energies.}

\begin{figure}[!tbp]
\centering
\includegraphics[width=\textwidth]{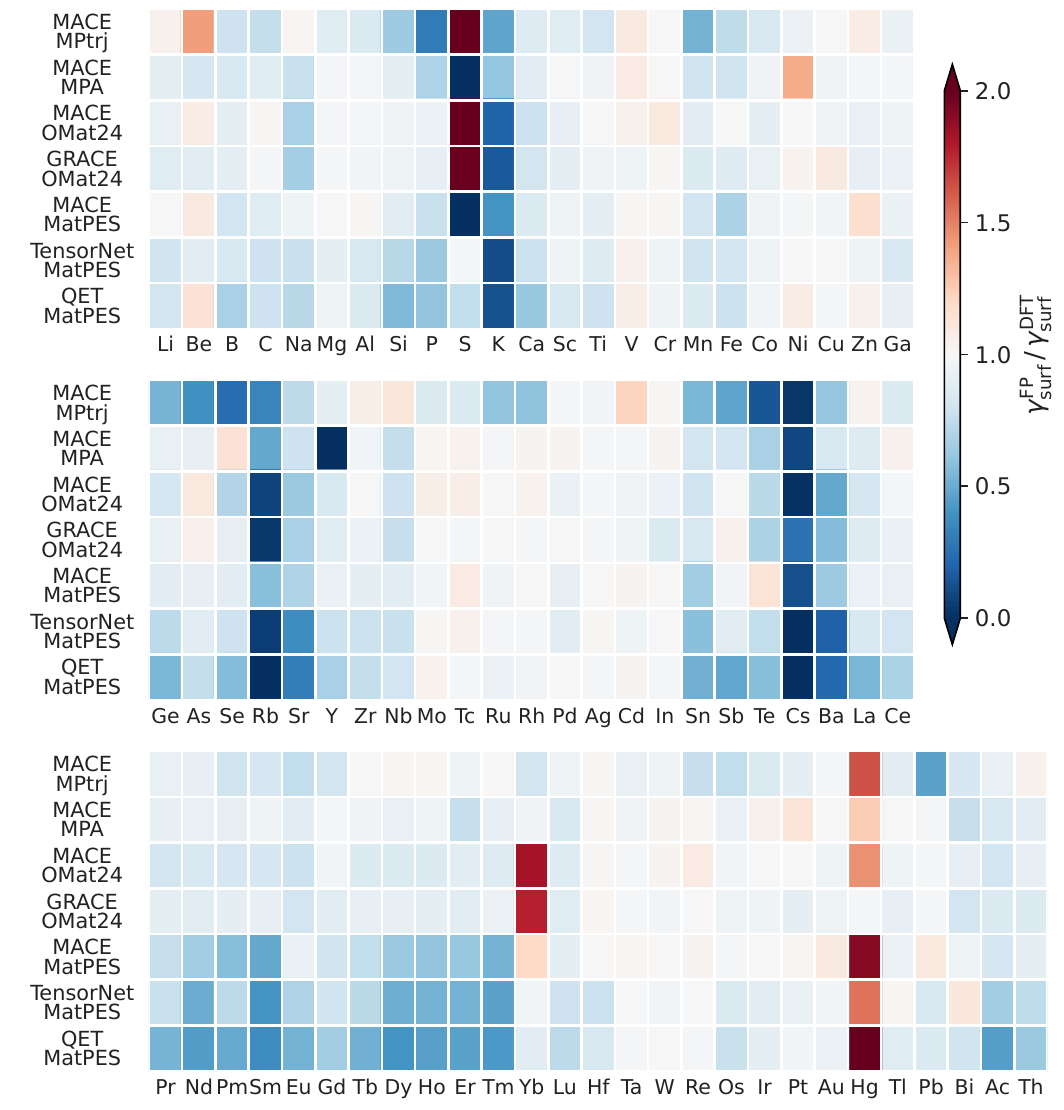}
\caption{\textbf{Surface energies from PBE-trained foundation potentials.} Ratio of the predicted to the reference surface energy $\gamma_\mathrm{surf}$ for each element, obtained from the sums over all slabs of the element. The color scale is capped at 0 and 2.}
\label{fig:surface-heatmap}
\end{figure}

\begin{figure}[!tbp]
\centering
\includegraphics[width=\textwidth]{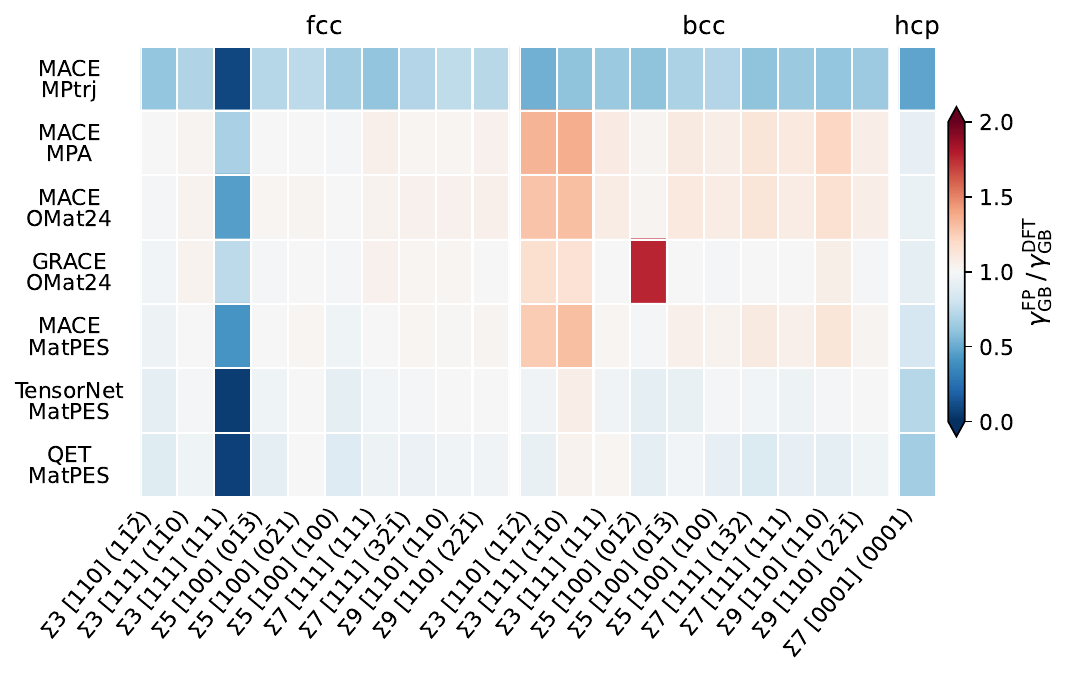}
\caption{\textbf{Grain boundary energies from PBE-trained foundation potentials.} Ratio of the predicted to the reference grain boundary energy $\gamma_\mathrm{GB}$ for each type of grain boundary in fcc, bcc, and hcp metals, obtained from the sums over all grain boundaries of the type. The color scale is capped at 0 and 2.}
\label{fig:gb-heatmap}
\end{figure}

MACE-OMat24 and GRACE-OMat24 give the lowest MAE in the surface energy $\gamma_\mathrm{surf}$ (0.12\,J\,m$^{-2}$), followed by the MatPES-trained FPs and MACE-MPtrj (0.14--0.25\,J\,m$^{-2}$; Table~\ref{tab:overview}), with errors concentrated in a few groups of elements (Fig.~\ref{fig:surface-heatmap}; see Supplementary Fig.~\ref{si:surface-hkl} for the results by Miller index). The MAE of MACE-MPA (52.66\,J\,m$^{-2}$) is dominated by Y: in 22 of 27 Y slabs, relaxation enters a spurious low-energy region of the PES and $\gamma_\mathrm{surf}$ turns negative, although bulk Y is described correctly (Supplementary Section~\ref{si:surface-y}). Without the Y slabs, the MAE of MACE-MPA is 0.14\,J\,m$^{-2}$. The ratio of predicted to reference $\gamma_\mathrm{surf}$ drops toward zero as the nearest-neighbor distance $d_\mathrm{NN}$ approaches the cutoff radius $r_\mathrm{c}$ (Supplementary Fig.~\ref{si:surface-cutoff}). For Cs ($d_\mathrm{NN}=5.3$\,\AA{}), the two FPs with $r_\mathrm{c}=5$\,\AA{} (TensorNet-MatPES-PBE and QET-MatPES-PBE) give $\gamma_\mathrm{surf}$ of zero for all 32 slabs, and the FPs with $r_\mathrm{c}=6$\,\AA{} have MAEs (0.06--0.08\,J\,m$^{-2}$) as large as the Cs surface energy itself (0.07\,J\,m$^{-2}$ on average; for the grain boundaries of Cs, see Supplementary Section~\ref{si:gb-cutoff}). Further deviations arise from differences between the pseudopotentials of the training data and of the reference. For Pr, Nd, Pm, Sm, Tb, Dy, Ho, Er, and Tm, the reference treats the 4f electrons as core states (\texttt{Ln\_3})\cite{tranSurfaceEnergiesElemental2016} and MatPES as valence states (\texttt{Ln\_h});\cite{kaplan2025matpes} the MatPES-trained FPs have MAEs of 0.36--0.54\,J\,m$^{-2}$ versus 0.07--0.14\,J\,m$^{-2}$ for the others. For Yb, the reference uses a divalent pseudopotential (\texttt{Yb\_2}), OMat24 a trivalent one (\texttt{Yb\_3}),\cite{barroso-luqueOpenMaterials20242024} and MatPES one with the 4f electrons in the valence (\texttt{Yb\_h}); the OMat24-trained FPs have MAEs of 0.38--0.41\,J\,m$^{-2}$ versus 0.02--0.10\,J\,m$^{-2}$ for the others.

MACE-MPA, MACE-MatPES-PBE, TensorNet-MatPES-PBE, and MACE-OMat24 give the lowest MAEs in the grain boundary energy $\gamma_\mathrm{GB}$ (0.06--0.07\,J\,m$^{-2}$), followed by GRACE-OMat24 and QET-MatPES-PBE (0.09\,J\,m$^{-2}$) and MACE-MPtrj (0.24\,J\,m$^{-2}$) (Table~\ref{tab:overview}). MACE-MPtrj underestimates $\gamma_\mathrm{GB}$ for all grain boundary types (Fig.~\ref{fig:gb-heatmap}; see Supplementary Fig.~\ref{si:gb-element-fig} for the results by element), which we attribute to the sparse coverage of atomic environments with compressed and stretched bonds in the MPtrj relaxation trajectories. For the other FPs, the largest systematic deviation is for the fcc coherent twin boundaries $\Sigma3[111](111)$, where the two FPs with a maximum angular order $l_\mathrm{max}=2$ (TensorNet-MatPES-PBE and QET-MatPES-PBE) give only 5\% and 7\% of the reference $\gamma_\mathrm{GB}$, respectively. The two sides of such a boundary differ only in the stacking of close-packed layers, which with $l_\mathrm{max}=2$ is resolved only through atoms beyond the adjacent layers (Supplementary Section~\ref{si:gb-angular}).
The same limitation appears at higher angular order: for constructed structures that differ only in the relative rotation of rings of six or eight atoms, the MACE models ($l_\mathrm{max}=3$) also give identical energies (Supplementary Section~\ref{si:gb-angular}), consistent with the limited expressive power of equivariant graph neural networks reported for point clouds.\cite{joshi2023expressive}
The cutoff and pseudopotential effects also appear here. The FPs with $r_\mathrm{c}=5$\,\AA{} give $\gamma_\mathrm{GB}$ of zero for Cs. For GRACE-OMat24, the Cs energy keeps decreasing with interatomic distance up to $r_\mathrm{c}=6$\,\AA{}, so that $d_\mathrm{NN}$ of the relaxed bulk (5.95\,\AA{}) is 12\% larger than in the reference. In grain boundary structures scaled to this lattice, some atoms have no neighbor within $r_\mathrm{c}$, and one grain boundary has $\gamma_\mathrm{GB}$ of 10.77\,J\,m$^{-2}$. For Yb, the MAEs are 0.06--0.10\,J\,m$^{-2}$ for the MatPES-trained FPs and 0.20--0.23\,J\,m$^{-2}$ for the OMat24-trained FPs, versus 0.01\,J\,m$^{-2}$ for the others.

\paragraph{Adsorption and interfacial energies.}

\begin{figure}[!tbp]
\centering
\includegraphics[width=\textwidth]{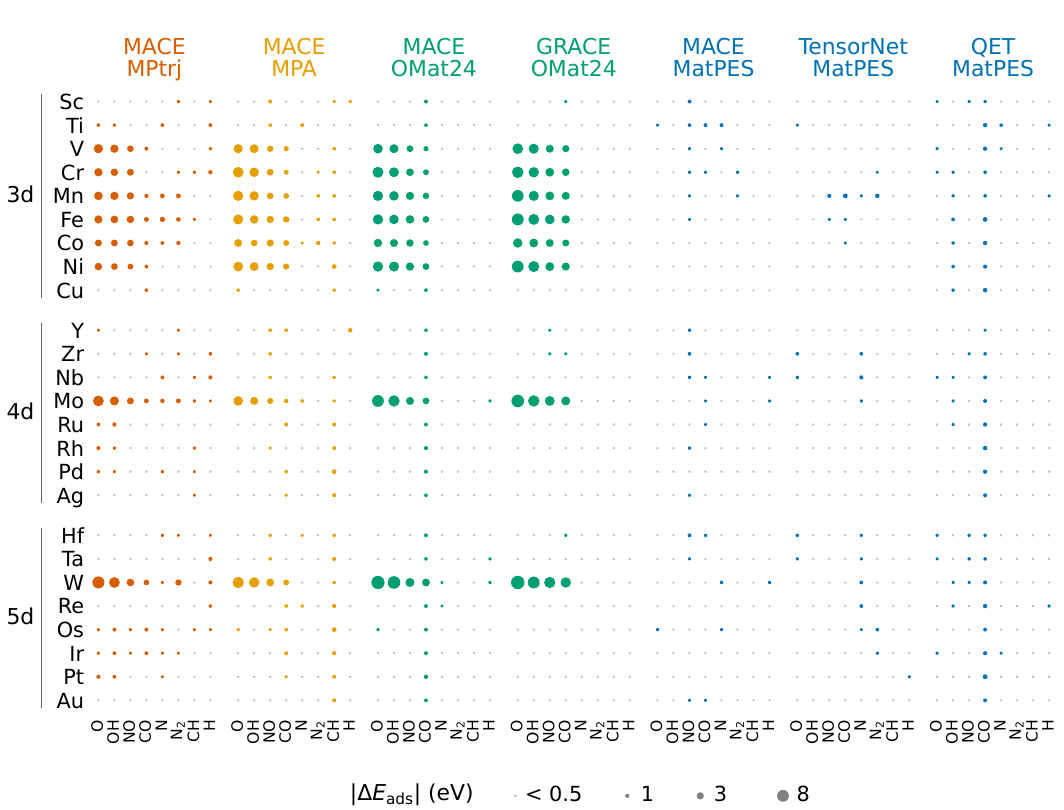}
\caption{\textbf{Adsorption energies from PBE-trained foundation potentials.} Absolute error of the adsorption energy $E_\mathrm{ads}$ for each metal and adsorbate. The area of each circle is proportional to the absolute error; errors below 0.5~eV are shown as gray dots.}
\label{fig:adsorption-bubbles}
\end{figure}

The three MatPES-trained FPs achieve MAEs of 0.24--0.36\,eV in the adsorption energy $E_\mathrm{ads}$, whereas MACE-MPtrj, MACE-MPA, MACE-OMat24, and GRACE-OMat24 have MAEs of 0.86--1.10\,eV (Table~\ref{tab:overview}). The large errors are confined to O, OH, NO, and CO on V, Cr, Mn, Fe, Co, Ni, Mo, and W (Fig.~\ref{fig:adsorption-bubbles}): for these 32 reactions, the latter four FPs give MAEs of 3.14--5.81\,eV and underestimate the binding in every case, as reported for O adsorbed on these metals,\cite{warfordBetterWithoutImpact2026} whereas for the other 168 reactions all seven FPs have MAEs of 0.21--0.43\,eV. The training data of these four FPs (MPtrj, Alexandria, and OMat24)\cite{dengCHGNetPretrainedUniversal2023,schmidt2024alexandria,barroso-luqueOpenMaterials20242024} follow the Hubbard $U$ scheme of the Materials Project, which applies $U$ to these eight metals only in compounds that contain O or F,\cite{jainFormationEnthalpiesMixing2011} so the same metal is described with PBE+$U$ when it is bonded to O and with PBE otherwise; MatPES uses PBE without $U$ for all structures.\cite{kaplan2025matpes} Recomputing the 72 affected reference cells with PBE+$U$ reduces the MAEs of these four FPs on the 32 reactions only from 3.14--5.81 to 2.63--5.25~eV (Supplementary Table~\ref{si:ads-pbeu-mae}), so the errors arise from the mixture of PBE and PBE+$U$ energies in the training data rather than from the choice of PBE for the reference. When the structures are relaxed, the underbinding can lead to desorption: with MACE-OMat24 and GRACE-OMat24, O desorbs from Ni(111) and Ni(100), and with MACE-OMat24 also CO from Ni(111) (Supplementary Section~\ref{si:ads-relax}). Against the r$^2$SCAN reference values, the r$^2$SCAN-trained FPs have MAEs of 0.26--0.39\,eV. The same trends hold for larger molecules on metal surfaces in ADS41\cite{mallikarjunsharadaAdsorptionTransitionMetal2019,trepteDatadrivenConstrainedOptimization2022} and for transition metal compounds from Catalysis-Hub\cite{wintherCatalysisHubOpenElectronic2019,yohannesCombinedHighThroughputDFT2023} (Supplementary Sections~\ref{si:ads41} and~\ref{si:ads-cathub}). Against the experimental values of ADS41, without dispersion corrections, the PBE-trained MatPES FPs (MAEs of 0.41--0.49\,eV) are close to the PBE functional itself (0.36\,eV), whereas the r$^2$SCAN-trained FPs (0.46--0.73\,eV) fall short of the r$^2$SCAN functional (0.32\,eV), mainly for physisorbed molecules (the effect of dispersion corrections is discussed in Supplementary Section~\ref{si:ads41}).

The same mismatch affects the metal/oxide interfaces. MACE-MPtrj, MACE-MPA, MACE-OMat24, and GRACE-OMat24 have MAEs of 2.06--4.71\,J\,m$^{-2}$ in $W_\mathrm{adh}$ and 2.25--4.89\,J\,m$^{-2}$ in $\gamma_\mathrm{int}$, compared with 0.17--0.29 and 0.20--0.24\,J\,m$^{-2}$ for the MatPES-trained FPs (Table~\ref{tab:overview}); their errors are again confined to the interfaces of V, Cr, Mn, Fe, Co, and Ni (Supplementary Fig.~\ref{si:interface-bubbles}).

\subsubsection{Comparison with experiment}

Benchmarks against DFT assess how well an FP reproduces its training functional, but materials scientists ultimately need agreement with experiment, which also includes the error of the functional. MatCalc-Bench includes experimental benchmark datasets for eight bulk properties, compiled from the sources listed in Supplementary Table~\ref{si:bench-tasks}. Figure~\ref{fig:expt-summary} shows, for each property, the median relative error against experiment, averaged over the PBE-trained and over the r$^2$SCAN-trained FPs. The corresponding median absolute relative errors are shown in Supplementary Fig.~\ref{si:expt-unsigned}.

On average, the PBE-trained FPs overestimate $a$ by 1.0\% and $\alpha_V$ by 19\%, predict $\Delta H_\mathrm{f}$ that are 10\% less negative than experiment, and underestimate $K$ and $G$ by 13\%, $K_T$ by 9\%, and $\kappa$ by 22\%. The larger $a$ and the smaller moduli are the known errors of PBE for solids,\cite{haasCalculationLatticeConstant2009,zhangPerformanceVariousDensityfunctional2018} and the less negative $\Delta H_\mathrm{f}$ reflects the underbinding of the PBE functional.\cite{jainFormationEnthalpiesMixing2011,stevanovicCorrectingDensityFunctional2012} The larger $\alpha_V$ is consistent with the smaller $K_T$, since $\alpha_V=\gamma C_V/(K_T V)$, and the smaller $\kappa$ with the lower phonon frequencies of a softer PES. The median error in $\gamma$ is small (4\%) and differs in sign between the FPs. Averaged over the r$^2$SCAN-trained FPs, the median errors of all eight properties are within 7\%, from $-6.7\%$ for $\kappa$ to $+6.9\%$ for $\alpha_V$. The shift also appears within each of the three pairs of MatPES-trained FPs with the same architecture, for all properties except $\gamma$, for example in $\kappa$ from $-30\%$ (TensorNet-MatPES-PBE) to $-2\%$ (TensorNet-MatPES-r2SCAN) and in $\alpha_V$ from $+24\%$ (QET-MatPES-PBE) to $+2\%$ (QET-MatPES-r2SCAN). Because the two FPs of each pair differ mainly in the functional of the training data, the smaller deviations of the r$^2$SCAN-trained FPs can be attributed to the underlying functional, which describes the lattice parameters, moduli, and formation enthalpies of solids more accurately than PBE.\cite{furnessAccurateNumericallyEfficient2020,kingsburyFlexibleScalableScheme2022} Conversely, for properties for which r$^2$SCAN offers no advantage over PBE, or an advantage smaller than the error of the FPs relative to r$^2$SCAN, the r$^2$SCAN-trained FPs show no advantage either, as for the adsorption energies of ADS41 (Supplementary Section~\ref{si:ads41}).

\begin{figure}[htbp]
\centering
\includegraphics[width=0.8\textwidth]{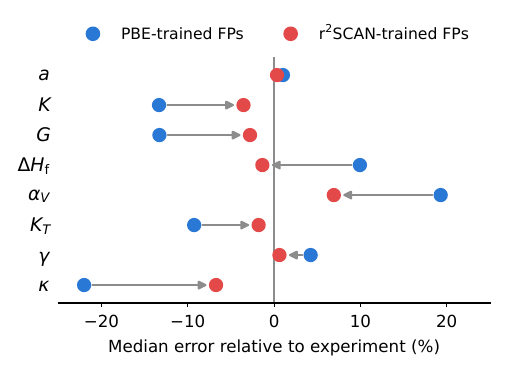}
\caption{\textbf{Systematic deviations of the FPs from experiment.} Median relative error against experiment of the lattice parameter $a$, the bulk and shear moduli $K$ and $G$, the formation enthalpy $\Delta H_\mathrm{f}$, the volumetric thermal expansion coefficient $\alpha_V$, the isothermal bulk modulus $K_T$, the Gr\"{u}neisen parameter $\gamma$, and the lattice thermal conductivity $\kappa$; $\alpha_V$, $K_T$, $\gamma$, and $\kappa$ are at 300\,K. The median is taken over the materials for each FP and then averaged over the PBE-trained (blue) and r$^2$SCAN-trained (red) FPs. Arrows point from the PBE-trained to the r$^2$SCAN-trained FPs.}
\label{fig:expt-summary}
\end{figure}

\subsection{Computational cost}

Besides accuracy, the choice of an FP depends on its computational cost, which we measured for the seven PBE-trained FPs on one NVIDIA A100 GPU (Methods and Supplementary Section~\ref{si:cost}). For MD of 1,000 atoms, a step takes 31--89\,ms using the accelerated implementations of MACE (cuEquivariance), TensorNet (Warp), and QET (Warp). TensorNet-MatPES-PBE and QET-MatPES-PBE are the fastest and GRACE-OMat24 the slowest. Depending on the FP, a complete \texttt{ElasticityCalc} and \texttt{QHACalc} workflow takes 0.3--2.3\,s and 45--81\,s, respectively.

\section{Discussion}

We may summarize several key findings from the preceding benchmarks.

First, state-of-the-art pre-trained FPs now deliver property-level accuracy without system-specific fitting. For example, the errors of the OMat24- and MatPES-trained FPs lie within the range of the element-specific MLIPs of \citet{zuoPerformanceCostAssessment2020} for lattice parameters, elastic constants, and defect energetics, and the element-specific MLIPs retain a clear advantage only in $\Delta_\mathrm{EOS}$. Hence, for most properties, the question is no longer whether to use an FP, but which one to use, fine-tune, or distill.

Second, the training data matter more than architecture or dataset size. Models trained on OMat24\cite{barroso-luqueOpenMaterials20242024} or MatPES\cite{kaplan2025matpes} generally outperform MACE-MPtrj on structural fidelity, shear moduli, lattice thermal conductivity, ionic conductivity, and grain boundary energies, which we attribute to their coverage of off-equilibrium configurations. The DFT settings of the training data matter as much as their coverage. The FPs trained on MPtrj, Alexandria, and OMat24 inherit the Materials Project mixture of PBE and PBE+$U$ energies and underbind O-containing adsorbates on eight transition metals by 3--6~eV and misdescribe metal/oxide interfaces by 2--5~J~m$^{-2}$, whereas the MatPES-trained FPs, trained on PBE without $U$, do not. Pseudopotential choices (lanthanides with 4f electrons in the core or valence, Yb) cause further, chemically localized errors. Architecture still matters: FPs with $l_\mathrm{max}=2$ fail to distinguish the fcc coherent twin boundary, FPs with $r_\mathrm{c}\leq6$\,\AA{} have large errors for the surfaces of Cs, and FPs with non-conservative forces have larger errors for properties that depend on higher-order derivatives of the PES, such as $K_\mathrm{VRH}$ and $\kappa$ (Supplementary Section~\ref{si:nonconservative}). Benchmarks of bulk energetics and stability alone therefore cannot establish how a model performs in practical property calculations.

Third, curation can also substitute for scale.  MatPES is more than two orders of magnitude smaller than OMat24 (433K versus 100.8M structures), yet the MatPES-trained models match or exceed the OMat24-trained models on thermal expansion coefficients, isothermal bulk moduli, and grain boundary energies and, owing to their uniform PBE settings, on adsorption and interfacial energies; the OMat24-trained models lead on relaxed structures, heat capacities, surface energies, and lattice thermal conductivity. Because the MACE-MatPES FPs were fine-tuned from MACE-OMat24, this comparison is cleanest for TensorNet and QET, which were trained on MatPES alone.

Fourth, given that nearly all state-of-the-art architectures reproduce the reference training data closely, the level of theory or exchange-correlation functional of that reference sets the accuracy floor against experiment. The r$^2$SCAN-trained FPs reduce the median errors against experiment of all eight bulk properties to within 7\%, compared with up to 22\% for the PBE-trained FPs; for example, the volumetric thermal expansion coefficient is overestimated by $+6.9\%$ rather than $+19\%$, and the MAPE of the elemental lattice parameters falls from 0.58--1.07\% to 0.33--0.59\%. This comparison is possible only because MatCalc-Bench provides r$^2$SCAN and experimental reference data alongside PBE. We therefore strongly recommend that the community use r$^2$SCAN-trained FPs over the PBE-trained counterparts for most systems.

Finally, reliability is an important dimension of model quality that most benchmarks miss. For example, QET-MatPES-PBE has the lowest $E_\mathrm{m}$ MAE but fails on 139 of 1,681 migration paths, against 95 for MACE-MatPES-PBE, and MACE-OMat24 has the lowest $\log_{10}\sigma$ MAE but 14 failed MD simulations, against 2 for QET-MatPES-PBE. Over the seven FPs, the failures range from 91 to 144 of the 1,681 migration paths and from 2 to 20 of the 748 MD simulations. For many simulations and high-throughput screening, a model that converges reliably at slightly higher error is often preferable, and benchmarks should report failure rates alongside MAEs.

Both MatCalc and MatCalc-Bench are being extended. \texttt{PropCalc} and \texttt{Benchmark} classes for further properties are under development (Supplementary Table~\ref{si:tab-more-classes}). In MatCalc-Bench, each property is currently compared with a single set of DFT reference values, whose settings differ from those of the training data of some FPs (Supplementary Table~\ref{si:wbm-references}), and the benchmarks of $\alpha_V$, $K_T$, and $\kappa$ comprise few materials; future versions will include more materials for these properties and reference values computed with r$^2$SCAN and with the settings of each training set.

\section{Conclusions}

In summary, we have introduced MatCalc, an open-source Python framework that reduces property calculations with any PES to a uniform \texttt{calc(structure) $\rightarrow$ dict} interface executed via \texttt{PropCalc} classes with either ASE or LAMMPS as the simulation backend. Using MatCalc's \texttt{Benchmark} classes, we have developed MatCalc-Bench, a property-level FP evaluation suite spanning 17 properties, from relaxed structures and elastic moduli to lattice thermal conductivities, migration barriers, ionic conductivities, and surface, grain boundary, adsorption, and interfacial energies, and the first to provide r$^2$SCAN and experimental as well as PBE reference values across multiple properties. Applied to ten FPs, MatCalc-Bench shows that no single FP leads on more than a few properties, that inconsistent DFT settings in the training data, such as the mixing of PBE and PBE+$U$, cause errors of several eV that benchmarks of bulk energetics do not probe, and that training on r$^2$SCAN data reduces the median errors against experiment of eight bulk properties to within 7\%. By collapsing the marginal cost of property-level evaluation, MatCalc makes benchmark-then-deploy a practical default for MLIP-driven materials simulation. As FPs proliferate, we anticipate that property-resolved evaluation of the kind enabled here, extensible to new properties through the \texttt{PropCalc} interface and tracked publicly at \url{https://bench.matcalc.ai}, will be as central to the field's progress as the models themselves.

\section{Methods}

\subsection{Property calculators and benchmark classes}

Each property workflow is implemented as a \texttt{PropCalc} (Table~\ref{tab:propcalc_summary}). By default, a \texttt{PropCalc} first relaxes the input structure with \texttt{RelaxCalc}. The workflows rely on pymatgen\cite{ongPythonMaterialsGenomics2013} for strained structures, slabs, grain boundaries, interfaces, adsorption sites, and the analysis of diffusion, on phonopy and phono3py\cite{togoFirstPrinciplesPhonon2015,togoDistributionsPhononLifetimes2015} for force constants and the properties derived from them, and on ASE\cite{larsenAtomicSimulationEnvironment2017} for optimization, the nudged elastic band method, and molecular dynamics. The \texttt{Benchmark} classes combine a \texttt{PropCalc} with a reference dataset; the protocols used in this work are given below.  Further classes, which are under development or not used in this work, are listed in Supplementary Table~\ref{si:tab-more-classes}.

\begingroup
\footnotesize
\setlength{\tabcolsep}{5pt}
\renewcommand{\arraystretch}{1.08}
\setlength{\LTcapwidth}{\linewidth}
\begin{longtable}{@{}>{\raggedright\arraybackslash}p{0.30\linewidth}>{\raggedright\arraybackslash}p{0.66\linewidth}@{}}
\caption{\textbf{Property workflows and benchmarks in MatCalc.}}
\label{tab:propcalc_summary}\\
\toprule
Class & Result \\
\midrule
\endfirsthead
\toprule
Class & Result \\
\midrule
\endhead
\bottomrule
\endfoot
\multicolumn{2}{@{}l}{\textit{PropCalc}}\\*
\texttt{RelaxCalc} & Relaxed structure, energy, forces, stress, and lattice parameters \\
\texttt{EnergeticsCalc} & Formation and cohesive energies \\
\texttt{EOSCalc} & Energy--volume curve and Birch--Murnaghan bulk modulus \\
\texttt{ElasticityCalc} & Elastic tensor; Voigt--Reuss--Hill bulk and shear moduli; Young's modulus \\
\texttt{PhononCalc} & Phonon frequencies; harmonic free energy, entropy, and heat capacity \\
\texttt{QHACalc} & Thermal expansion coefficient, Gibbs free energy, isothermal bulk modulus, heat capacity at constant pressure, and Gr\"uneisen parameter \\
\texttt{Phonon3Calc} & Lattice thermal conductivity from second- and third-order force constants \\
\texttt{NEBCalc} & Minimum energy path, migration barrier, and reaction energy \\
\texttt{MDCalc} & Molecular dynamics trajectory and averaged energies in NVE, NVT, and NpT ensembles \\
\texttt{DiffusionCalc} & Tracer and charge diffusivities, ionic conductivity, and Haven ratio \\
\texttt{DefectCalc} & Formation energies of neutral point defects \\
\texttt{SurfaceCalc} & Surface energy \\
\texttt{GBCalc} & Grain boundary energy \\
\texttt{AdsorptionCalc} & Adsorption configurations and adsorption energy \\
\texttt{InterfaceCalc} & Interfacial energy and work of adhesion of coherent interfaces \\
\texttt{OrderCalc} & Lowest-energy orderings of a disordered structure from Metropolis Monte Carlo \\
\texttt{ChainedCalc} & Sequence of \texttt{PropCalc}s that pass their results to the next \\
\addlinespace
\midrule
\multicolumn{2}{@{}l}{\textit{Benchmark}}\\*
\texttt{EquilibriumBenchmark} & Relaxed structures and formation energies \\
\texttt{ElasticityBenchmark} & Voigt--Reuss--Hill bulk and shear moduli \\
\texttt{PhononBenchmark} & Harmonic heat capacity at 300~K \\
\texttt{QHABenchmark} & Volumetric thermal expansion coefficient and isothermal bulk modulus at 300~K \\
\texttt{KappaBenchmark} & Lattice thermal conductivity at 300~K \\
\texttt{BarrierBenchmark} & Migration barriers \\
\texttt{DiffusionBenchmark} & Ionic conductivities from molecular dynamics \\
\texttt{SurfaceBenchmark} & Surface energies \\
\texttt{GBBenchmark} & Grain boundary energies \\
\texttt{AdsorptionBenchmark} & Adsorption energies \\
\texttt{InterfaceBenchmark} & Interfacial energies and works of adhesion \\
\texttt{BenchmarkSuite} & Several benchmarks for several models, with checkpoints \\
\end{longtable}
\endgroup

\subsection{Interatomic potentials}

MatCalc accepts any ASE calculator. Its model registry provides over 100 pretrained FPs by name, from nine packages: MatGL\cite{Ko2025MatGL} (TensorNet, QET, M3GNet, and CHGNet), MACE, SevenNet, GRACE, Orb, MatterSim\cite{yangMatterSimDeepLearning2024}, FAIRChem (UMA), UPET (PET), and DeePMD-kit (DPA\cite{zhangDPA2LargeAtomic2024}). Potentials trained by the user are loaded from their files, including MatGL models, GAP, MTP, NNP, SNAP, and qSNAP models through maml, and ACE, NequIP, and Deep Potential models. Each architecture is described in Supplementary Table~\ref{si:tab-supported}.

The benchmarks in this work used the seven PBE-trained and three r$^2$SCAN-trained FPs listed in Table~\ref{tab:models}, with the published pretrained weights.\cite{Batatia2025MACEMP,Batatia2022MACE,Simeon2023TensorNet,ko2026,Lysogorskiy2026GRACE,kaplan2025matpes} All ten obtain forces as the gradient of the predicted energy. The MACE-MatPES FPs were fine-tuned on MatPES from a model pretrained on OMat24, whereas the TensorNet and QET FPs were trained on MatPES only. Details of the architectures are given in Supplementary Table~\ref{si:tab-model-table}.

\begin{table}[htbp]
\centering
\footnotesize
\setlength{\tabcolsep}{3.5pt}
\caption{\textbf{Foundation potentials evaluated in this work.} $r_\mathrm{c}$, cutoff radius (\AA); $l_\mathrm{max}$, maximum angular order, given for both layers of GRACE-OMat24.}
\label{tab:models}
\resizebox{\textwidth}{!}{%
\begin{tabular}{@{}llllrcc@{}}
\toprule
Model & Name in the model registry & Training data & Structures & Parameters & $r_\mathrm{c}$ & $l_\mathrm{max}$\\
\midrule
\multicolumn{7}{@{}l}{\textit{PBE-trained}}\\
MACE-MPtrj & \texttt{MACE-MPtrj-0a-medium} & MPtrj & 1.58M & 4.69M & 6 & 3\\
MACE-MPA & \texttt{MACE-MPA-0-medium} & MPtrj + sAlex & 12.0M & 9.06M & 6 & 3\\
MACE-OMat24 & \texttt{MACE-OMat24-0-medium} & OMat24 & 100.8M & 9.06M & 6 & 3\\
GRACE-OMat24 & \texttt{GRACE-OMat24-2L-large-base} & OMat24 & 100.8M & 26.4M & 6 & 4/3\\
MACE-MatPES-PBE & \texttt{MACE-OMat24-ft-MatPES-PBE-0} & MatPES-PBE & 435K & 9.06M & 6 & 3\\
TensorNet-MatPES-PBE & \texttt{TensorNet-MatPES-PBE-2025.2} & MatPES-PBE & 433K & 837K & 5 & 2\\
QET-MatPES-PBE & \texttt{QET-MatPES-PBE-2025.2} & MatPES-PBE & 433K & 905K & 5 & 2\\
\addlinespace[2pt]
\multicolumn{7}{@{}l}{\textit{r$^2$SCAN-trained}}\\
MACE-MatPES-r2SCAN & \texttt{MACE-OMat24-ft-MatPES-r2SCAN-0} & MatPES-r$^2$SCAN & 388K & 9.06M & 6 & 3\\
TensorNet-MatPES-r2SCAN & \texttt{TensorNet-MatPES-r2SCAN-2025.2} & MatPES-r$^2$SCAN & 387K & 837K & 5 & 2\\
QET-MatPES-r2SCAN & \texttt{QET-MatPES-r2SCAN-2025.2} & MatPES-r$^2$SCAN & 387K & 905K & 5 & 2\\
\bottomrule
\end{tabular}}
\end{table}

\subsection{Benchmark design}

Each property is computed with the same \texttt{PropCalc} subclass that a user would run, so that the evaluation includes the relaxations, deformations, atomic displacements, path optimizations, and MD simulations that the property requires. For each property and reference dataset, all FPs are evaluated on the same structures with the same protocol. Where the reference quantity depends on the simulation setup, such as the supercell, $q$-point mesh, volume grid, or temperature, the setup of the reference calculation is adopted.

\paragraph{Reference data.} The error of an FP against a DFT reference combines the error of the model with the effect of any difference in the functional and the numerical settings between its training data and the reference. Each FP is therefore compared with reference data computed with the functional of its training data: PBE-trained FPs with PBE references, and r$^2$SCAN-trained FPs with r$^2$SCAN references, which exist for a subset of the properties. Where several references exist, each FP is compared with the one whose settings are closest to those of its training data; for example, the FPs trained on MatPES-PBE are compared with references for the WBM structures re-relaxed at the MatPES settings. The adsorption and interface references were computed in this work. Comparisons with experiment, which also include the error of the functional, are reported separately.

\paragraph{Structures and energies.} Bulk, surface, and grain boundary structures are relaxed with each FP, and the elemental, bulk, and gas-phase reference energies are computed with the same FP. Adsorption and interface structures are evaluated at the geometries of the reference calculations, so that the FP and reference energies correspond to the same configurations.

\paragraph{Metrics.} Errors are reported as MAEs (of $\log_{10}\kappa$ and $\log_{10}\sigma$ for the two conductivities), except for relaxed structures, which are compared by the CrystalNN fingerprint distance. For migration barriers and ionic conductivities, a property-specific failure criterion is defined, the MAE is computed over the calculations that did not fail, and the number of failures is reported separately.

\subsection{Comparison with element-specific MLIPs}

The comparison covers fcc Ni and Cu, bcc Li and Mo, and diamond-cubic Si and Ge, using the GAP, MTP, NNP, SNAP, and quadratic SNAP (qSNAP) models of \citet{zuoPerformanceCostAssessment2020}. MatCalc loads these models through \texttt{maml} and evaluates them alongside the FPs with the same property calculators and settings. For each model and element, the calculators were composed into a \texttt{ChainedCalc} pipeline. All calculations started from the DFT structures of the Materials Project, whose lattice parameters coincide with the DFT values of \citet{zuoPerformanceCostAssessment2020}, and all structural relaxations used a force threshold of 0.01~eV~\AA$^{-1}$. All properties were compared with the DFT-PBE values of \citet{zuoPerformanceCostAssessment2020}. The lattice parameters of the three r$^2$SCAN-trained FPs in Supplementary Table~\ref{si:tab-elemental-lattice} were compared with r$^2$SCAN values from the Materials Project.\cite{jainCommentaryMaterialsProject2013,kingsburyFlexibleScalableScheme2022}

Lattice parameters were obtained with \texttt{RelaxCalc}, which relaxes atomic positions and cell parameters with the FIRE optimizer\cite{bitzekStructuralRelaxationMade2006} and a Fr\'echet cell filter. Elastic constants were obtained with \texttt{ElasticityCalc} by relaxing the structure, applying normal strains of $\pm0.5\%$ and $\pm1\%$ and shear strains of $\pm3\%$ and $\pm6\%$, and fitting the resulting stresses linearly; $C_{11}$, $C_{12}$, and $C_{44}$ were taken from the cubic elastic tensor, and $K_\mathrm{VRH}$ from its Voigt--Reuss--Hill average.\cite{hillElasticBehaviourCrystalline1952}

Vacancy energetics were computed in $3\times3\times3$ conventional supercells for the fcc and bcc elements (108 and 54 atoms) and a $2\times2\times2$ conventional supercell for the diamond-cubic elements (64 atoms). The perfect supercell was relaxed with respect to atomic positions and cell parameters, a single atom was removed, and the defective supercell was relaxed at fixed cell with \texttt{DefectCalc}. The vacancy formation energy was defined as
\begin{equation}
E_v=E_{N-1}^{\mathrm{vac}}-\frac{N-1}{N}E_N^{\mathrm{bulk}},
\end{equation}
where the two total energies refer to the defective and perfect supercells evaluated with the same potential.

The vacancy migration barrier $E_m$ was obtained with \texttt{NEBCalc} using climbing-image nudged elastic band (CI-NEB) calculations\cite{henkelmanClimbingImageNudged2000} with the improved tangent estimate.\cite{henkelmanImprovedTangentEstimate2000} The initial and final states, a vacancy at a lattice site and at a nearest-neighbor site of the relaxed perfect supercell, were relaxed at fixed cell; seven intermediate images were generated by linear interpolation and optimized with BFGS to a force threshold of 0.01~eV~\AA$^{-1}$ (at most 1,000 steps), with a spring constant of 0.1~eV~\AA$^{-2}$. $E_m$ is the energy of the highest image relative to the initial state, and the vacancy-mediated activation energy is $E_a=E_v+E_m$.

Equations of state were computed with \texttt{EOSCalc}. The structure was relaxed, scaled isotropically to 11 volumes spanning $\pm10\%$ of the relaxed volume, with atomic positions relaxed at each volume, and the energies were fitted to the third-order Birch--Murnaghan equation of state.\cite{birch1947} The DFT reference was fitted in the same way to the DFT data points within $\pm10\%$ of the DFT equilibrium volume. Following Lejaeghere et al.,\cite{lejaeghere2014,lejaeghereReproducibilityDensityFunctional2016} $\Delta_\mathrm{EOS}$ is the root-mean-square difference between the model and DFT curves, each referenced to its own minimum energy, over a $\pm6\%$ volume window centered on the mean of the two equilibrium volumes. The energy--volume curves in Supplementary Fig.~\ref{si:eos-full} were computed using the same procedure with lattice parameters scaled from 0.90 to 1.10 times their equilibrium values.

\subsection{MatCalc-Bench protocols}

Each potential was evaluated with the default MatCalc-Bench settings or the defaults of the corresponding property calculator for directly constructed workflows.

\paragraph{Relaxed structures and formation energies.}
Relaxed structures and formation energies were evaluated with \texttt{EquilibriumBenchmark} for 1,000 structures from the WBM dataset\cite{wangPredictingStableCrystalline2021} against PBE reference values for the PBE-trained FPs, and for 1,000 structures from the GNoME dataset\cite{merchantScalingDeepLearning2023} against r$^2$SCAN reference values for the r$^2$SCAN-trained FPs. Each atom was displaced in a random direction by a distance drawn uniformly between 0 and 0.1~\AA\ before joint relaxation of the atomic coordinates and cell using FIRE, with a force threshold of 0.05~eV~\AA$^{-1}$ and a maximum of 500 steps. This was repeated five times with different random perturbations to reduce sampling noise. Relaxations that did not converge within 500 steps (8 of 49,805) were retained. Relaxed and DFT-reference structures were compared through the Euclidean distance between their CrystalNN fingerprints,\cite{zimmermannAssessingLocalStructure2017,zimmermannLocalStructureOrder2020,panBenchmarkingCoordinationNumber2021} each comprising the mean, standard deviation, minimum, and maximum of the site-resolved local order-parameter features; $\bar{d}_\mathrm{fp}$ is the mean distance over all evaluated structures and runs. The formation energy per atom was calculated as
\begin{equation}
E_\mathrm{f}^{0}=\frac{E-\sum_i n_i\varepsilon_i}{N},
\end{equation}
where $E$ is the energy of the relaxed structure, $n_i$ is its number of atoms of element $i$, $N=\sum_i n_i$, and $\varepsilon_i$ is the energy per atom of the elemental reference structure of element $i$ relaxed with the same FP, and reported in meV\,atom$^{-1}$. The elemental reference structures are those distributed with MatCalc for the functional and settings of the DFT reference data; where several structures are available for an element, the one with the lowest energy is used. The superscript 0 denotes static-lattice energies at 0\,K, without zero-point or thermal contributions.

\paragraph{Elastic moduli.}
Elastic moduli were evaluated with \texttt{ElasticityBenchmark} for 12,122 materials from the Materials Project against the PBE reference values.\cite{jainCommentaryMaterialsProject2013, dejongChartingCompleteElastic2015} The parent crystal was relaxed to 0.05~eV~\AA$^{-1}$ using FIRE with a maximum of 500 steps, and independent normal deformations with amplitudes $\pm0.005$ and $\pm0.01$ and shear deformations with amplitudes $\pm0.03$ and $\pm0.06$ were applied. The stress response, including that of the undeformed structure, was fitted to obtain the elastic tensor. Bulk and shear moduli were calculated as Voigt--Reuss--Hill averages and reported in GPa.

\paragraph{Constant-volume heat capacities.}
Constant-volume heat capacities $C_V$ were evaluated with \texttt{PhononBenchmark} for 1,170 binary compounds against PBE reference values from the Alexandria phonon database.\cite{loewUniversalMachineLearning2025} Each structure was relaxed to 0.05~eV~\AA$^{-1}$ using FIRE with a maximum of 5,000 steps, and harmonic force constants were computed with atomic displacements of 0.015~\AA\ in supercells at least 15~\AA\ long along each lattice vector. $C_V$ at 300~K was normalized per atom and reported in J~mol-atom$^{-1}$~K$^{-1}$.

\paragraph{Thermal expansion coefficients and isothermal bulk moduli.}
Volumetric thermal expansion coefficients $\alpha_V$ and isothermal bulk moduli $K_T$ were evaluated with \texttt{QHABenchmark} for 47 materials against PBE reference values.\cite{fan2022bicationic,sun2019thermal,fekete2023thermalshock} Each structure was relaxed to $10^{-4}$~eV~\AA$^{-1}$ using FIRE with a maximum of 5,000 steps. At each point of the reference volume grid, the relaxed cell was scaled isotropically and the atomic positions were relaxed to the same threshold, and harmonic phonons were computed with the supercell and $q$-point mesh of the reference calculation and atomic displacements of 0.01~\AA. The free energy at each temperature was fitted with the Vinet equation of state,\cite{vinetUniversalEquationState1986} and $\alpha_V$ and $K_T$ at 300~K were reported in ppm~K$^{-1}$ and GPa.

\paragraph{Lattice thermal conductivities.}
Lattice thermal conductivities $\kappa$ were evaluated with \texttt{KappaBenchmark} for 103 compounds against PBE reference values computed from the force sets of the phonondb dataset,\cite{togoDistributionsPhononLifetimes2015,sekoPredictionLowThermalConductivityCompounds2015} the same compounds as in the kSRME benchmark.\cite{Pota2024kSRME} Each structure was relaxed to 0.05~eV~\AA$^{-1}$ using FIRE with a maximum of 5,000 steps, and second- and third-order force constants were computed with atomic displacements of 0.03~\AA\ in the supercells of the reference calculations. The thermal conductivity was obtained on the $q$-point mesh of the reference calculations from the Wigner transport equation in the relaxation-time approximation, including isotope scattering,\cite{simoncelliUnifiedTheoryThermal2019} and the mean of its three diagonal components at 300~K was reported in W~m$^{-1}$~K$^{-1}$. The error was computed as the MAE of $\log_{10}\kappa$, because the reference values span more than three orders of magnitude.

\paragraph{Migration barriers and ionic conductivities.}
Li-ion migration barriers $E_\mathrm{m}$ were evaluated with \texttt{BarrierBenchmark} for 1,681 migration paths against PBE reference values from the nebDFT2k dataset.\cite{dembitskiy2025benchmarking} Each path comprises two endpoints and three or five intermediate images taken from the dataset. Following the protocol of the nebDFT2k benchmark, the endpoints were relaxed at fixed cell to 0.1~eV~\AA$^{-1}$ using FIRE with a maximum of 100 steps, and the path was then optimized with the climbing-image nudged elastic band method\cite{henkelmanClimbingImageNudged2000} with the improved tangent estimate\cite{henkelmanImprovedTangentEstimate2000} and a spring constant of 5~eV~\AA$^{-2}$, using FIRE with the same threshold and maximum number of steps. A path was counted as a failure if the largest force component over all images and both endpoints was not below 0.1~eV~\AA$^{-1}$ at the end of the optimization. $E_\mathrm{m}$ was taken as the difference between the highest and the lowest energy along the path and reported in eV, and its MAE was computed over the converged paths.

Ionic conductivities $\sigma$ (in mS\,cm$^{-1}$) were evaluated with \texttt{DiffusionBenchmark} for 748 AIMD simulations of Li and Na conductors against PBE reference values from the MVL-batt dataset.\cite{kaplan2025matpes} As in the reference simulations, each structure was simulated in the NVT ensemble at the same temperature (300--2,100~K) for 110~ps, of which the first 10~ps were used for equilibration and discarded. The simulations used a time step of 1~fs, a Nos\'e--Hoover chain thermostat\cite{noseUnifiedFormulationConstant1984,hooverCanonicalDynamicsEquilibrium1985,martynaNoseHooverChains1992} with a damping time of 100~fs, and initial velocities drawn from the Maxwell--Boltzmann distribution. The diffusivity of the mobile ion was obtained from its mean-squared displacement with pymatgen-analysis-diffusion\cite{ongPhaseStabilityElectrochemical2013,moFirstPrinciplesStudy2012} and converted to $\sigma$ with the Nernst--Einstein relation. A simulation was counted as a failure if the fitted slope of the mean-squared displacement was not positive, in which case the code returns the numerical floor of the fit, $1.667\times10^{-17}$~cm$^2$~s$^{-1}$. The MAE of $\log_{10}\sigma$ was computed over the remaining simulations.

\paragraph{Surface and grain boundary energies.}
Surface energies $\gamma_\mathrm{surf}$ were evaluated with \texttt{SurfaceBenchmark} for 1,497 slabs of 73 elements against PBE reference values from the Crystalium database.\cite{tranSurfaceEnergiesElemental2016,jainCommentaryMaterialsProject2013} For each material, the bulk structure was relaxed in both atomic positions and cell to 0.05~eV~\AA$^{-1}$ using FIRE with a maximum of 500 steps. The atomic positions of the constructed slab structure were relaxed at fixed cell with the same settings. The surface energy was calculated as
\begin{equation}
\gamma_\mathrm{surf}=\frac{E_\mathrm{slab}-Ne_\mathrm{bulk}}{2A},
\end{equation}
where $E_\mathrm{slab}$ is the energy of the relaxed slab, $N$ is its number of atoms, $e_\mathrm{bulk}$ is the energy per atom of the relaxed bulk, and $A$ is the surface area, and reported in J~m$^{-2}$.

Grain boundary energies $\gamma_\mathrm{GB}$ were evaluated with \texttt{GBBenchmark} for 327 grain boundaries of 58 elements against PBE reference values from the Crystalium database.\cite{zhengGrainBoundaryProperties2020,jainCommentaryMaterialsProject2013} The bulk and the constructed grain boundary structures were relaxed in the same way as for the surface energies. The grain boundary energy was calculated as
\begin{equation}
\gamma_\mathrm{GB}=\frac{E_\mathrm{GB}-Ne_\mathrm{bulk}}{2A},
\end{equation}
where $E_\mathrm{GB}$ is the energy of the relaxed grain boundary structure, $N$ is its number of atoms, $e_\mathrm{bulk}$ is the energy per atom of the relaxed bulk, and $A$ is the grain boundary area, and reported in J~m$^{-2}$.

\paragraph{Adsorption energies.}
Adsorption energies $E_\mathrm{ads}$ were evaluated with \texttt{AdsorptionBenchmark} for 200 adsorption reactions from the CMR adsorption database\cite{schmidtBenchmarkDatabaseTransition2018} against PBE and r$^2$SCAN reference values computed in this work. The reactions involve eight adsorbates (H, O, N, \ce{N2}, CO, NO, CH, and OH) on 25 transition metals, each modeled by one adsorbate on a three-layer slab with a $1\times1$ surface cell. The reference calculations were performed with VASP\cite{kresseEfficientIterativeSchemes1996} and used a plane-wave cutoff of 680~eV, Gaussian smearing of 0.05~eV, a $k$-point spacing of 0.22~\AA$^{-1}$, and spin polarization, inherited from \texttt{MatPESStaticSet}. The r$^2$SCAN reference values were computed with the same settings. All reference calculations are single points on the CMR structures, in which the slabs are separated by a vacuum of about 10~\AA. The gas-phase references are \ce{H2}, \ce{O2}, and \ce{N2} for H, O, and N, the molecule itself for \ce{N2}, CO, and NO, \ce{CH4} and \ce{H2} for CH, and \ce{H2O} and \ce{H2} for OH. For the structures that contain V, Cr, Mn, Fe, Co, Ni, Mo, or W, PBE+$U$ reference values were also computed with the $U$ values of the Materials Project (Supplementary Section~\ref{si:ads-pbeu}). The CMR structures were evaluated with each FP at fixed geometry. The adsorption energy was calculated as
\begin{equation}
E_\mathrm{ads}=E_\mathrm{slab+ads}-E_\mathrm{slab}-\sum_i \nu_i E_i,
\end{equation}
where $E_\mathrm{slab+ads}$ and $E_\mathrm{slab}$ are the energies of the slab with and without the adsorbate and $E_i$ is the energy of the gas-phase molecule $i$ with stoichiometric coefficient $\nu_i$, and reported in eV.

\paragraph{Interfacial energies and works of adhesion.}
Interfacial energies $\gamma_\mathrm{int}$ and works of adhesion $W_\mathrm{adh}$ were evaluated with \texttt{InterfaceBenchmark} for 213 coherent interfaces between the 3d transition metals and their oxides against PBE reference values computed in this work. The 22 metal and 144 oxide phases are stable and metastable phases within 50~meV~atom$^{-1}$ of the convex hull in the Materials Project.\cite{jainCommentaryMaterialsProject2013} The interfaces were constructed with pymatgen\cite{ongPythonMaterialsGenomics2013} by matching slabs of the metal and the oxide\cite{zurLatticeMatchApplication1984}, with the oxide strained to the in-plane cell of the metal by at most 5\%. The cells contain 49--196 atoms and a vacuum of 18~\AA. The reference calculations were performed with VASP and used the PBE functional without a Hubbard $U$, spin polarization, a plane-wave cutoff of 520~eV, Gaussian smearing of 0.05~eV, $\Gamma$-centered $k$-point grids with a reciprocal density of 64~\AA$^{-3}$, and a fixed cell. The atoms of each interface were relaxed, and the metal and the oxide slabs were then taken from the relaxed interface and relaxed separately in the same cell. The relaxed interface and slab structures were evaluated with each FP at fixed geometry. The work of adhesion was calculated as
\begin{equation}
W_\mathrm{adh}=\frac{E_\mathrm{metal}+E_\mathrm{oxide}-E_\mathrm{int}}{A},
\end{equation}
where $E_\mathrm{int}$, $E_\mathrm{metal}$, and $E_\mathrm{oxide}$ are the energies of the interface, the metal slab, and the oxide slab, and $A$ is the interface area. The interfacial energy was calculated as
\begin{equation}
\gamma_\mathrm{int}=\gamma_\mathrm{metal}+\gamma_\mathrm{oxide}-W_\mathrm{adh},
\end{equation}
where $\gamma_\mathrm{metal}$ and $\gamma_\mathrm{oxide}$ are the surface energies of the metal and the oxide slabs, calculated as for the surface energies with the energy per atom of the unstrained bulk phase. Both quantities are reported in J~m$^{-2}$.

\paragraph{Computational cost.}
Timings were measured on one NVIDIA A100-SXM4-40GB GPU with 16 cores of an AMD EPYC 7763 processor, in single precision and in a new process for each test. MD used \texttt{MDCalc} in the NVT ensemble at 300\,K with a time step of 1\,fs for diamond Si with 1,000, 4,096, and 10,648 atoms; 50 steps were timed after 50 steps of warm-up. Each configuration was run at least twice in separate jobs, and medians are reported. CPU timings used 32 cores of a node with two Intel Xeon Gold 6354 processors.

\begin{acknowledgement}

This work was intellectually led by the U.S. Department of Energy, Office of Science, Office of Basic Energy Sciences, Materials Sciences and Engineering Division under contract No. DE-AC02-05-CH11231 (Materials Project program KC23MP). T. W. K. acknowledges the support of the Eric and Wendy Schmidt AI in Science Postdoctoral Fellowship, a Schmidt Futures program. A. S. R. acknowledges support from the Wilke 1989 School of Engineering and Applied Science Innovation Fund at Princeton University. S. P. O. also acknowledges support from the National University of Singapore under WBS \# A-0010359-00-00. Computational resources were provided by the National Supercomputing Centre (NSCC), Singapore, and by NERSC, a U.S. Department of Energy Office of Science User Facility.

During the preparation of this manuscript, the authors used Claude (Anthropic; Claude Opus 5.5) to proof-read and suggest edits, phrasing and literature references. The authors reviewed and edited all AI suggestions, and take full responsibility for the content of this publication. AI tools were not used to generate data or produce figures.

\end{acknowledgement}

\section*{Code and Data Availability}

MatCalc is released under the BSD 3-Clause license and is available from PyPI (\texttt{pip install matcalc}) and from the public GitHub repository at \url{https://github.com/materialyzeai/matcalc}. Full API documentation and tutorials are hosted at \url{https://matcalc.ai}, and the current MatCalc-Bench leaderboard at \url{https://bench.matcalc.ai}. All reference data of MatCalc-Bench are distributed with the benchmark in the form used in this work via the Hugging Face dataset repository \texttt{materialyze/matcalc-bench}.

\section*{Author Contributions}

S.P.O.\ conceived and supervised the project and designed the API and architecture of MatCalc. R.L.\ led the implementation of the \texttt{PESCalculator}, \texttt{PropCalc}, and \texttt{Benchmark} classes, developed MatCalc-Bench, performed the benchmark and reference calculations, and analyzed the results. A.C.T.\ contributed to the surface and grain boundary calculators, M.K.P.\ to the adsorption calculator, S.K.\ to the interface calculator, J.Q.\ to the NEB calculator, and A.S.R.\ to the phonon and QHA calculators. T.W.K.\ and A.C.T.\ contributed to \texttt{PESCalculator}. R.L.\ and S.P.O.\ wrote the manuscript. All authors reviewed the manuscript.

\section*{Competing Interests}

The authors declare no competing interests.

\begin{suppinfo}

Potentials supported by MatCalc and evaluated in this work; MatCalc-Bench reference data and element coverage; detailed results for the elemental crystals, formation energies, heat capacities, surface, grain boundary, adsorption, and interfacial energies; analyses of the effects of the cutoff radius, the maximum angular order, the Hubbard $U$ correction, and the negative Y surface energies of MACE-MPA; and a comparison of conservative and direct-force models (PDF).

\end{suppinfo}

\bibliography{references,new_refs}

\clearpage

\end{document}


\maketitle
\setstretch{1.12}
{\small\linespread{0.92}\selectfont\tableofcontents}
\clearpage

\section{Further property workflows and benchmarks in MatCalc}
\label{si:more-classes}

Table~\ref{si:tab-more-classes} lists the classes of MatCalc that are under development or not used in this work.

\begin{table}[htbp]
\centering\small
\caption{\textbf{Further property workflows and benchmarks in MatCalc.} $^\dagger$Under development.}
\label{si:tab-more-classes}
\begin{tabular}{@{}>{\raggedright\arraybackslash}p{0.30\linewidth}>{\raggedright\arraybackslash}p{0.64\linewidth}@{}}
\toprule
Class & Result \\
\midrule
\texttt{PheasyCalc}$^\dagger$ & Second- and higher-order force constants with pheasy \\
\texttt{AlamodeCalc}$^\dagger$ & Second- and higher-order force constants with ALAMODE \\
\texttt{FourPhononCalc}$^\dagger$ & Lattice thermal conductivity with three- and four-phonon scattering \\
\texttt{MCCalc}$^\dagger$ & Metropolis Monte Carlo sampling with user-defined moves \\
\texttt{IntercalationCalc}$^\dagger$ & Configurations and energies as a function of the concentration of an intercalating ion \\
\texttt{VoltageCalc}$^\dagger$ & Intercalation voltage profile, capacity, and energy density \\
\texttt{SofteningBenchmark} & Softening of the PES in high-energy configurations \\
\bottomrule
\end{tabular}
\end{table}

\section{Interatomic potentials}
\label{si:potentials}

\subsection{Potentials supported by MatCalc}
\label{si:supported}

MatCalc evaluates the PES through an ASE calculator, so that any potential for which an ASE calculator exists can be used. Table~\ref{si:tab-supported} lists the potentials that can be used with MatCalc, in three groups. For the first group, the model registry of MatCalc assigns names to over 100 pretrained FPs from nine packages, which are loaded by name; the complete list of names is maintained at \url{https://matcalc.ai/models.html}. Other models of these packages are loaded by the identifier or file that the package itself accepts. For the second group, potentials trained by the user are loaded from their files; for MatGL, this includes fine-tuned models, multi-fidelity models, and architectures without a pretrained FP, such as SO3Net. The third group gives examples of potentials without a dedicated loader, whose ASE calculators are passed to MatCalc directly. Besides FPs whose forces are the gradient of the predicted energy, the registry contains FPs that predict forces directly (Orb-v2 and the direct variants of Orb-v3), and the PET models can be evaluated in either way.

\begingroup
\footnotesize
\setlength{\tabcolsep}{4pt}
\renewcommand{\arraystretch}{1.08}
\setlength{\LTcapwidth}{\linewidth}
\begin{longtable}{@{}>{\raggedright\arraybackslash}p{0.155\linewidth}>{\raggedright\arraybackslash}p{0.13\linewidth}c>{\raggedright\arraybackslash}p{0.60\linewidth}@{}}
\caption{\textbf{Interatomic potentials supported by MatCalc.} Package names link to the source code. FPs is the number of pretrained foundation potentials of the architecture in the model registry of MatCalc.}
\label{si:tab-supported}\\
\toprule
Package & Potential & FPs & Description \\
\midrule
\endfirsthead
\toprule
Package & Potential & FPs & Description \\
\midrule
\endhead
\bottomrule
\endfoot
\multicolumn{4}{@{}l}{\textit{Pretrained foundation potentials, loaded by name from the model registry}}\\*
\href{https://github.com/materialyzeai/matgl}{matgl} & TensorNet & 4 & O(3)-equivariant message passing with Cartesian tensor features.\cite{Simeon2023TensorNet} \\
 & QET & 2 & TensorNet with an analytically solved charge equilibration that scales linearly with system size.\cite{ko2026} \\
 & M3GNet & 2 & Graph network with three-body interactions.\cite{chenUniversalGraphDeep2022} \\
 & CHGNet & 4 & Graph network with three-body interactions, trained on magnetic moments in addition to energies, forces, and stresses.\cite{dengCHGNetPretrainedUniversal2023} \\
\href{https://github.com/ACEsuit/mace}{mace} & MACE & 16 & Equivariant message passing with higher-body-order messages; includes multi-head variants.\cite{Batatia2022MACE} \\
\href{https://github.com/MDIL-SNU/SevenNet}{sevenn} & SevenNet & 7 & Equivariant message passing based on the NequIP architecture; includes multi-fidelity variants.\cite{park2024} \\
\href{https://github.com/ICAMS/grace-tensorpotential}{tensorpotential} & GRACE & 26 & Atomic cluster expansion extended by graph basis functions for semilocal interactions.\cite{bochkarevGraphAtomicCluster2024a} \\
\href{https://github.com/orbital-materials/orb-models}{orb-models} & Orb & 13 & Attention-based graph network without built-in rotational equivariance; forces predicted directly or as energy gradients.\cite{neumann2024,rhodesOrbv3AtomisticSimulation2025} \\
\href{https://github.com/microsoft/mattersim}{mattersim} & MatterSim & 2 & M3GNet architecture trained on structures sampled over wide ranges of temperature and pressure.\cite{yangMatterSimDeepLearning2024} \\
\href{https://github.com/facebookresearch/fairchem}{fairchem} & UMA & 4 & Equivariant network with a mixture of linear experts, trained jointly on molecules, materials, and catalysts and conditioned on the task.\cite{Wood2025UMA} \\
\href{https://github.com/lab-cosmo/upet}{upet} & PET & 12 & Point-edge transformer without built-in rotational equivariance.\cite{pozdnyakovSmoothExactRotational2023} \\
\href{https://github.com/deepmodeling/deepmd-kit}{deepmd-kit} & DPA2 & 1 & Large atomic model pretrained as a multi-task learner.\cite{zhangDPA2LargeAtomic2024} \\
 & DPA3 & 3 & Multilayer graph network on line graph series, with multi-task training.\cite{Zhang2026DPA3} \\
 & DPA4 & 5 & SE(3)-equivariant network with SO(2) convolutions in edge-local frames.\cite{liDPA4PushingAccuracyCost2026} \\
\midrule
\multicolumn{4}{@{}l}{\textit{Potentials trained by the user, loaded from their files}}\\*
\href{https://github.com/materialyzeai/matgl}{matgl} & SO3Net & -- & SO(3)-equivariant message passing with spherical harmonics and Clebsch--Gordan tensor products.\cite{Schutt2023SchNetPack} \\
 & Multi-fidelity M3GNet & -- & M3GNet trained on data of several levels of theory, with the level given as a global state label.\cite{koDataEfficientConstructionHighFidelity2024} \\
\href{https://github.com/deepmodeling/deepmd-kit}{deepmd-kit} & Deep Potential & -- & Many-body potential from a deep neural network that preserves the symmetries of the system.\cite{zhangDeepPotentialMolecular2018}\cite{zengDeePMDkitV2Software2023} \\
\href{https://github.com/materialyzeai/maml}{maml} & GAP & -- & Gaussian process regression on descriptors of the local atomic environment.\cite{Bartok2010GAP} \\
 & MTP & -- & Systematically improvable basis of invariants built from moment tensors of the local atomic environment.\cite{Shapeev2016MTP} \\
 & NNP & -- & Atomic neural networks on atom-centered symmetry functions.\cite{Behler2007NNP} \\
 & SNAP & -- & Linear model in the bispectrum components of the local neighbor density.\cite{Thompson2015SNAP} \\
 & qSNAP & -- & SNAP with additional quadratic terms in the bispectrum components.\cite{woodExtendingAccuracySNAP2018} \\
\href{https://github.com/ICAMS/python-ace}{python-ace} & ACE & -- & Expansion in local, complete basis functions of many-atom interactions.\cite{Drautz2019ACE} \\
\href{https://github.com/mir-group/nequip}{nequip} & NequIP & -- & E(3)-equivariant convolutions acting on geometric tensors.\cite{batznerEquivariantGraphNeural2022} \\
\midrule
\multicolumn{4}{@{}l}{\textit{Examples of other potentials, passed as ASE calculators}}\\*
\href{https://github.com/facebookresearch/fairchem}{fairchem} (v1) & eSEN & -- & Equivariant network designed for a smooth potential energy surface that conserves energy in molecular dynamics.\cite{Fu2025ESEN} \\
 & EquiformerV2 & -- & Equivariant transformer with eSCN convolutions for higher-degree representations.\cite{Liao2024EquiformerV2} \\
\href{https://github.com/mir-group/nequip}{nequip} & NequIP-OAM, Allegro-OAM & -- & Pretrained FPs of the NequIP and Allegro architectures, compiled with the package and used through its ASE calculator.\cite{kavanaghFastAccurateEquivariant2026} \\
ASE &  & -- & Empirical potentials and interfaces to electronic-structure codes.\cite{larsenAtomicSimulationEnvironment2017} \\
LAMMPS &  & -- & Any LAMMPS force field, through the \texttt{LAMMPSCalculator} of MatCalc.\cite{Thompson2022LAMMPS} \\
\end{longtable}
\endgroup

\subsection{Foundation potentials evaluated in this work}
\label{si:evaluated}

Table~\ref{si:tab-model-table} lists the architectures and training settings of the FPs evaluated in this work.

\begin{center}
\rotatebox{90}{%
\begin{minipage}{0.96\textheight}
\centering
\captionof{table}{\textbf{Foundation potentials evaluated in this work.} All models are conservative and cover the same 89 elements. $r_\mathrm{c}$, cutoff radius; $T$, message-passing layers. MACE-MatPES-PBE and MACE-MatPES-r2SCAN were fine-tuned on MatPES from a model pretrained on OMat24.}
\label{si:tab-model-table}
\setstretch{1.0}
\fontsize{8}{9.6}\selectfont
\setlength{\tabcolsep}{4pt}
\renewcommand{\arraystretch}{1.0}
\begin{tabular}{@{}L{3.45cm}L{1.3cm}L{2.5cm}rccL{2.1cm}L{2.9cm}L{2.25cm}L{2.7cm}@{}}
\toprule
Model & Training structures & Functional; PAW set & Params. & $r_\mathrm{c}$\,(\AA) & $T$ & Channels & Angular and body order & Radial basis & Readout; activation\\
\midrule
\multicolumn{10}{@{}l}{\textit{MPtrj}}\\
MACE-MPtrj & 1.58M & PBE/PBE+$U$; PBE & 4.69M & 6.0 & 2 & 128 (128x0e + 128x1o) & $l_\mathrm{max}=3$; $L=1$; correlation 3 (4-body) & 10 Bessel, fixed; MLP $64^3$ & Linear + MLP(16); SiLU\\
\addlinespace[3pt]
\multicolumn{10}{@{}l}{\textit{MPA (MPtrj + sAlex)}}\\
MACE-MPA & 12.0M & PBE/PBE+$U$; PBE & 9.06M & 6.0 & 2 & 128 (128x0e + 128x1o) & $l_\mathrm{max}=3$; $L=1$; correlation 3 (4-body) & 10 Bessel, fixed; MLP $64^3$ & Linear + MLP(16); SiLU\\
\addlinespace[3pt]
\multicolumn{10}{@{}l}{\textit{OMat24}}\\
MACE-OMat24 & 100.8M & PBE/PBE+$U$; PBE54 & 9.06M & 6.0 & 2 & 128 (128x0e + 128x1o) & $l_\mathrm{max}=3$; $L=1$; correlation 3 (4-body) & 10 Bessel, fixed; MLP $64^3$ & Linear + MLP(16); SiLU\\
\addlinespace[2pt]
GRACE-OMat24 & 100.8M & PBE/PBE+$U$; PBE54 & 26.4M & 6.0 & 2 & Embedding 128; radial 42/32; product 42/64 & $l_\mathrm{max}=4/3$ by layer; indicator $L=3$; product order 4 & 10 Chebyshev; MLP $64^2$ & 1 linear + 16 densities via MLP(64); SiLU, tanh\\
\addlinespace[3pt]
\multicolumn{10}{@{}l}{\textit{MatPES-PBE}}\\
MACE-MatPES-PBE & 435K & PBE; PBE64 & 9.06M & 6.0 & 2 & 128 (128x0e + 128x1o) & $l_\mathrm{max}=3$; $L=1$; correlation 3 (4-body) & 10 Bessel, fixed; MLP $64^3$ & Linear + MLP(16); SiLU\\
\addlinespace[2pt]
TensorNet-MatPES-PBE & 433K & PBE; PBE64 & 837K & 5.0 & 2 & 128 & Rank-2 Cartesian ($l\leq 2$) & 3 spherical Bessel & Gated MLP(128, 128); SiLU\\
\addlinespace[2pt]
QET-MatPES-PBE & 433K & PBE; PBE64 & 905K & 5.0 & 2 & 128 & Rank-2 Cartesian ($l\leq 2$) & 3 spherical Bessel & Gated MLP on features, $q_i$ and $V_i$; SiLU, softplus\\
\addlinespace[3pt]
\multicolumn{10}{@{}l}{\textit{MatPES-r$^2$SCAN}}\\
MACE-MatPES-r2SCAN & 388K & r$^2$SCAN; PBE64 & 9.06M & 6.0 & 2 & 128 (128x0e + 128x1o) & $l_\mathrm{max}=3$; $L=1$; correlation 3 (4-body) & 10 Bessel, trainable; MLP $64^3$ & Linear + MLP(16); SiLU\\
\addlinespace[2pt]
TensorNet-MatPES-r2SCAN & 387K & r$^2$SCAN; PBE64 & 837K & 5.0 & 2 & 128 & Rank-2 Cartesian ($l\leq 2$) & 3 spherical Bessel & Gated MLP(128, 128); SiLU\\
\addlinespace[2pt]
QET-MatPES-r2SCAN & 387K & r$^2$SCAN; PBE64 & 905K & 5.0 & 2 & 128 & Rank-2 Cartesian ($l\leq 2$) & 3 spherical Bessel & Gated MLP on features, $q_i$ and $V_i$; SiLU, softplus\\
\bottomrule
\end{tabular}
\end{minipage}%
}
\end{center}

\section{MatCalc-Bench data}
\label{si:bench}

\begin{figure}[!htbp]
\centering
\includegraphics[width=\textwidth]{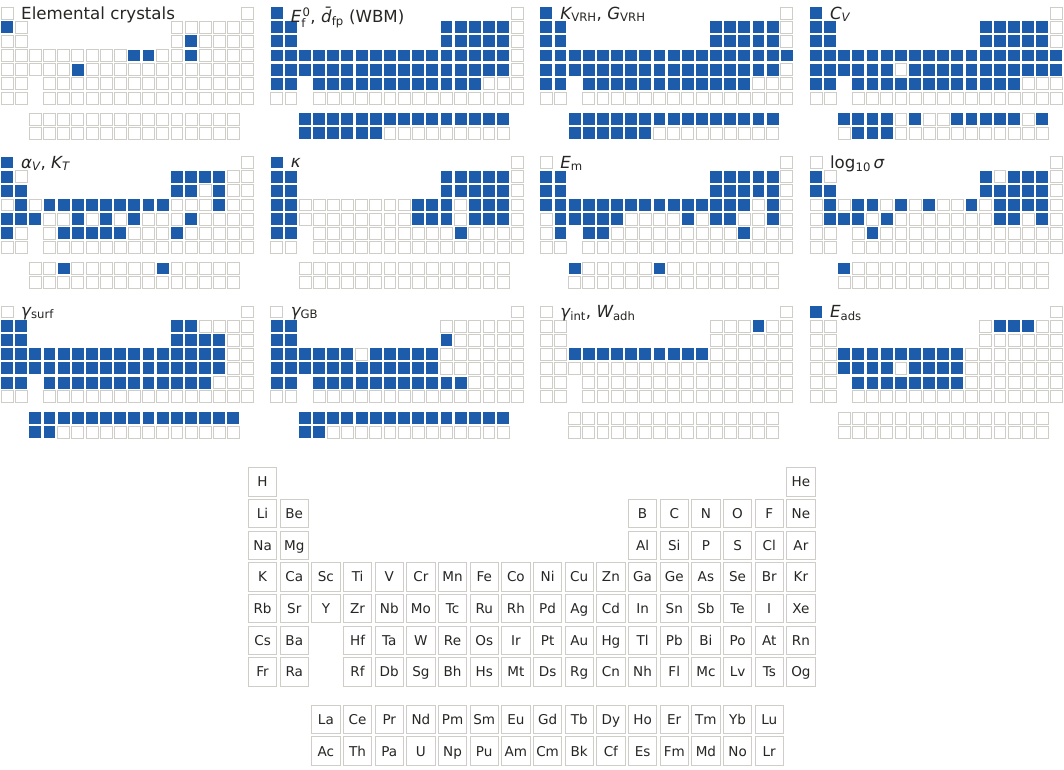}
\caption{\textbf{Element coverage of the MatCalc-Bench properties.} Each small periodic table marks in blue the elements present in the structures of one benchmark: the elemental crystals, formation energies and structural similarity (WBM), elastic moduli, heat capacities, thermal expansion coefficients and isothermal bulk moduli, lattice thermal conductivities, Li migration barriers, ionic conductivities, surface energies, grain boundary energies, interfacial energies and works of adhesion, and adsorption energies (CMR). The table at the bottom gives the element positions.}
\label{si:element-coverage}
\end{figure}

Table~\ref{si:bench-tasks} lists the reference data of each MatCalc-Bench property, and Fig.~\ref{si:element-coverage} shows the elements covered by each benchmark.

\begin{table}[htbp]
\centering
\caption{\textbf{Structures and reference data of MatCalc-Bench.} Structures: origin of the structures. Values: origin of the reference values, given only where it differs from that of the structures. $N$: number of materials, structures, paths, trajectories, reactions, or interfaces. $^{*}$References of Table~1 of the main text. Shaded: benchmarks with more than one reference for the same structures. WBM: original relaxations at the Materials Project settings\cite{wangPredictingStableCrystalline2021} (PBE/PBE+$U$, PBE\_54 pseudopotentials, 520~eV cutoff; 1,000 structures) and re-relaxations at the MatPES settings\cite{kaplan2025matpes} (PBE, PBE\_64 pseudopotentials, 680~eV cutoff; 987 structures), used for the FPs not trained and trained on MatPES, respectively (Supplementary Section~\ref{si:wbm}).}
\label{si:bench-tasks}
\footnotesize
\setlength{\tabcolsep}{4.5pt}
\renewcommand{\arraystretch}{1.1}
\definecolor{multiref}{gray}{0.92}
\resizebox{\textwidth}{!}{%
\begin{tabular}{@{}ll lr lr lr@{}}
\toprule
 & & \multicolumn{2}{c}{PBE} & \multicolumn{2}{c}{r$^2$SCAN} & \multicolumn{2}{c}{Expt.} \\
\cmidrule(lr){3-4}\cmidrule(lr){5-6}\cmidrule(l){7-8}
Property & Structures & Values & $N$ & Values & $N$ & Values & $N$ \\
\midrule
\multicolumn{8}{@{}l}{\textit{Relaxation and elasticity}} \\
\rowcolor{multiref} Structural similarity ($\bar{d}_\mathrm{fp}$) & WBM & Refs.~\citenum{wangPredictingStableCrystalline2021}/\citenum{kaplan2025matpes} & 1,000/987$^{*}$ & & & &  \\
 & GNoME\cite{merchantScalingDeepLearning2023} & & & Ref.~\citenum{kingsburyFlexibleScalableScheme2022} & 1,000 & &  \\
\rowcolor{multiref} Formation energy ($E_\mathrm{f}^{0}$) & WBM & Refs.~\citenum{wangPredictingStableCrystalline2021}/\citenum{kaplan2025matpes} & 1,000/987$^{*}$ & & & &  \\
 & GNoME & & & Ref.~\citenum{kingsburyFlexibleScalableScheme2022} & 1,000 & &  \\
Formation enthalpy ($\Delta H_\mathrm{f}$) & MP\cite{jainCommentaryMaterialsProject2013} & & & & & Refs.~\citenum{wangFrameworkQuantifyingUncertainty2021,kubaschewskiMaterialsThermochemistry1993,kimExperimentalFormationEnthalpies2017,chase1998janaf} & 1,723  \\
Lattice parameter ($a$) & Ref.~\citenum{tranOrbitalfreeApproximationsKineticenergy2018} & & & & & Refs.~\citenum{schimkaImprovedHybridFunctional2011,haasCalculationLatticeConstant2009,lejaeghere2014} & 44  \\
Bulk modulus ($K_\mathrm{VRH}$) & MP & Ref.~\citenum{dejongChartingCompleteElastic2015} & 12,122$^{*}$ & & & &  \\
Shear modulus ($G_\mathrm{VRH}$) & MP & Ref.~\citenum{dejongChartingCompleteElastic2015} & 12,122$^{*}$ & & & &  \\
Bulk modulus ($K$) & MP & & & & & Ref.~\citenum{toherCombiningAFLOWGIBBS2017} & 62  \\
Shear modulus ($G$) & MP & & & & & Ref.~\citenum{toherCombiningAFLOWGIBBS2017} & 52  \\
\addlinespace
\multicolumn{8}{@{}l}{\textit{Vibration and thermodynamics}} \\
Heat capacity ($C_V$) & Alexandria\cite{loewUniversalMachineLearning2025} & & 1,170$^{*}$ & & & &  \\
\rowcolor{multiref}  & phonondb\cite{togoDistributionsPhononLifetimes2015} & This work & 95 & This work & 95 & &  \\
Thermal expansion ($\alpha_V$) & Refs.~\citenum{fan2022bicationic,sun2019thermal,fekete2023thermalshock} & & 47$^{*}$ & & & &  \\
 & MP & & & & & Refs.~\citenum{mochizuki2024thermal,lide2005,touloukianThermalExpansionNonmetallic1977,feiThermalExpansion1995,swensonRecommendedValuesThermal1983,arblasterCrystallographicPropertiesRuthenium2013,kirbySRM736Copper1990,kirbySRM737Tungsten1976} & 71  \\
Isothermal bulk modulus ($K_T$) & Refs.~\citenum{fan2022bicationic,sun2019thermal,fekete2023thermalshock} & & 47$^{*}$ & & & &  \\
 & MP & & & & & Refs.~\citenum{mochizuki2024thermal,kittel2005,gschneidner1964,toherCombiningAFLOWGIBBS2017,nathHighthroughputPredictionFinitetemperature2016,andersonElasticConstantsMantle1995,hollandImprovedExtendedInternally2011,lejaeghere2014,ledbetterElasticPropertiesMetals1974,featherstonElasticConstantsTantalum1963,suzukiFreeOscillationElastically1992,palkoElasticConstantsYttria2001,alkhatatbehPhaseDiagram1052010,alkhatatbehPhaseRelationsHardness2010} & 61  \\
Gr\"{u}neisen parameter ($\gamma$) & MP & & & & & Refs.~\citenum{mochizuki2024thermal,gschneidner1964,gruneisenTheorieFestenZustandes1912,hasegawa1980,andersonElasticConstantsMantle1995,toherHighthroughputComputationalScreening2014,toherCombiningAFLOWGIBBS2017} & 48  \\
Thermal conductivity ($\kappa$) & phonondb\cite{togoDistributionsPhononLifetimes2015,sekoPredictionLowThermalConductivityCompounds2015} & & 103$^{*}$ & & & &  \\
\rowcolor{multiref}  & phonondb & & 50 & Ref.~\citenum{kimSevenNetOmni2026} & 50 & Refs.~\citenum{togoDistributionsPhononLifetimes2015,chenMachineLearningModels2019,toherCombiningAFLOWGIBBS2017,kangExperimentalObservationHigh2018,liHighThermalConductivity2018,tianUnusualHighThermal2018a,chengGoodSolidStateElectrolytes2021,liEnhancedThermoelectricPerformance2014,strzalkowskiAlloyDisorderingEffects2023} & 50  \\
\addlinespace
\multicolumn{8}{@{}l}{\textit{Migration and diffusion}} \\
Li migration barrier ($E_\mathrm{m}$) & nebDFT2k\cite{dembitskiy2025benchmarking} & & 1,681$^{*}$ & & & &  \\
Ionic conductivity ($\log_{10}\sigma$) & MVL-batt\cite{kaplan2025matpes} & & 748$^{*}$ & & & &  \\
\addlinespace
\multicolumn{8}{@{}l}{\textit{Surfaces and grain boundaries}} \\
Surface energy ($\gamma_\mathrm{surf}$) & Crystalium\cite{tranSurfaceEnergiesElemental2016,jainCommentaryMaterialsProject2013} & & 1,497$^{*}$ & & & &  \\
Grain boundary energy ($\gamma_\mathrm{GB}$) & Crystalium\cite{zhengGrainBoundaryProperties2020,jainCommentaryMaterialsProject2013} & & 327$^{*}$ & & & &  \\
\addlinespace
\multicolumn{8}{@{}l}{\textit{Adsorption and interfaces}} \\
\rowcolor{multiref} Adsorption energy ($E_\mathrm{ads}$) & CMR\cite{schmidtBenchmarkDatabaseTransition2018} & This work & 200$^{*}$ & This work & 200 & &  \\
\rowcolor{multiref}  & ADS41\cite{trepteDatadrivenConstrainedOptimization2022} & & 41 & & 41 & Refs.~\citenum{wellendorffBenchmarkDatabaseAdsorption2015,mallikarjunsharadaAdsorptionTransitionMetal2019} & 41  \\
 & Catalysis-Hub\cite{wintherCatalysisHubOpenElectronic2019,yohannesCombinedHighThroughputDFT2023} & & 12,935 & & & &  \\
Interfacial energy ($\gamma_\mathrm{int}$) & This work & & 213$^{*}$ & & & &  \\
Work of adhesion ($W_\mathrm{adh}$) & This work & & 213$^{*}$ & & & &  \\
\bottomrule
\end{tabular}}
\end{table}

\section{Elemental crystals}
\label{si:elemental}

\begin{figure}[htbp]
\centering
\includegraphics[width=0.8\textwidth]{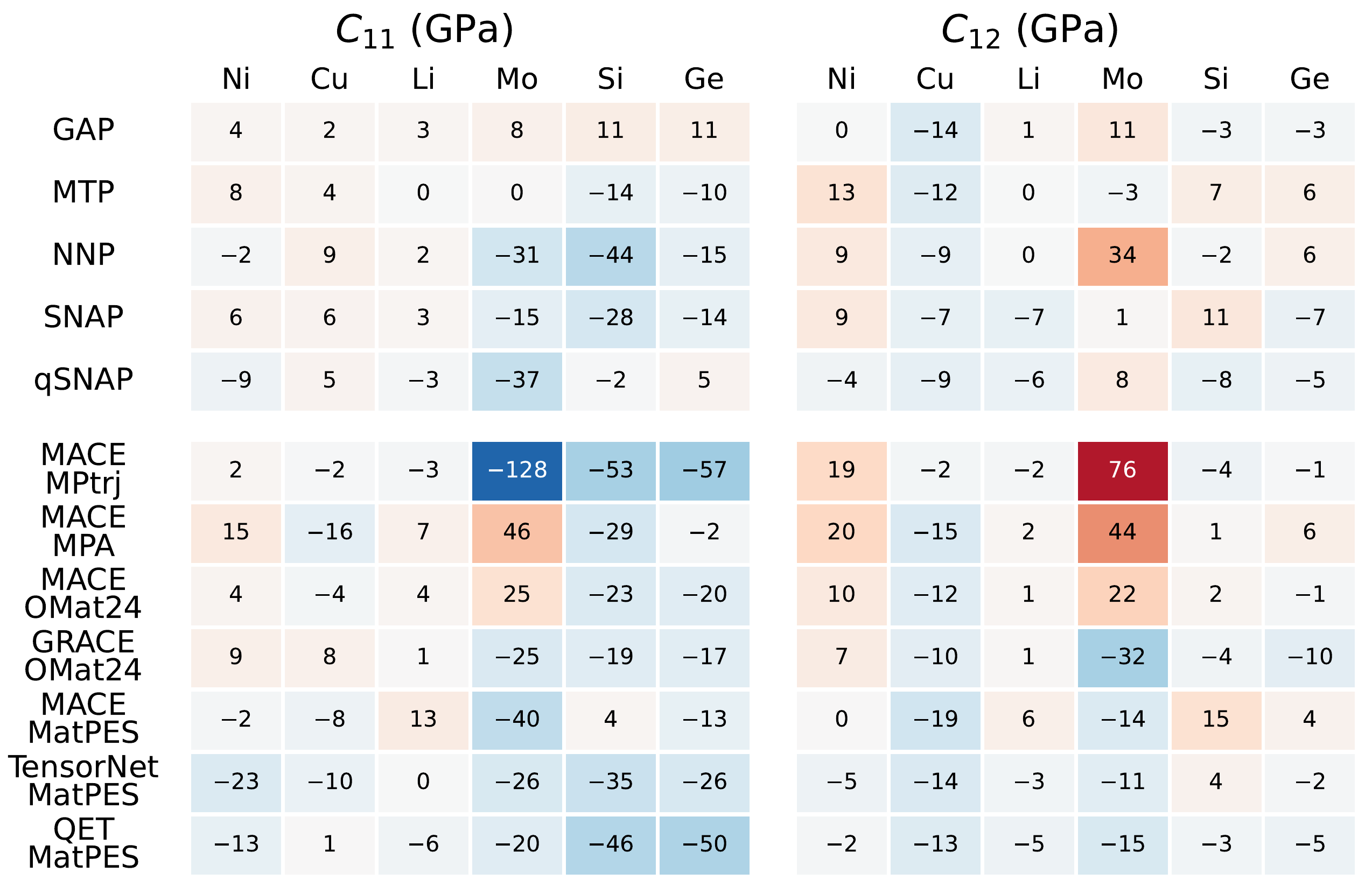}
\caption{\textbf{Per-element errors of the elastic constants $C_{11}$ and $C_{12}$ (GPa).} Each cell gives the signed error (model $-$ DFT-PBE) for one element; blue indicates underestimation and red overestimation, with a color scale symmetric about zero and limited by the largest absolute error of that property. References are the DFT-PBE values of \citet{zuoPerformanceCostAssessment2020}.}
\label{si:elem-c11-c12}
\end{figure}

\begin{figure}[htbp]
\centering
\includegraphics[width=0.8\textwidth]{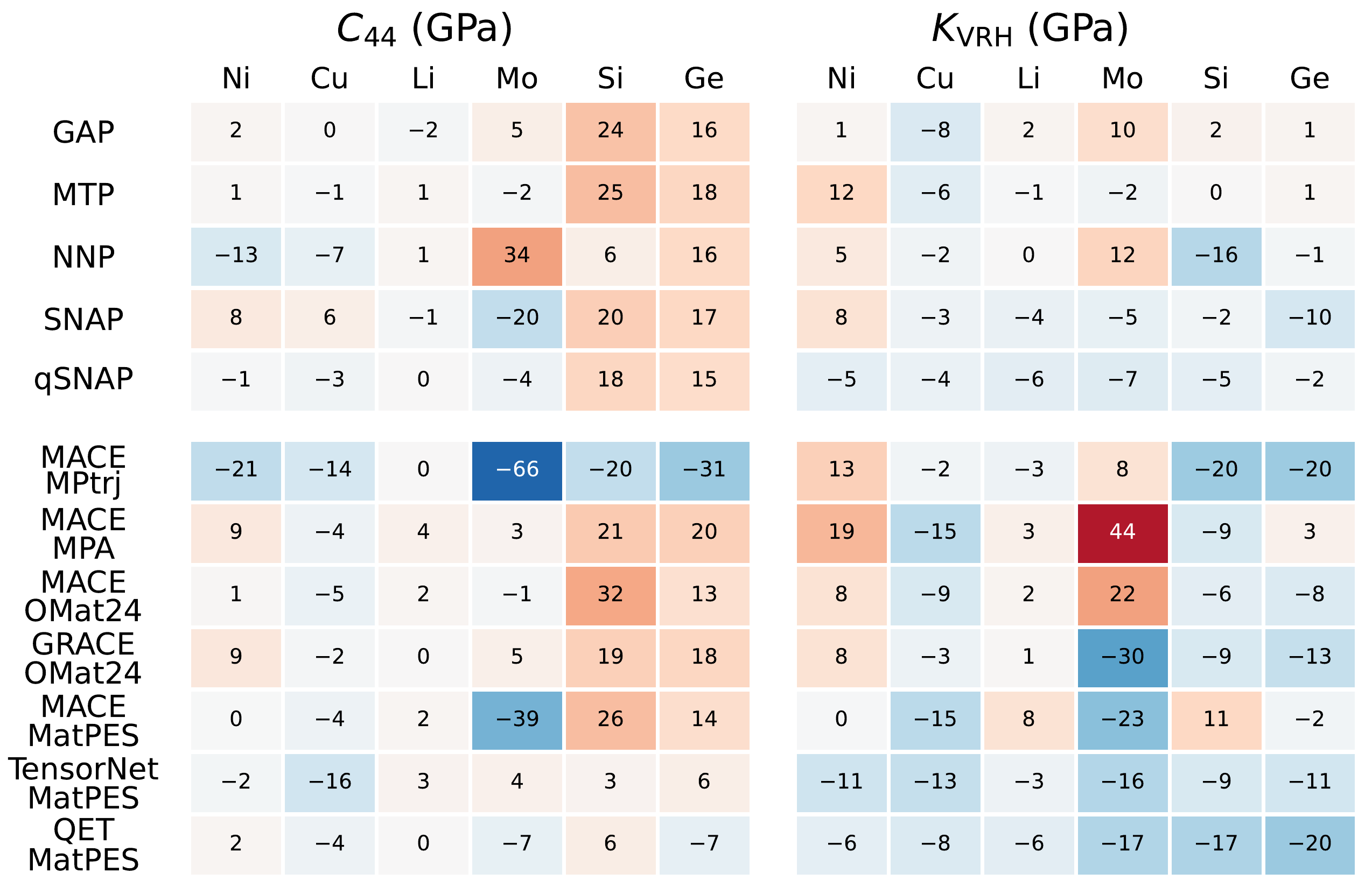}
\caption{\textbf{Per-element errors of the elastic constant $C_{44}$ and the Voigt--Reuss--Hill bulk modulus $K_\mathrm{VRH}$ (GPa).} Each cell gives the signed error (model $-$ DFT-PBE) for one element; blue indicates underestimation and red overestimation, with a color scale symmetric about zero and limited by the largest absolute error of that property. References are the DFT-PBE values of \citet{zuoPerformanceCostAssessment2020}.}
\label{si:elem-c44-bvrh}
\end{figure}

\begin{figure}[htbp]
\centering
\includegraphics[width=0.8\textwidth]{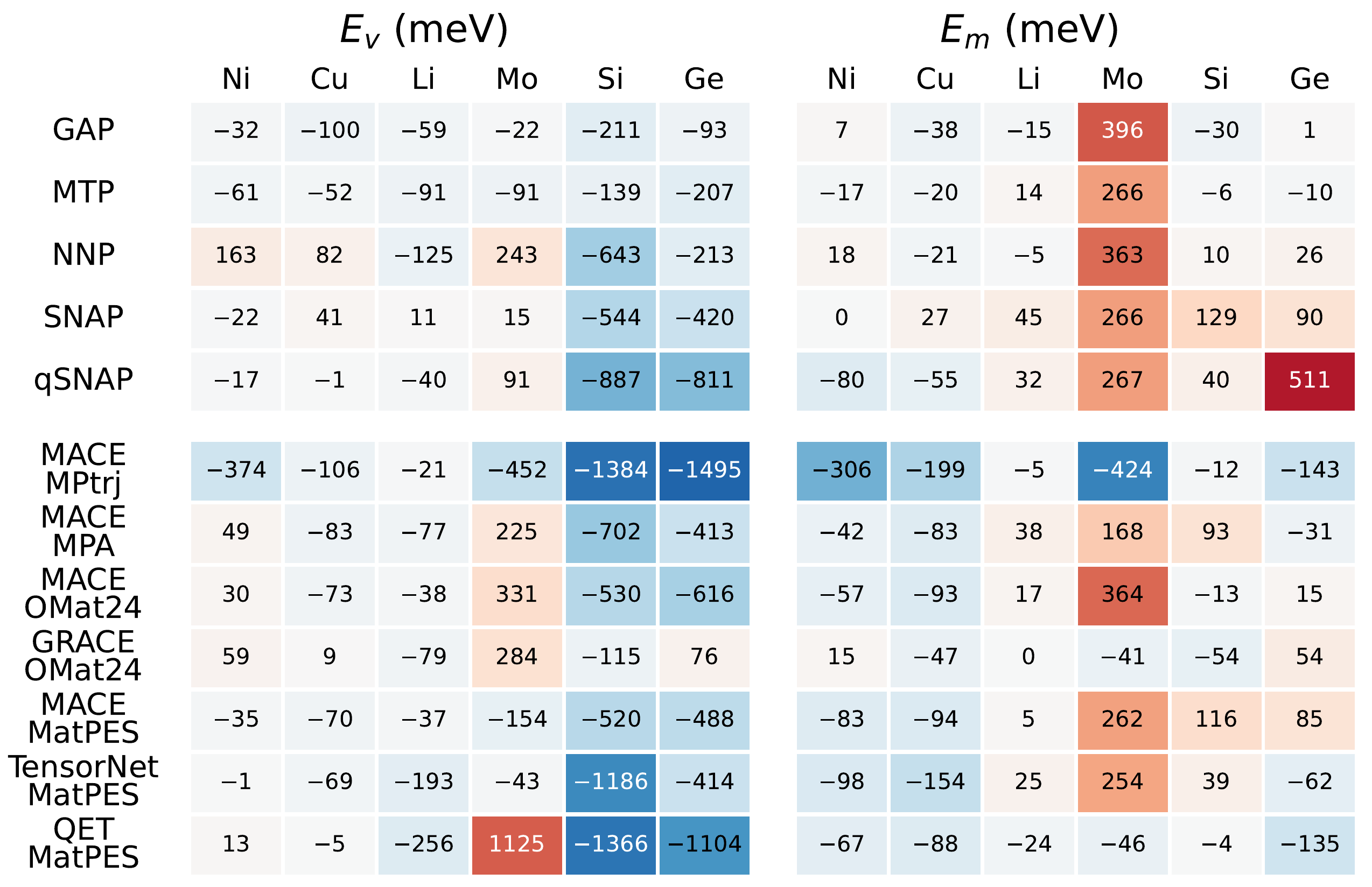}
\caption{\textbf{Per-element errors of the vacancy formation energy $E_v$ and the vacancy migration barrier $E_m$ (meV).} Each cell gives the signed error (model $-$ DFT-PBE) for one element; blue indicates underestimation and red overestimation, with a color scale symmetric about zero and limited by the largest absolute error of that property. References are the DFT-PBE values of \citet{zuoPerformanceCostAssessment2020}.}
\label{si:elem-ev-em}
\end{figure}

\begin{figure}[htbp]
\centering
\includegraphics[width=0.8\textwidth]{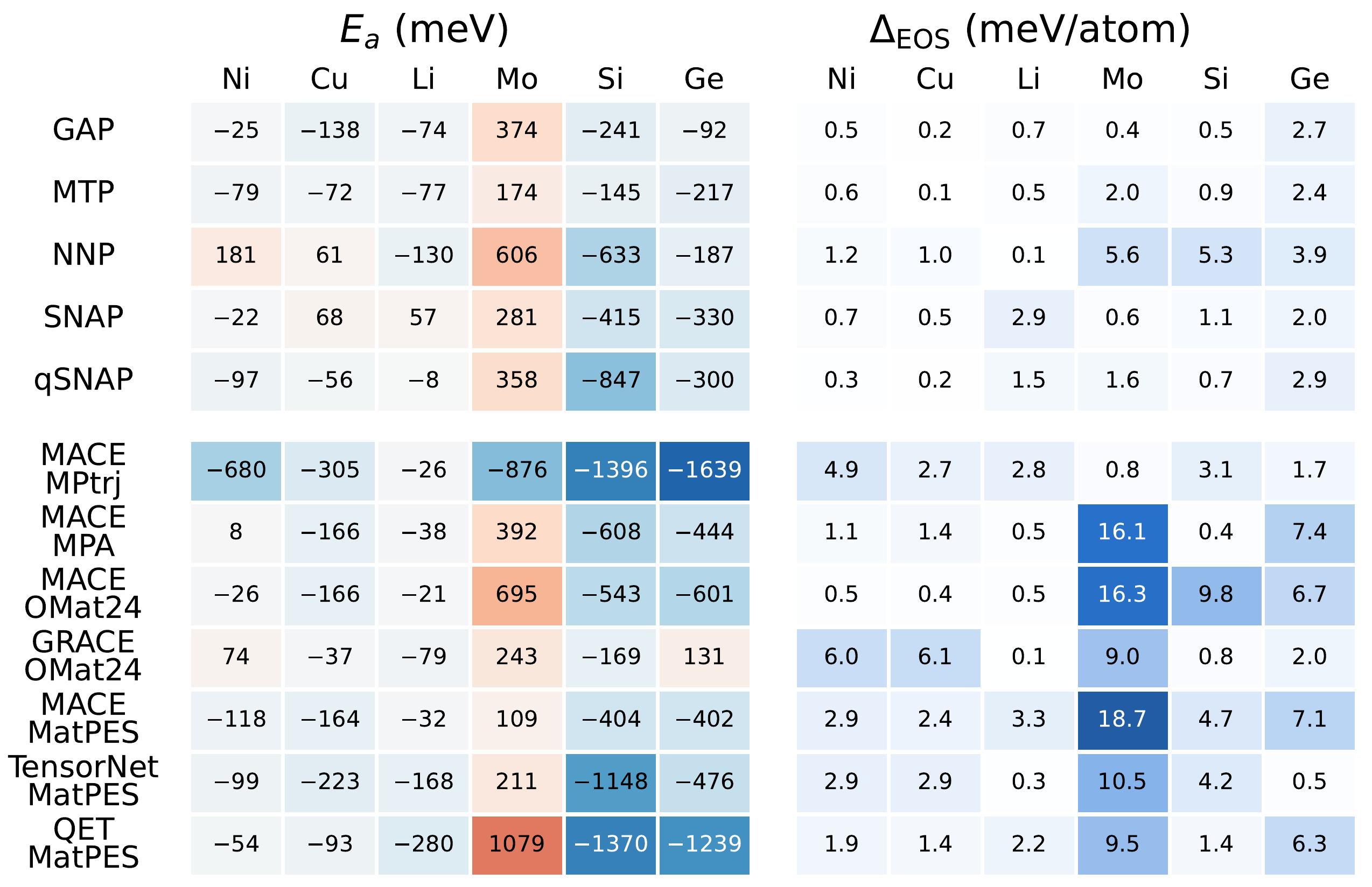}
\caption{\textbf{Per-element errors of the activation energy $E_a = E_v + E_m$ (meV) and the equation-of-state deviation $\Delta_\mathrm{EOS}$ (meV\,atom$^{-1}$).} Left: signed errors of $E_a$ relative to the DFT-PBE values of \citet{zuoPerformanceCostAssessment2020}, colored as in Fig.~\ref{si:elem-c11-c12}. Right: $\Delta_\mathrm{EOS}$, which is non-negative, relative to the PBE equation of state, computed over a $\pm6\%$ volume window about the mean of the DFT and model equilibrium volumes.\cite{lejaeghere2014}}
\label{si:elem-ea-deos}
\end{figure}

\begin{figure}[htbp]
\centering
\includegraphics[width=0.8\textwidth]{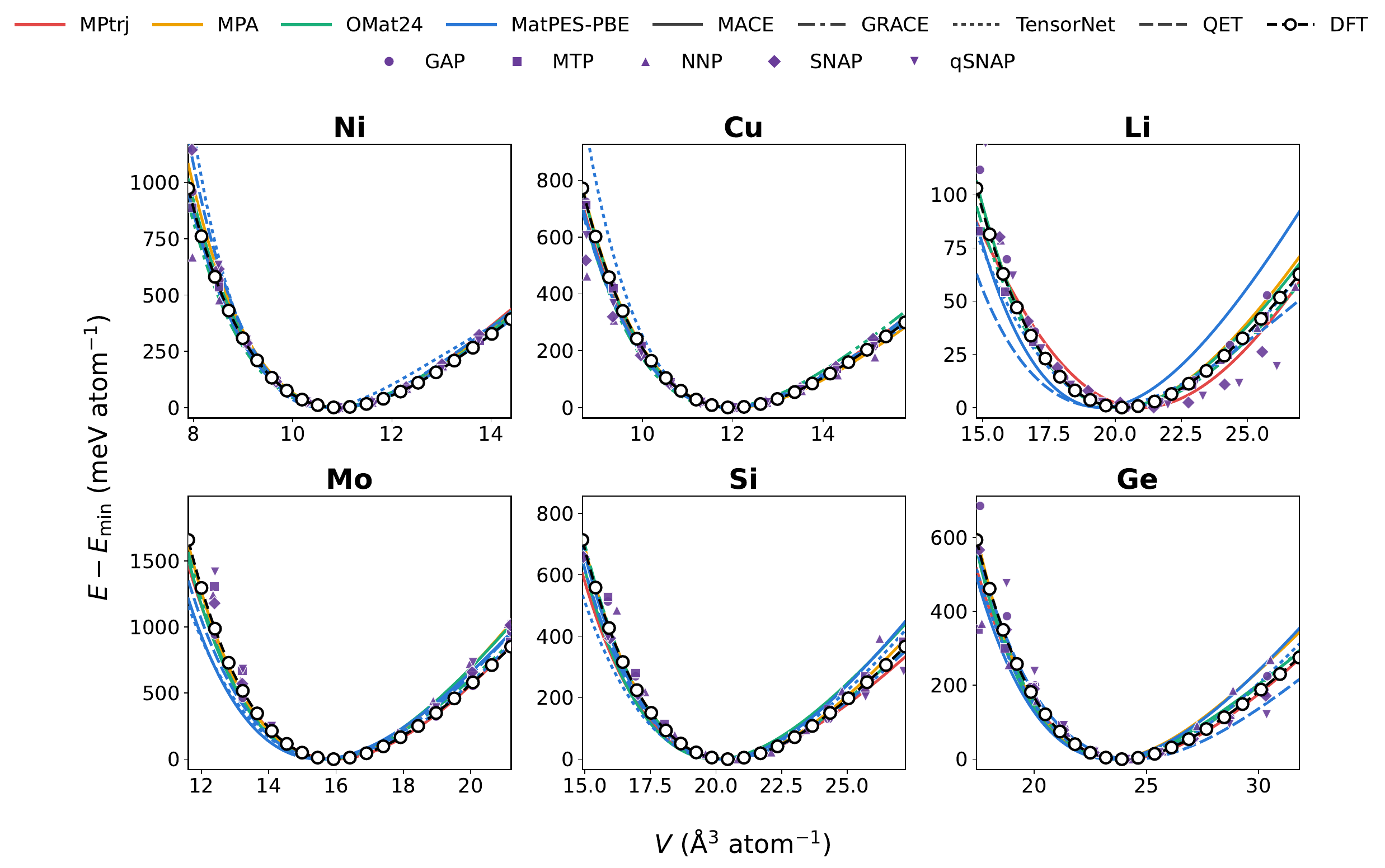}
\caption{\textbf{Energy--volume curves of the six elemental crystals for linear strains of up to $\pm$10\%.} Black open circles and dashed line: PBE data and a cubic spline through them. Colored lines: third-order Birch--Murnaghan fits to the computed points of the FPs, extended across each panel; color indicates the training dataset and line style the architecture. Purple markers: computed points of the custom MLIPs. Each curve is shifted by its own minimum energy.}
\label{si:eos-full}
\end{figure}

This section gives the per-element results of the six-element benchmark of the main text. Table~\ref{si:tab-elemental-lattice} lists the lattice parameters of all models, including three r$^2$SCAN-trained FPs, and compares them with DFT and 0\,K experimental values. Figures~\ref{si:elem-c11-c12}--\ref{si:elem-ea-deos} give the signed errors for each element, and Fig.~\ref{si:eos-full} shows the energy--volume curves.

\begin{table}[H]
\centering
\caption{\textbf{Equilibrium lattice parameters $a$ of the six elemental crystals.} Structures were relaxed with \texttt{RelaxCalc} ($f_\mathrm{max}=0.01$~eV\,\AA$^{-1}$). For each model, the first row gives $a$ (\AA{}) and the second gives the percentage errors (DFT\,/\,Expt.), where DFT is PBE~\cite{zuoPerformanceCostAssessment2020} for PBE-trained models and r$^2$SCAN~\cite{jainCommentaryMaterialsProject2013,kingsburyFlexibleScalableScheme2022} for r$^2$SCAN-trained models. Experimental values are room-temperature data from Ref.\,\citenum{simmons1971} corrected to 0\,K following Ref.\,\citenum{lejaeghere2014}. The last column gives the MAE (\AA{}) and MAPE (\%) over six elements for DFT and five for experiment. Bold: lowest error among the PBE-trained models and among the r$^2$SCAN-trained models. $^{\ddagger}$\,bcc Li transforms to a 9R phase below 80\,K~\cite{PRBLi1986}, so no 0\,K experimental value exists~\cite{PRBLiPhonon1976}.}
\label{si:tab-elemental-lattice}
\scriptsize
\setlength{\tabcolsep}{3.0pt}
\renewcommand{\arraystretch}{1.08}
\begin{tabular}{cccccccccc}
\toprule
Architecture & Training Data & Functional & Ni & Cu & Li & Mo & Si & Ge & MAE\,(\AA) \\
 & & & & & & & & & MAPE\,(\%) \\
\midrule
\multicolumn{3}{l}{DFT [PBE]~\cite{zuoPerformanceCostAssessment2020}\,(\AA)} & 3.508 & 3.621 & 3.427 & 3.168 & 5.469 & 5.763 & --- \\
\multicolumn{3}{l}{DFT [r$^2$SCAN]~\cite{jainCommentaryMaterialsProject2013,kingsburyFlexibleScalableScheme2022}\,(\AA)} & 3.475 & 3.577 & 3.439 & 3.168 & 5.444 & 5.675 & --- \\
\multicolumn{3}{l}{Expt.\ [0\,K]~\cite{lejaeghere2014}\,(\AA)} & 3.508 & 3.596 & ---$^{\ddagger}$ & 3.141 & 5.430 & 5.646 & --- \\
\midrule
\multicolumn{10}{l}{\textit{Element-specific MLIPs}} \\
\multirow{2}{*}{GAP} & \multirow{2}{*}{--} & \multirow{2}{*}{PBE} & 3.524 & 3.634 & 3.448 & 3.170 & 5.465 & 5.776 & 0.012\,/\,0.050 \\
& & & ($+$0.5\,/\,$+$0.5) & ($+$0.4\,/\,$+$1.0) & ($+$0.6\,/\,---) & ($+$0.1\,/\,$+$0.9) & ($-$0.1\,/\,$+$0.6) & ($+$0.2\,/\,$+$2.3) & 0.30\,/\,1.07 \\
\multirow{2}{*}{MTP} & \multirow{2}{*}{--} & \multirow{2}{*}{PBE} & 3.522 & 3.635 & 3.443 & 3.169 & 5.465 & 5.770 & 0.009\,/\,0.048 \\
& & & ($+$0.4\,/\,$+$0.4) & ($+$0.4\,/\,$+$1.1) & ($+$0.5\,/\,---) & (\textbf{$+$0.0}\,/\,$+$0.9) & ($-$0.1\,/\,$+$0.6) & ($+$0.1\,/\,$+$2.2) & 0.25\,/\,1.04 \\
\multirow{2}{*}{NNP} & \multirow{2}{*}{--} & \multirow{2}{*}{PBE} & 3.524 & 3.637 & 3.428 & 3.165 & 5.501 & 5.788 & 0.016\,/\,0.059 \\
& & & ($+$0.5\,/\,$+$0.5) & ($+$0.5\,/\,$+$1.1) & (\textbf{$+$0.0}\,/\,---) & ($-$0.1\,/\,$+$0.8) & ($+$0.6\,/\,$+$1.3) & ($+$0.4\,/\,$+$2.5) & 0.34\,/\,1.24 \\
\multirow{2}{*}{SNAP} & \multirow{2}{*}{--} & \multirow{2}{*}{PBE} & 3.522 & 3.634 & 3.506 & 3.169 & 5.473 & 5.771 & 0.020\,/\,0.050 \\
& & & ($+$0.4\,/\,$+$0.4) & ($+$0.4\,/\,$+$1.1) & ($+$2.3\,/\,---) & (\textbf{$+$0.0}\,/\,$+$0.9) & ($+$0.1\,/\,$+$0.8) & ($+$0.1\,/\,$+$2.2) & 0.55\,/\,1.07 \\
\multirow{2}{*}{qSNAP} & \multirow{2}{*}{--} & \multirow{2}{*}{PBE} & 3.521 & 3.635 & 3.464 & 3.170 & 5.465 & 5.774 & 0.014\,/\,0.049 \\
& & & ($+$0.4\,/\,$+$0.4) & ($+$0.4\,/\,$+$1.1) & ($+$1.1\,/\,---) & ($+$0.1\,/\,$+$0.9) & ($-$0.1\,/\,$+$0.7) & ($+$0.2\,/\,$+$2.3) & 0.36\,/\,1.06 \\
\midrule
\multicolumn{10}{l}{\textit{FPs}} \\
\multirow{2}{*}{MACE} & \multirow{2}{*}{MPtrj} & \multirow{2}{*}{PBE} & 3.510 & 3.626 & 3.497 & 3.170 & 5.455 & 5.772 & 0.017\,/\,0.042 \\
& & & ($+$0.1\,/\,$+$0.1) & (\textbf{$+$0.1}\,/\,$+$0.8) & ($+$2.0\,/\,---) & ($+$0.1\,/\,$+$0.9) & ($-$0.3\,/\,$+$0.5) & ($+$0.2\,/\,$+$2.2) & 0.45\,/\,0.90 \\
\multirow{2}{*}{MACE} & \multirow{2}{*}{MPA} & \multirow{2}{*}{PBE} & 3.521 & 3.639 & 3.427 & 3.151 & 5.469 & 5.718 & 0.015\,/\,0.035 \\
& & & ($+$0.4\,/\,$+$0.4) & ($+$0.5\,/\,$+$1.2) & (\textbf{$+$0.0}\,/\,---) & ($-$0.5\,/\,$+$0.3) & (\textbf{$+$0.0}\,/\,$+$0.7) & ($-$0.8\,/\,\textbf{$+$1.3}) & 0.36\,/\,0.77 \\
\multirow{2}{*}{MACE} & \multirow{2}{*}{OMat24} & \multirow{2}{*}{PBE} & 3.522 & 3.633 & 3.427 & 3.151 & 5.423 & 5.720 & 0.022\,/\,0.029 \\
& & & ($+$0.4\,/\,$+$0.4) & ($+$0.3\,/\,$+$1.0) & (\textbf{$+$0.0}\,/\,---) & ($-$0.5\,/\,$+$0.3) & ($-$0.8\,/\,\textbf{$-$0.1}) & ($-$0.7\,/\,\textbf{$+$1.3}) & 0.48\,/\,0.64 \\
\multirow{2}{*}{GRACE} & \multirow{2}{*}{OMat24} & \multirow{2}{*}{PBE} & 3.508 & 3.614 & 3.430 & 3.159 & 5.466 & 5.750 & \textbf{0.006}\,/\,0.035 \\
& & & (\textbf{$+$0.0}\,/\,\textbf{$+$0.0}) & ($-$0.2\,/\,\textbf{$+$0.5}) & ($+$0.1\,/\,---) & ($-$0.3\,/\,$+$0.6) & ($-$0.1\,/\,$+$0.7) & ($-$0.2\,/\,$+$1.8) & \textbf{0.14}\,/\,0.72 \\
\multirow{2}{*}{MACE} & \multirow{2}{*}{MatPES} & \multirow{2}{*}{PBE} & 3.515 & 3.627 & 3.380 & 3.148 & 5.447 & 5.719 & 0.024\,/\,\textbf{0.027} \\
& & & ($+$0.2\,/\,$+$0.2) & ($+$0.2\,/\,$+$0.8) & ($-$1.4\,/\,---) & ($-$0.6\,/\,\textbf{$+$0.2}) & ($-$0.4\,/\,$+$0.3) & ($-$0.8\,/\,\textbf{$+$1.3}) & 0.59\,/\,\textbf{0.58} \\
\multirow{2}{*}{TensorNet} & \multirow{2}{*}{MatPES} & \multirow{2}{*}{PBE} & 3.516 & 3.644 & 3.430 & 3.158 & 5.452 & 5.762 & 0.010\,/\,0.042 \\
& & & ($+$0.2\,/\,$+$0.2) & ($+$0.6\,/\,$+$1.3) & ($+$0.1\,/\,---) & ($-$0.3\,/\,$+$0.5) & ($-$0.3\,/\,$+$0.4) & (\textbf{$+$0.0}\,/\,$+$2.1) & 0.27\,/\,0.92 \\
\multirow{2}{*}{QET} & \multirow{2}{*}{MatPES} & \multirow{2}{*}{PBE} & 3.519 & 3.630 & 3.397 & 3.159 & 5.467 & 5.807 & 0.018\,/\,0.052 \\
& & & ($+$0.3\,/\,$+$0.3) & ($+$0.2\,/\,$+$0.9) & ($-$0.9\,/\,---) & ($-$0.3\,/\,$+$0.6) & (\textbf{$+$0.0}\,/\,$+$0.7) & ($+$0.8\,/\,$+$2.9) & 0.42\,/\,1.07 \\
\multirow{2}{*}{MACE} & \multirow{2}{*}{MatPES} & \multirow{2}{*}{r$^2$SCAN} & 3.486 & 3.575 & 3.433 & 3.136 & 5.441 & 5.643 & 0.014\,/\,\textbf{0.013} \\
& & & ($+$0.3\,/\,\textbf{$-$0.6}) & (\textbf{$-$0.1}\,/\,\textbf{$-$0.6}) & ($-$0.2\,/\,---) & ($-$1.0\,/\,\textbf{$-$0.2}) & (\textbf{$-$0.1}\,/\,$+$0.2) & ($-$0.6\,/\,\textbf{$+$0.0}) & 0.36\,/\,\textbf{0.33} \\
\multirow{2}{*}{TensorNet} & \multirow{2}{*}{MatPES} & \multirow{2}{*}{r$^2$SCAN} & 3.471 & 3.572 & 3.439 & 3.164 & 5.436 & 5.669 & \textbf{0.005}\,/\,0.022 \\
& & & (\textbf{$-$0.1}\,/\,$-$1.0) & (\textbf{$-$0.1}\,/\,$-$0.7) & (\textbf{$+$0.0}\,/\,---) & (\textbf{$-$0.1}\,/\,$+$0.7) & (\textbf{$-$0.1}\,/\,\textbf{$+$0.1}) & (\textbf{$-$0.1}\,/\,$+$0.4) & \textbf{0.11}\,/\,0.59 \\
\multirow{2}{*}{QET} & \multirow{2}{*}{MatPES} & \multirow{2}{*}{r$^2$SCAN} & 3.477 & 3.575 & 3.422 & 3.157 & 5.420 & 5.689 & 0.012\,/\,0.024 \\
& & & (\textbf{$+$0.1}\,/\,$-$0.9) & (\textbf{$-$0.1}\,/\,\textbf{$-$0.6}) & ($-$0.5\,/\,---) & ($-$0.3\,/\,$+$0.5) & ($-$0.4\,/\,$-$0.2) & ($+$0.2\,/\,$+$0.8) & 0.27\,/\,0.58 \\
\bottomrule
\end{tabular}
\end{table}

\section{Confidence intervals of the errors in Table~1}
\label{si:ci}

Table~\ref{si:table1-ci} gives the 95\% confidence intervals of the values in Table~1 of the main text. For each benchmark, the entries were resampled with replacement 10,000 times, the MAE of each FP was computed for each resample over the entries on which the FP did not fail, and the 2.5th and 97.5th percentiles of the resampled MAEs were taken as the confidence interval. All FPs were evaluated on the same resamples, so that the confidence interval of the difference between two FPs accounts for the correlation of their errors. For $\bar{d}_\mathrm{fp}$ and $E_\mathrm{f}^{0}$, each entry is the average over the five perturbed relaxations of a structure.

\begin{table}[htbp]
\centering
\caption{\textbf{95\% bootstrap confidence intervals of the values in Table~1 of the main text.} $^{\dagger}$Not significantly different from the lowest value of the row (the 95\% confidence interval of the difference includes zero).}
\label{si:table1-ci}
\footnotesize
\setlength{\tabcolsep}{3pt}\renewcommand{\arraystretch}{1.35}
\resizebox{\textwidth}{!}{%
\begin{tabular}{l ccccccc}
\toprule
\multirow{2}{*}{Property} & \multicolumn{1}{c}{MPtrj} & \multicolumn{1}{c}{MPA} & \multicolumn{2}{c}{OMat24} & \multicolumn{3}{c}{MatPES-PBE} \\
\cmidrule(lr){2-2}\cmidrule(lr){3-3}\cmidrule(lr){4-5}\cmidrule(lr){6-8}
 & MACE & MACE & MACE & GRACE & MACE & TensorNet & QET \\
\midrule
$\bar{d}_\mathrm{fp}$ & [0.365, 0.420] & [0.239, 0.284] & [0.221, 0.268] & [0.195, 0.239] & [0.249, 0.294] & [0.283, 0.332] & [0.301, 0.348] \\
$E_\mathrm{f}^{0}$ (meV\,atom$^{-1}$) & [71.8, 84.7] & [29.3, 38.6]$^{\dagger}$ & [31.6, 41.2] & [26.7, 36.5] & [48.4, 57.7] & [51.9, 60.9] & [72.7, 82.9] \\
$K_\mathrm{VRH}$ (GPa) & [15.9, 16.7] & [18.1, 18.9] & [16.5, 17.3] & [15.5, 16.3] & [17.8, 18.5] & [19.5, 20.5] & [17.2, 18.0] \\
$G_\mathrm{VRH}$ (GPa) & [22.1, 28.1] & [15.3, 19.5] & [16.1, 17.3] & [15.7, 23.5] & [13.5, 14.4] & [14.9, 18.3] & [15.0, 16.5] \\
\addlinespace[3pt]
$C_V$ (J\,mol-atom$^{-1}$\,K$^{-1}$) & [0.597, 0.681] & [0.324, 0.390] & [0.204, 0.269] & [0.155, 0.201] & [0.566, 0.643] & [0.627, 0.704] & [0.480, 0.558] \\
$\alpha_V$ (ppm\,K$^{-1}$) & [11.5, 26.2] & [7.6, 14.4] & [5.8, 23.0]$^{\dagger}$ & [4.6, 25.6]$^{\dagger}$ & [7.7, 15.2]$^{\dagger}$ & [7.7, 18.1]$^{\dagger}$ & [9.7, 20.6]$^{\dagger}$ \\
$K_T$ (GPa) & [18.0, 30.6]$^{\dagger}$ & [19.4, 30.3] & [14.9, 99.0]$^{\dagger}$ & [13.7, 24.5]$^{\dagger}$ & [19.6, 33.1] & [13.9, 25.1]$^{\dagger}$ & [13.2, 24.5] \\
$\log_{10}\kappa$ & [0.166, 0.232] & [0.077, 0.107] & [0.043, 0.062] & [0.023, 0.033] & [0.114, 0.173] & [0.145, 0.236] & [0.130, 0.204] \\
\addlinespace[3pt]
$E_\mathrm{m}$ (eV) & [0.152, 0.176] & [0.148, 0.177]$^{\dagger}$ & [0.143, 0.171]$^{\dagger}$ & [0.147, 0.175]$^{\dagger}$ & [0.206, 0.230] & [0.174, 0.196] & [0.143, 0.162] \\
$\log_{10}\sigma$ & [0.336, 0.398] & [0.168, 0.217]$^{\dagger}$ & [0.152, 0.195] & [0.162, 0.213]$^{\dagger}$ & [0.181, 0.229] & [0.182, 0.225] & [0.197, 0.247] \\
\addlinespace[3pt]
$\gamma_\mathrm{surf}$ (J\,m$^{-2}$) & [0.209, 0.246] & [0.197, 157.4] & [0.115, 0.134]$^{\dagger}$ & [0.109, 0.134] & [0.131, 0.156] & [0.186, 0.213] & [0.235, 0.266] \\
$\gamma_\mathrm{GB}$ (J\,m$^{-2}$) & [0.202, 0.274] & [0.056, 0.073] & [0.063, 0.080]$^{\dagger}$ & [0.051, 0.159]$^{\dagger}$ & [0.062, 0.079]$^{\dagger}$ & [0.065, 0.082]$^{\dagger}$ & [0.082, 0.104] \\
\addlinespace[3pt]
$E_\mathrm{ads}$ (eV) & [0.696, 1.045] & [0.741, 1.129] & [0.755, 1.251] & [0.814, 1.417] & [0.229, 0.289]$^{\dagger}$ & [0.214, 0.277] & [0.317, 0.400] \\
$\gamma_\mathrm{int}$ (J\,m$^{-2}$) & [1.903, 2.613] & [2.905, 3.903] & [4.014, 5.395] & [4.184, 5.607] & [0.209, 0.265] & [0.182, 0.224] & [0.195, 0.243]$^{\dagger}$ \\
$W_\mathrm{adh}$ (J\,m$^{-2}$) & [1.742, 2.403] & [2.727, 3.617] & [3.843, 5.114] & [4.056, 5.380] & [0.153, 0.192] & [0.170, 0.213]$^{\dagger}$ & [0.258, 0.317] \\
\bottomrule
\end{tabular}}
\end{table}

\section{Relaxed structures and formation energies}
\label{si:relaxation}

\subsection{Formation energies by element}
\label{si:wbm-ef}

\begin{figure}[!htbp]
\centering
\includegraphics[width=\textwidth]{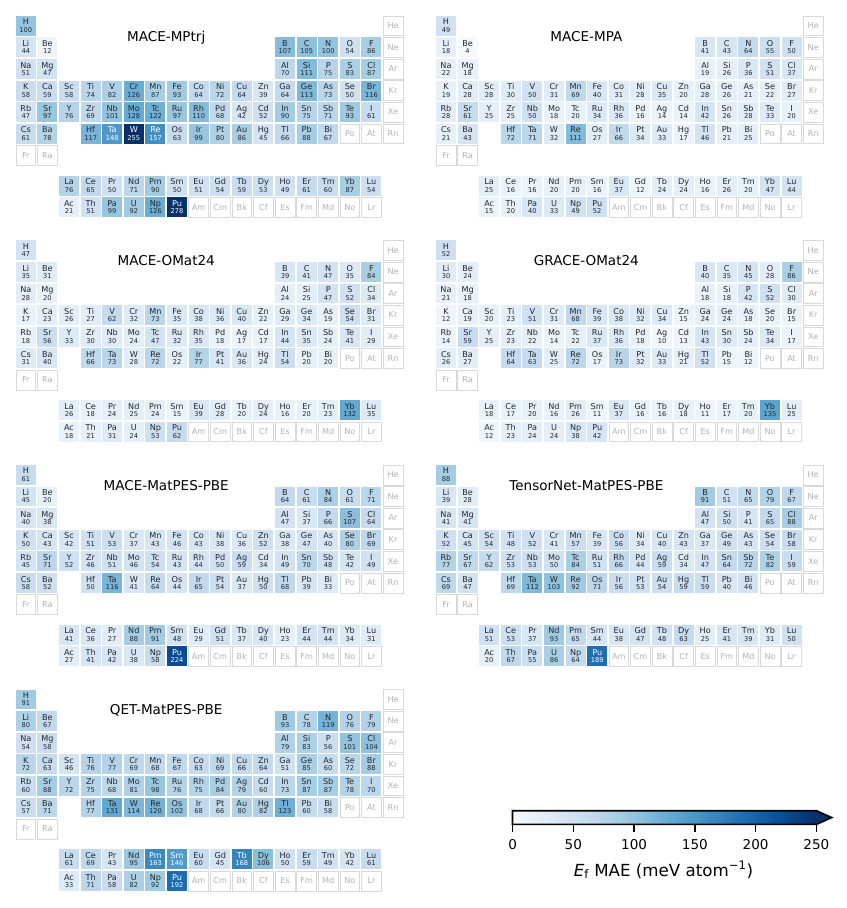}
\caption{\textbf{Errors of the formation energies by element.} MAE of the formation energy $E_\mathrm{f}^{0}$ of the WBM structures for each PBE-trained FP and element. The absolute error of each structure is distributed among its elements in proportion to their atomic fractions. The color scale is capped at 250~meV~atom$^{-1}$.}
\label{si:wbm-ef-ptable}
\end{figure}

Figure~\ref{si:wbm-ef-ptable} shows the errors of the formation energies of the WBM structures by element.

\subsection{Sensitivity to the reference data}
\label{si:wbm}

Table~\ref{si:wbm-references} compares $\bar{d}_\mathrm{fp}$ and $E_\mathrm{f}^{0}$ against the original WBM reference and the re-relaxation at the MatPES settings.

\begin{table}[htbp]
\centering
\caption{\textbf{Sensitivity of the equilibrium benchmark to the reference data.} Wang et al.\ report the original WBM results at settings compatible with the Materials Project.\cite{wangPredictingStableCrystalline2021} Kaplan et al.\ re-relaxed the WBM structures at the tighter MatPES settings (PBE\_64 pseudopotentials, 680~eV cutoff, no Hubbard $U$).\cite{kaplan2025matpes} All comparisons use the structures common to both references, which exclude the structures that did not converge at the MatPES settings. $\bar{d}_\mathrm{fp}$ and $E_\mathrm{f}^{0}$ MAE are computed for each of five runs with different random perturbations and then averaged.}
\label{si:wbm-references}
\footnotesize
\begin{tabular}{lcccc}
\toprule
 & \multicolumn{2}{c}{$\bar{d}_\mathrm{fp}$} & \multicolumn{2}{c}{$E_\mathrm{f}^{0}$ MAE (meV\,atom$^{-1}$)}\\
\cmidrule(lr){2-3}\cmidrule(lr){4-5}
Model & Wang et al. & Kaplan et al. & Wang et al. & Kaplan et al.\\
\midrule
MACE-MPtrj & 0.39 & 0.40 & 77 & 84\\
MACE-MPA & 0.26 & 0.28 & 33 & 56\\
MACE-OMat24 & 0.24 & 0.25 & 35 & 50\\
GRACE-OMat24 & 0.22 & 0.23 & 30 & 48\\
MACE-MatPES-PBE & 0.30 & 0.27 & 79 & 53\\
TensorNet-MatPES-PBE & 0.33 & 0.31 & 76 & 56\\
QET-MatPES-PBE & 0.34 & 0.32 & 96 & 78\\
\bottomrule
\end{tabular}
\end{table}

\section{Elastic moduli}
\label{si:elastic}

\begin{figure}[!htbp]
\centering
\includegraphics[width=\textwidth]{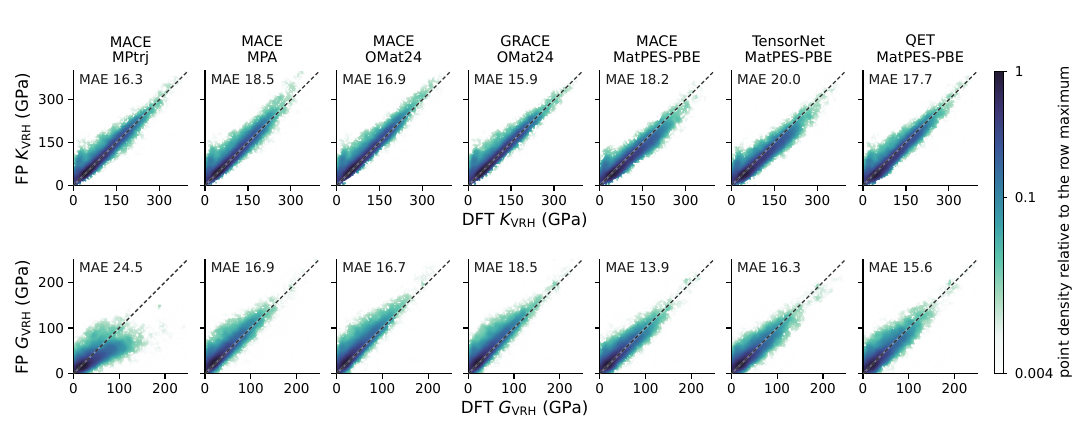}
\caption{\textbf{Elastic moduli of the PBE-trained FPs.} $K_\mathrm{VRH}$ (top) and $G_\mathrm{VRH}$ (bottom) predicted by each FP against the PBE reference values. The color shows the local density of the materials, and the MAE in GPa is given in each panel.}
\label{si:elastic-parity}
\end{figure}

Figure~\ref{si:elastic-parity} compares $K_\mathrm{VRH}$ and $G_\mathrm{VRH}$ of the seven PBE-trained FPs with the PBE reference values.

\section{Heat capacities and thermal conductivities}
\label{si:cv-r2scan}

\begin{figure}[!htbp]
\centering
\includegraphics[width=0.72\textwidth]{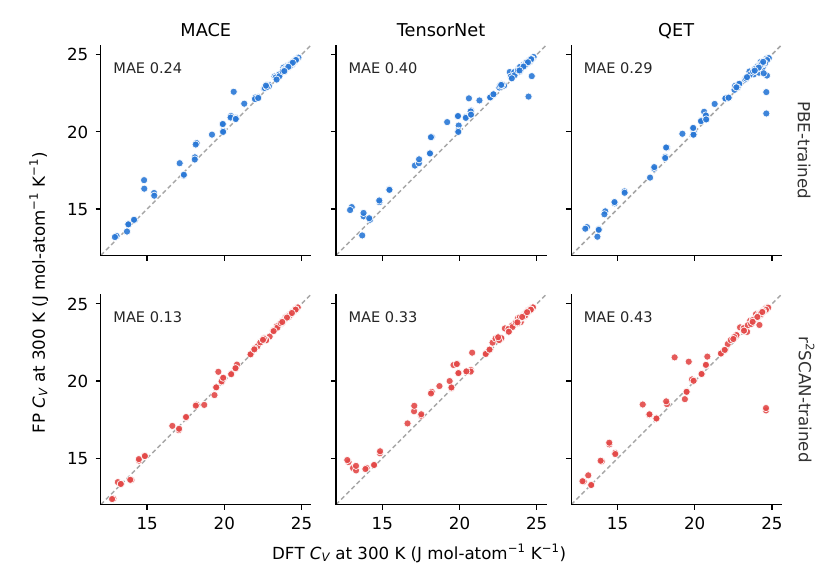}
\caption{\textbf{Heat capacities of the PBE-trained and r$^2$SCAN-trained FPs.} $C_V$ at 300\,K predicted by the MatPES-trained FPs against the DFT reference of their training functional. Top: PBE-trained FPs against PBE; bottom: r$^2$SCAN-trained FPs against r$^2$SCAN. The MAE in J\,mol-atom$^{-1}$\,K$^{-1}$ is given in each panel.}
\label{si:cv-parity-r2scan}
\end{figure}

\begin{figure}[!htbp]
\centering
\includegraphics[width=0.72\textwidth]{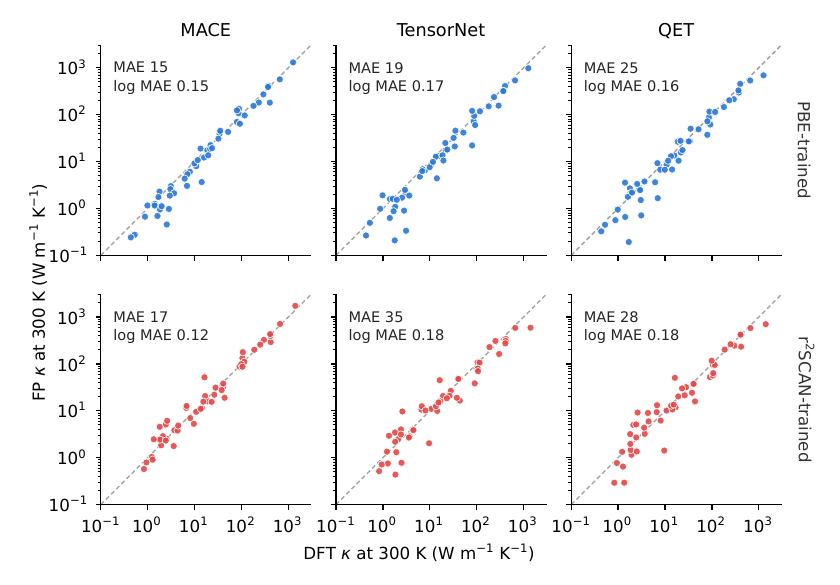}
\caption{\textbf{Lattice thermal conductivities of the PBE-trained and r$^2$SCAN-trained FPs.} $\kappa$ at 300\,K predicted by the MatPES-trained FPs against the DFT reference of their training functional, for the compounds with PBE, r$^2$SCAN, and experimental reference values. Top: PBE-trained FPs against PBE; bottom: r$^2$SCAN-trained FPs against r$^2$SCAN. The MAE in W\,m$^{-1}$\,K$^{-1}$ and the MAE of $\log_{10}\kappa$ are given in each panel.}
\label{si:kappa-parity-r2scan}
\end{figure}

Figure~\ref{si:cv-parity-r2scan} compares $C_V$ at 300\,K of the MatPES-trained FPs with the DFT reference of their training functional, for the compounds for which both PBE and r$^2$SCAN reference values are available. Figure~\ref{si:kappa-parity-r2scan} makes the same comparison for $\kappa$ at 300\,K, for the compounds for which PBE, r$^2$SCAN, and experimental reference values are available. For both properties, the PBE-trained and the r$^2$SCAN-trained FPs reproduce the functional they were trained on with comparable accuracy: the MAEs are 0.24--0.40 and 0.13--0.43~J\,mol-atom$^{-1}$\,K$^{-1}$ for $C_V$, and 0.15--0.17 and 0.12--0.18 for $\log_{10}\kappa$. The PBE reference values were computed from the force constants of phonondb;\cite{togoDistributionsPhononLifetimes2015,sekoPredictionLowThermalConductivityCompounds2015} the r$^2$SCAN reference values of $\kappa$ are those of Kim et al.,\cite{kimSevenNetOmni2026} and those of $C_V$ were computed from their phonon frequencies.

\section{Migration barriers and ionic conductivities}
\label{si:migration}

The MAEs of $E_\mathrm{m}$ and $\log_{10}\sigma$ in Table~1 of the main text are computed for each FP over the NEB calculations that converged and the MD simulations in which diffusion was measured, so that each FP is evaluated on a different set. Table~\ref{si:migration-failed} shows that the paths and simulations on which an FP fails are also more difficult for the other FPs: on these, the MAEs of the other six FPs are 30--50\% higher for $E_\mathrm{m}$ and three to seven times higher for $\log_{10}\sigma$ than on the paths and simulations on which the FP succeeds. An FP that fails more often is therefore evaluated on an easier set. Table~\ref{si:migration-common} gives the MAEs on the 1,347 of 1,681 paths and the 711 of 748 simulations on which all seven FPs succeed. These MAEs are lower than those in Table~1 by 0.003--0.013\,eV for $E_\mathrm{m}$ and by 0.012--0.026 for $\log_{10}\sigma$.

\begin{table}[!htbp]
\centering\small
\caption{\textbf{Errors of the other FPs on the calculations on which an FP fails.} For each FP, the table gives the number of failed calculations and the MAE of the other six FPs on the calculations on which this FP fails and on those on which it succeeds. $\sigma$ is in mS\,cm$^{-1}$.}
\label{si:migration-failed}
\begin{tabular}{lcccccc}
\toprule
 & \multicolumn{3}{c}{$E_\mathrm{m}$ (eV)} & \multicolumn{3}{c}{$\log_{10}\sigma$} \\
\cmidrule(lr){2-4}\cmidrule(lr){5-7}
 & & \multicolumn{2}{c}{MAE of the other FPs} & & \multicolumn{2}{c}{MAE of the other FPs} \\
FP & Failed & on failed & on succeeded & Failed & on failed & on succeeded \\
\midrule
MACE-MPtrj & 93 & 0.22 & 0.17 & 20 & 0.72 & 0.19 \\
MACE-MPA & 129 & 0.24 & 0.17 & 14 & 0.74 & 0.22 \\
MACE-OMat24 & 144 & 0.26 & 0.17 & 14 & 0.86 & 0.22 \\
GRACE-OMat24 & 126 & 0.25 & 0.17 & 15 & 0.65 & 0.22 \\
MACE-MatPES-PBE & 95 & 0.24 & 0.16 & 10 & 0.95 & 0.22 \\
TensorNet-MatPES-PBE & 91 & 0.22 & 0.17 & 5 & 1.61 & 0.22 \\
QET-MatPES-PBE & 139 & 0.24 & 0.17 & 2 & 0.94 & 0.22 \\
\bottomrule
\end{tabular}
\end{table}

\begin{table}[!htbp]
\centering\small
\caption{\textbf{MAEs on the calculations on which all seven FPs succeed.} ``Own'' denotes the calculations on which the FP itself succeeds, as in Table~1 of the main text, and ``common'' those on which all seven FPs succeed. $\sigma$ is in mS\,cm$^{-1}$.}
\label{si:migration-common}
\begin{tabular}{lcccc}
\toprule
 & \multicolumn{2}{c}{$E_\mathrm{m}$ (eV)} & \multicolumn{2}{c}{$\log_{10}\sigma$} \\
\cmidrule(lr){2-3}\cmidrule(lr){4-5}
FP & Own & Common & Own & Common \\
\midrule
MACE-MPtrj & 0.164 & 0.151 & 0.366 & 0.354 \\
MACE-MPA & 0.162 & 0.151 & 0.191 & 0.179 \\
MACE-OMat24 & 0.157 & 0.149 & 0.172 & 0.158 \\
GRACE-OMat24 & 0.161 & 0.152 & 0.186 & 0.166 \\
MACE-MatPES-PBE & 0.218 & 0.215 & 0.203 & 0.182 \\
TensorNet-MatPES-PBE & 0.184 & 0.176 & 0.202 & 0.183 \\
QET-MatPES-PBE & 0.152 & 0.143 & 0.220 & 0.194 \\
\bottomrule
\end{tabular}
\end{table}

\section{Surface and grain boundary energies}
\label{si:surface-gb}

\subsection{Surface energies by Miller index}
\label{si:surface-miller}

\begin{figure}[!htbp]
\centering
\includegraphics[width=\textwidth]{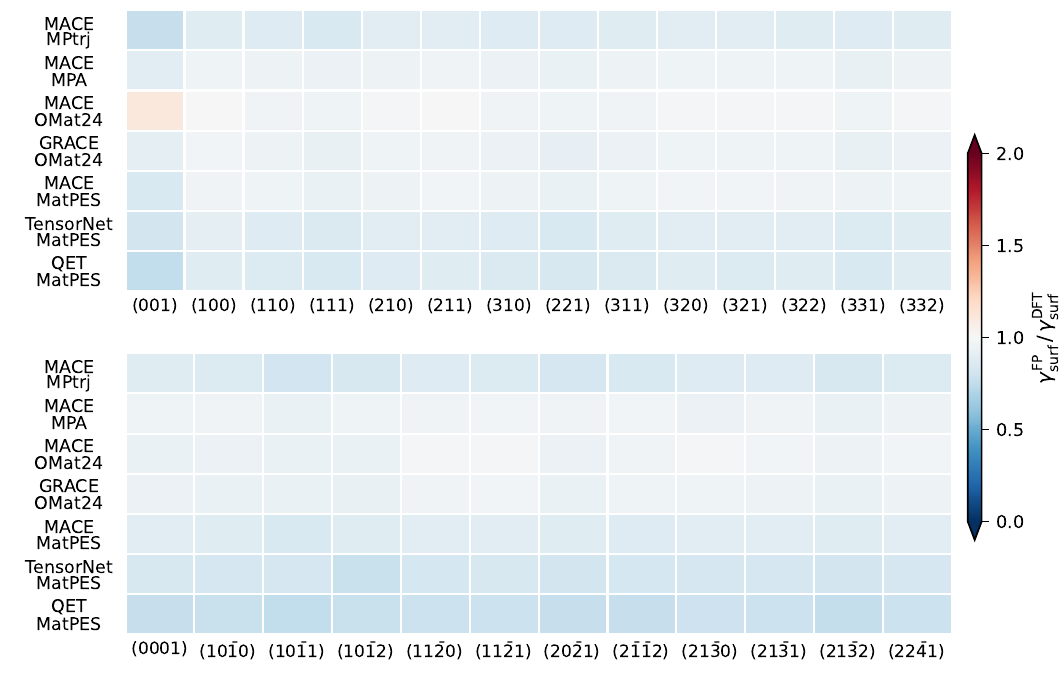}
\caption{\textbf{Surface energies by Miller index.} Ratio of the predicted to the reference $\gamma_\mathrm{surf}$ for each Miller index and FP, obtained from the sums over all slabs with this index. Upper panel: three-index notation; lower panel: four-index notation of hexagonal crystals. The Y slabs are not included for MACE-MPA.}
\label{si:surface-hkl}
\end{figure}

Figure~\ref{si:surface-hkl} shows the ratio of the predicted to the reference $\gamma_\mathrm{surf}$ for each Miller index with at least 15 slabs. For each FP, the ratio varies little between Miller indices: 0.77--0.89 for MACE-MPtrj, 0.90--0.97 for MACE-MPA, 0.93--1.10 for MACE-OMat24, 0.90--0.98 for GRACE-OMat24, 0.85--0.97 for MACE-MatPES-PBE, 0.78--0.90 for TensorNet-MatPES-PBE, and 0.76--0.88 for QET-MatPES-PBE.

\subsection{Negative surface energies of Y with MACE-MPA}
\label{si:surface-y}

\begin{figure}[!htbp]
\centering
\includegraphics[width=0.9\textwidth]{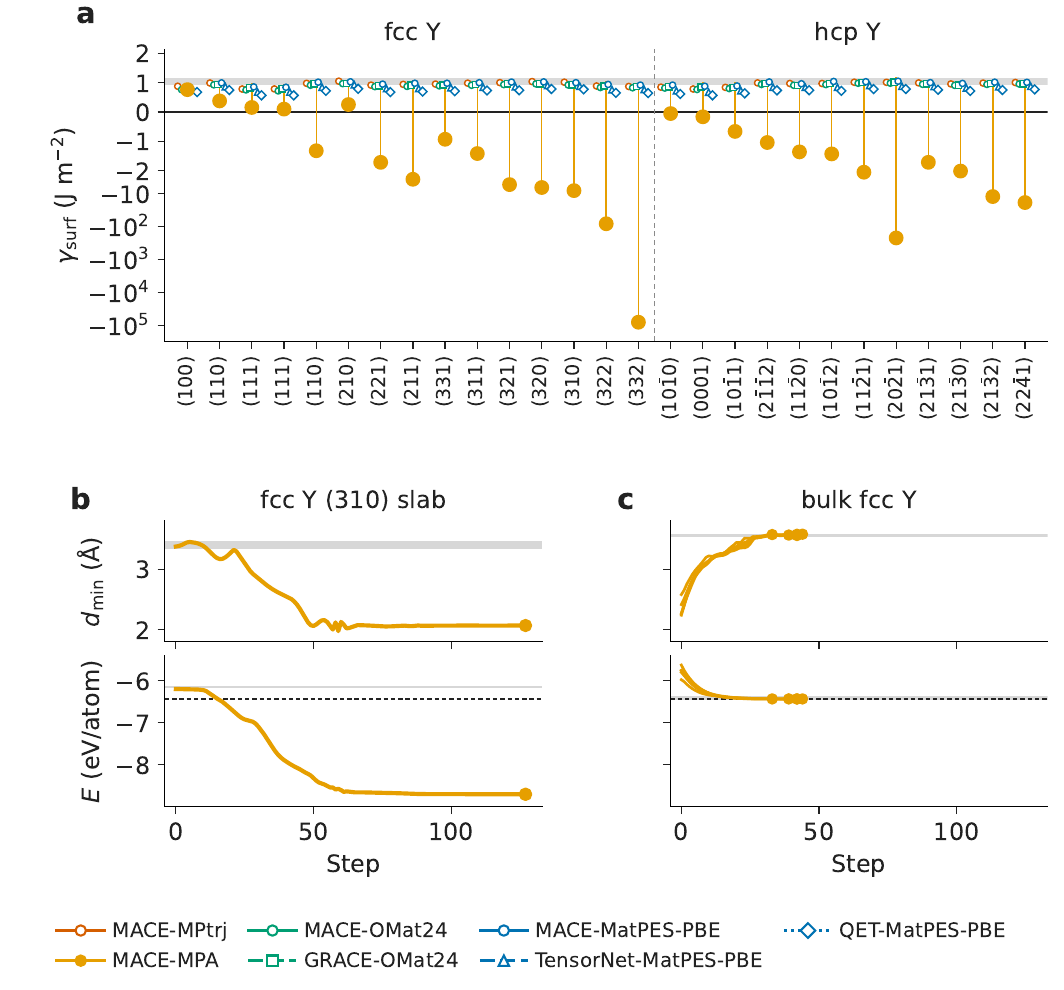}
\caption{\textbf{Surface energies of Y.} (a) $\gamma_\mathrm{surf}$ of the Y slabs for the seven FPs; the gray band is the range of the reference values. (b) Shortest interatomic distance $d_\mathrm{min}$ and energy per atom $E$ during the relaxation of the fcc Y(310) slab with MACE-MPA. (c) The same quantities during the relaxation of bulk fcc Y with randomly displaced atoms with MACE-MPA. In (b) and (c), the gray bands are the ranges of the final values of the other six FPs, and the dashed line is the energy per atom of bulk fcc Y with MACE-MPA.}
\label{si:surface-y-fig}
\end{figure}

Figure~\ref{si:surface-y-fig}a shows $\gamma_\mathrm{surf}$ of the 27 Y slabs. With MACE-MPA, $\gamma_\mathrm{surf}$ is negative for 22 slabs; the other six FPs give values close to the reference for all 27. Figure~\ref{si:surface-y-fig}b shows the relaxation of the fcc Y(310) slab with MACE-MPA. The outermost atoms first move away from the surface, and the energy per atom falls below that of bulk Y while the shortest interatomic distance is still above 3\,\AA{}. The remaining atoms are then compressed, and the shortest distance decreases to 2.1\,\AA{}. The relaxation of bulk fcc Y with MACE-MPA from structures with randomly displaced atoms returns to the equilibrium structure (Figure~\ref{si:surface-y-fig}c). Without the Y slabs, the MAE of MACE-MPA is 0.14\,J\,m$^{-2}$.

\subsection{Surface energies and the cutoff radius}
\label{si:surface-rc}

\begin{figure}[!htbp]
\centering
\includegraphics[width=\textwidth]{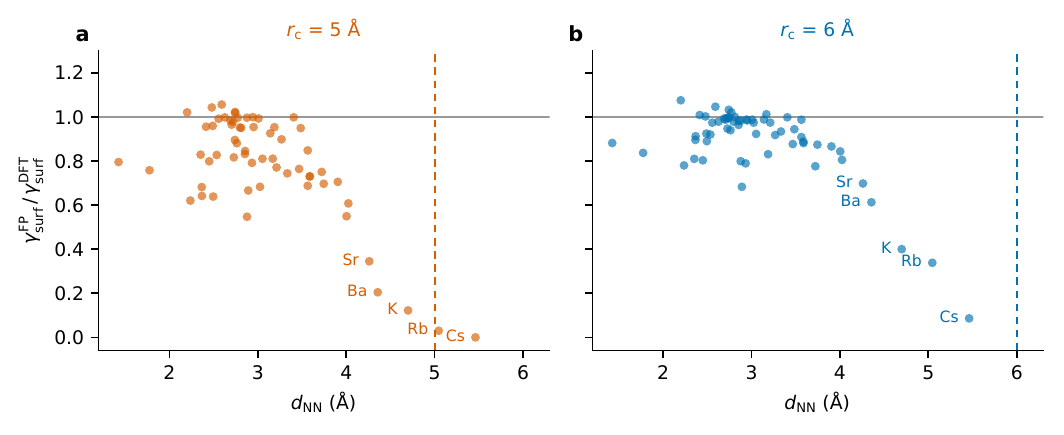}
\caption{\textbf{Surface energies and the cutoff radius.} Ratio of the predicted to the reference $\gamma_\mathrm{surf}$ against the nearest-neighbor distance $d_\mathrm{NN}$ in the reference structure for the FPs with a cutoff radius $r_\mathrm{c}$ of (a) 5\,\AA{} and (b) 6\,\AA{}. Each point is one element: the ratio is obtained from the sums over all slabs of the element, and the median over the FPs of the panel is shown. Dashed lines mark $r_\mathrm{c}$.}
\label{si:surface-cutoff}
\end{figure}

Figure~\ref{si:surface-cutoff} shows the ratio of the predicted to the reference $\gamma_\mathrm{surf}$ for each element as a function of the nearest-neighbor distance $d_\mathrm{NN}$. For both cutoff radii, the ratio decreases as $d_\mathrm{NN}$ approaches $r_\mathrm{c}$. For Sr, Ba, K, Rb, and Cs, the ratios are 0.34, 0.20, 0.12, 0.03, and 0.00 with a cutoff radius of 5\,\AA{} and 0.70, 0.61, 0.40, 0.34, and 0.09 with a cutoff radius of 6\,\AA{}.

\subsection{Grain boundary energies by element}
\label{si:gb-element}

\begin{figure}[!htbp]
\centering
\includegraphics[width=\textwidth]{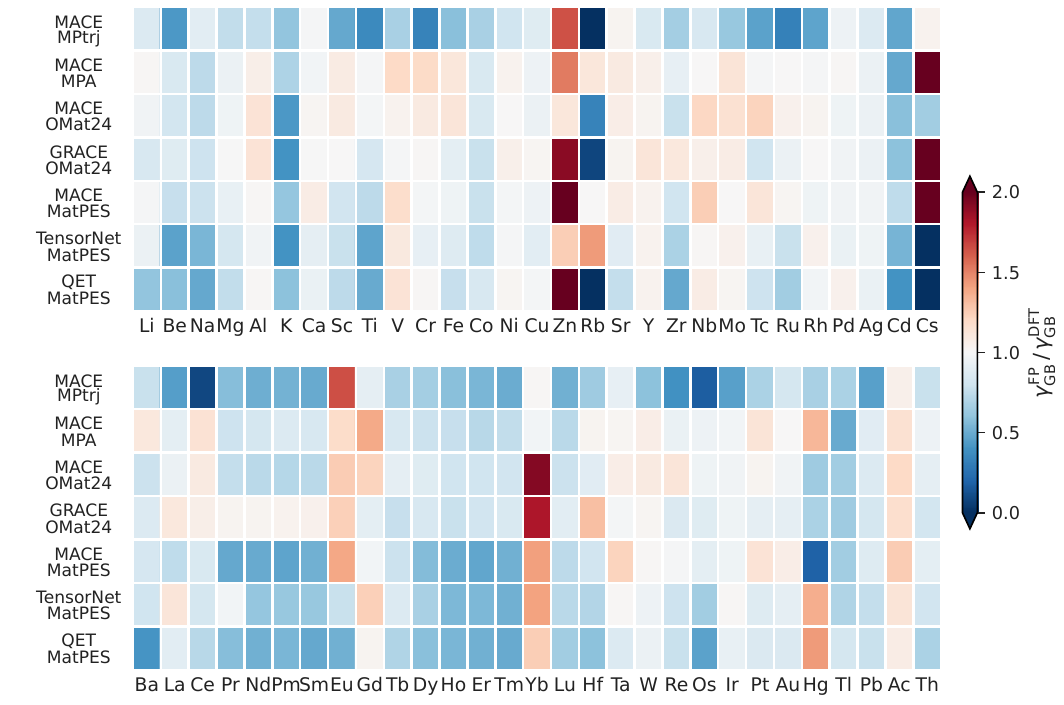}
\caption{\textbf{Grain boundary energies by element.} Ratio of the predicted to the reference grain boundary energy $\gamma_\mathrm{GB}$ for each element, obtained from the sums over all grain boundaries of the element. The color scale is capped at 0 and 2.}
\label{si:gb-element-fig}
\end{figure}

Figure~\ref{si:gb-element-fig} shows the ratio of the predicted to the reference $\gamma_\mathrm{GB}$ for each element.

\subsection{Grain boundary energies of Cs}
\label{si:gb-cutoff}

\begin{figure}[!htbp]
\centering
\includegraphics[width=0.7\textwidth]{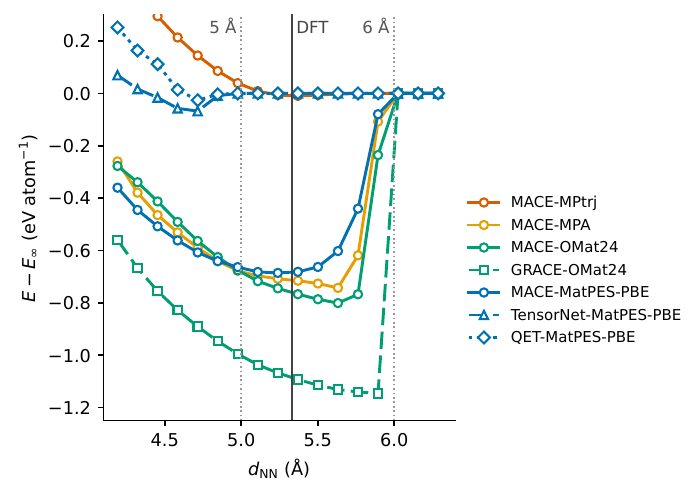}
\caption{\textbf{Energy of bcc Cs as a function of the nearest-neighbor distance.} Energy per atom relative to its value at large $d_\mathrm{NN}$. Dotted lines mark the cutoff radii of 5 and 6\,\AA{}, and the solid line marks $d_\mathrm{NN}$ of the reference.}
\label{si:gb-cs}
\end{figure}

Figure~\ref{si:gb-cs} shows the energy of bcc Cs as a function of the nearest-neighbor distance $d_\mathrm{NN}$. With the FPs with a cutoff radius of 5\,\AA{} (TensorNet-MatPES-PBE and QET-MatPES-PBE), the energy does not depend on $d_\mathrm{NN}$ above 5\,\AA{}, and $\gamma_\mathrm{GB}$ is zero for all ten grain boundaries of Cs. With GRACE-OMat24, the energy decreases with increasing $d_\mathrm{NN}$ up to the cutoff radius of 6\,\AA{}, and $d_\mathrm{NN}$ of the relaxed bulk is 5.95\,\AA{}, compared with 5.33\,\AA{} in the reference. In the $\Sigma5[100](0\bar{1}\bar{2})$ grain boundary scaled to this lattice, 5 of the 38 atoms have no neighbor within the cutoff radius, and $\gamma_\mathrm{GB}$ is 10.77\,J\,m$^{-2}$, compared with 0.01\,J\,m$^{-2}$ in the reference. Without this grain boundary, the MAE of GRACE-OMat24 is 0.06\,J\,m$^{-2}$.

\subsection{Coherent twin boundaries and the maximum angular order}
\label{si:gb-angular}

The maximum angular order $l_\mathrm{max}$ of each FP is listed in Table~\ref{si:tab-model-table}: it is 2 for TensorNet-MatPES-PBE and QET-MatPES-PBE and at least 3 for the other FPs.

\paragraph{Dependence on $l_\mathrm{max}$ and the cutoff radius.} The two sides of a coherent twin boundary of an fcc metal differ only in the stacking sequence of the close-packed layers, ABC and CBA. For the nearest-neighbor distance $d_\mathrm{NN}$ and the layer spacing $d$ of the bulk, every atom of the twin boundary structure has the same number of neighbors at the same distances as in the bulk below $2d=1.633\,d_\mathrm{NN}$, the distance to the atoms two layers away, and its environment differs from that in the bulk only by a rotation of 60$^\circ$ of an adjacent layer about the stacking axis, which requires $l_\mathrm{max}\geq3$ to be distinguished. An FP with $l_\mathrm{max}=2$ and a cutoff radius smaller than $2d$ therefore gives $\gamma_\mathrm{GB}=0$ for this structure, also after message passing.

\begin{figure}[!htbp]
\centering
\includegraphics[width=\textwidth]{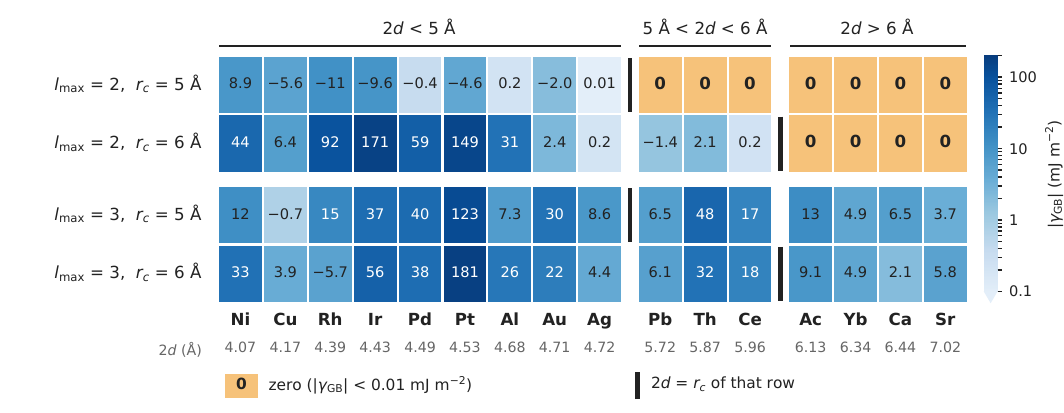}
\caption{\textbf{Energy of the coherent twin boundary for different $l_\mathrm{max}$ and cutoff radii $r_\mathrm{c}$.} Grain boundary energy $\gamma_\mathrm{GB}$ of the unrelaxed coherent twin boundary at the experimental lattice parameter, for the fcc metals with a coherent twin boundary in the benchmark in the order of increasing $2d$, the distance to the atoms two layers away. The rows are, from top to bottom, TensorNet-MatPES-PBE, UMA-S-1.2.1, SevenNet-MPtrj-l3i5, and MACE-MatPES-PBE. The numbers are $\gamma_\mathrm{GB}$ in mJ\,m$^{-2}$, and the color shows its magnitude.}
\label{si:gb-twin-lmax}
\end{figure}

Figure~\ref{si:gb-twin-lmax} shows $\gamma_\mathrm{GB}$ of this coherent twin boundary for the 16 fcc metals with a coherent twin boundary in the benchmark, calculated without relaxation at the experimental lattice parameter of each metal\cite{larsenAtomicSimulationEnvironment2017} in a cell of 12 layers with two boundaries. In the benchmark, the FPs with $l_\mathrm{max}=2$ (TensorNet-MatPES-PBE and QET-MatPES-PBE) both have a cutoff radius of 5\,\AA{} and those with $l_\mathrm{max}\geq3$ a cutoff radius of 6\,\AA{}, so that the two effects cannot be separated. Three further FPs were therefore evaluated: UMA-S-1.2.1\cite{Wood2025UMA} ($l_\mathrm{max}=2$, 6\,\AA{}), SevenNet-MPtrj-l3i5\cite{park2024} ($l_\mathrm{max}=3$, 5\,\AA{}), and SevenNet-MPtrj-0\cite{park2024} ($l_\mathrm{max}=2$, 5\,\AA{}), which has the same architecture and training data as SevenNet-MPtrj-l3i5. With the four FPs with $l_\mathrm{max}=2$, $\gamma_\mathrm{GB}$ vanishes ($|\gamma_\mathrm{GB}|<0.02$\,mJ\,m$^{-2}$) for all metals with $2d$ larger than the cutoff radius: seven metals with a cutoff radius of 5\,\AA{} and four with 6\,\AA{}. With $l_\mathrm{max}=3$, it is at least 2\,mJ\,m$^{-2}$ for the same metals with either cutoff radius. A larger cutoff radius thus reduces the number of affected metals, but only $l_\mathrm{max}\geq3$ distinguishes the twin boundary from the bulk. In the benchmark, $\gamma_\mathrm{GB}$ of the coherent twin boundaries of the seven metals is at most 4\,mJ\,m$^{-2}$ with TensorNet-MatPES-PBE and QET-MatPES-PBE; this remainder arises from the layer spacings of the reference structures, which differ from those of the bulk.

\begin{figure}[!htbp]
\centering
\includegraphics[width=\textwidth]{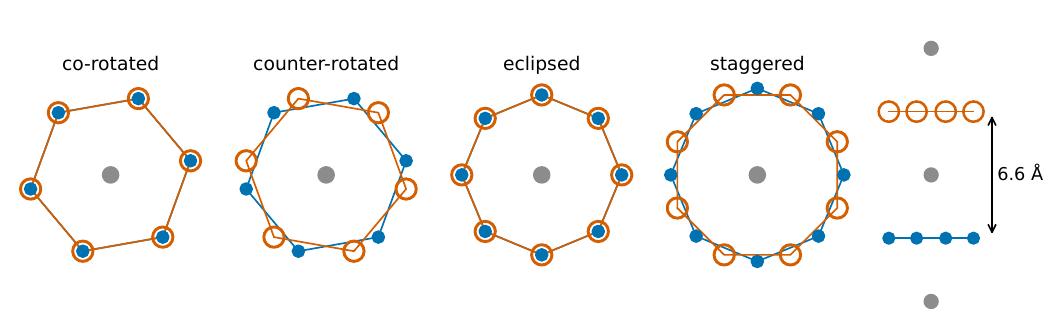}
\caption{\textbf{Structures of Cu with stacked rings.} Views along the ring axis of the co-rotated and counter-rotated six-membered rings and of the eclipsed and staggered eight-membered rings. Filled and open circles mark the rings in alternate layers, and gray circles the atoms on the axis between them. The rightmost panel shows a side view with the spacing between the ring layers.}
\label{si:gb-rings}
\end{figure}

\paragraph{Structures that require a higher angular order.} The same argument applies to groups of neighbors with higher rotational symmetry. Two pairs of structures of Cu were constructed in which rings of six or eight atoms are stacked along their axis at a spacing larger than the cutoff radius of all FPs. In the first pair, the six-membered rings in alternate layers are co-rotated or counter-rotated; in the second pair, the eight-membered rings are eclipsed or staggered. Within each pair, every atom has the same neighbor distances within 6\,\AA{} (Fig.~\ref{si:gb-rings}). For the eight-membered rings, all seven FPs give energy differences below $10^{-3}$\,meV\,atom$^{-1}$. For the six-membered rings, six FPs give energy differences below $10^{-3}$\,meV\,atom$^{-1}$ and GRACE-OMat24 gives 0.02\,meV\,atom$^{-1}$.

Without the 16 coherent twin boundaries of the fcc metals, the MAEs of TensorNet-MatPES-PBE and QET-MatPES-PBE in $\gamma_\mathrm{GB}$ remain 0.07 and 0.09\,J\,m$^{-2}$.

\section{Adsorption energies}
\label{si:ads}

\subsection{Adsorption energies against PBE+$U$ reference values}
\label{si:ads-pbeu}

To test whether the errors of MACE-MPtrj, MACE-MPA, MACE-OMat24, and GRACE-OMat24 on the CMR set arise from the choice of PBE for the reference, the 72 CMR cells that contain V, Cr, Mn, Fe, Co, Ni, Mo, or W were recomputed with PBE+$U$, using the $U$ values of the Materials Project for all of these cells, with and without O. The settings were otherwise those of the PBE reference calculations, with the lowest energy of several initial magnetic configurations. Nine of the 126 calculations did not reach electronic convergence; the reaction of N on V is therefore missing, and for CH on Cr the lowest converged energy may not be the ground state. Table~\ref{si:ads-pbeu-mae} gives the MAEs for the 32 reactions of O, OH, NO, and CO on these metals. Against the PBE+$U$ reference values, the MAEs of the four FPs decrease only slightly, from 3.14--5.81 to 2.63--5.25\,eV, whereas those of the MatPES-trained FPs increase from 0.26--0.53 to 0.51--0.88\,eV. The PBE+$U$ and PBE reference values of these reactions differ by 0.56\,eV on average.

\begin{table}[!htbp]
\centering
\caption{\textbf{MAEs of the adsorption energies of O, OH, NO, and CO on V, Cr, Mn, Fe, Co, Ni, Mo, and W against PBE and PBE+$U$ reference values (eV).}}
\label{si:ads-pbeu-mae}
\begin{tabular}{lcc}
\toprule
 & PBE & PBE+$U$ \\
\midrule
MACE-MPtrj & 3.14 & 2.63 \\
MACE-MPA & 3.71 & 3.16 \\
MACE-OMat24 & 4.66 & 4.10 \\
GRACE-OMat24 & 5.81 & 5.25 \\
MACE-MatPES-PBE & 0.27 & 0.57 \\
TensorNet-MatPES-PBE & 0.26 & 0.51 \\
QET-MatPES-PBE & 0.53 & 0.88 \\
\bottomrule
\end{tabular}
\end{table}

\subsection{Adsorption energies on the ADS41 dataset}
\label{si:ads41}

The ADS41 dataset\cite{trepteDatadrivenConstrainedOptimization2022} complements the CMR set used in the main text with four-layer slabs at low coverage, larger molecules, and experimental adsorption energies. It comprises molecular and dissociative adsorption of 18 molecules on 14 low-index surfaces of 10 transition metals. The structures and reference energies were taken from Ref.~\citenum{trepteDatadrivenConstrainedOptimization2022}, where they were relaxed separately with PBE and r$^2$SCAN. These references differ from those of the CMR set and of MatPES: they were computed with VASP~5.4.4, a plane-wave cutoff of 1000~eV, and spin polarization only for Co, Ni, and the O$_2$ and NO molecules. The experimental values are those of the ADS41 dataset,\cite{mallikarjunsharadaAdsorptionTransitionMetal2019} which extends the CE39 database,\cite{wellendorffBenchmarkDatabaseAdsorption2015} as collected in Ref.~\citenum{trepteDatadrivenConstrainedOptimization2022}. The PBE-trained and r$^2$SCAN-trained FPs were evaluated at fixed geometry on the PBE and r$^2$SCAN structures, respectively. Table~\ref{si:ads41-mae} lists the MAEs against the DFT and the experimental values.

\begin{table}[!htbp]
\centering\small
\caption{\textbf{MAEs of the adsorption energies on the ADS41 dataset (eV).} DFT denotes the functional used to train each FP. The first two rows give the MAEs of the DFT reference values against experiment.}
\label{si:ads41-mae}
\begin{tabular}{lcc}
\toprule
 & vs.\ DFT & vs.\ experiment \\
\midrule
PBE (DFT) & -- & 0.36 \\
r$^2$SCAN (DFT) & -- & 0.32 \\
\midrule
MACE-MPtrj & 0.64 & 0.73 \\
MACE-MPA & 1.06 & 1.01 \\
MACE-OMat24 & 2.13 & 2.06 \\
GRACE-OMat24 & 1.36 & 1.44 \\
MACE-MatPES-PBE & 0.23 & 0.41 \\
TensorNet-MatPES-PBE & 0.33 & 0.42 \\
QET-MatPES-PBE & 0.41 & 0.49 \\
\midrule
MACE-MatPES-r2SCAN & 0.46 & 0.73 \\
TensorNet-MatPES-r2SCAN & 0.37 & 0.46 \\
QET-MatPES-r2SCAN & 0.45 & 0.59 \\
\bottomrule
\end{tabular}
\end{table}

Against experiment, the r$^2$SCAN functional is only slightly more accurate than the PBE functional (MAEs of 0.32 and 0.36\,eV), whereas the r$^2$SCAN-trained FPs deviate from their own reference values by 0.37--0.46\,eV, more than the error of the functional itself, as molecules adsorbed on metal surfaces are largely absent from their training data. The small advantage of the functional is therefore lost, and the r$^2$SCAN-trained FPs are not more accurate against experiment than the PBE-trained FPs (0.46--0.73 versus 0.41--0.49\,eV for the MatPES-trained FPs). Further gains for such properties require both training data that cover the relevant configurations and a more accurate functional for these systems. The largest errors occur for physisorbed molecules: for benzene and naphthalene on Pt(111), the r$^2$SCAN functional gives adsorption energies of $-1.49$ and $-2.51$~eV, close to the experimental $-1.68$ and $-2.76$~eV, whereas the r$^2$SCAN-trained FPs give $-0.30$ and $-0.23$~eV on average.

Neither DFT reference includes a dispersion correction, and none was added to the FPs. Adding the D3 correction with Becke--Johnson damping\cite{grimmeConsistentAccurateInitio2010,grimmeEffectDampingFunction2011,ehlertR2SCAND4Dispersion2021} or the D4 correction\cite{caldeweyherGenerallyApplicableAtomiccharge2019,caldeweyherExtensionEvaluationD42020} at fixed geometry raises the MAE for chemisorption in every case (Table~\ref{si:ads41-disp}). For physisorption, it lowers the MAE of PBE from 0.48 to 0.16--0.18\,eV, whereas that of r$^2$SCAN, which already describes part of the intermediate-range van der Waals interaction\cite{sunAccurateFirstprinciplesStructures2016,furnessAccurateNumericallyEfficient2020} and has an MAE of 0.23\,eV without a correction, changes little (0.16--0.24\,eV). For the r$^2$SCAN-trained FPs, the MAE for physisorption is 0.55--0.69\,eV with either correction, compared with 0.51--0.87\,eV without.

\begin{table}[!htbp]
\centering\small
\caption{\textbf{Effect of dispersion corrections on the MAEs of the ADS41 adsorption energies against experiment (eV).} Chemisorption (Chem.) and physisorption (Phys.) reactions are classified as in Ref.~\citenum{trepteDatadrivenConstrainedOptimization2022}. D3(BJ) and D4 were added at fixed geometry; D3(BJ) includes two-body terms only, as in VASP, and D4 includes the three-body term.}
\label{si:ads41-disp}
\setlength{\tabcolsep}{4.5pt}
\begin{tabular}{l ccc ccc ccc}
\toprule
 & \multicolumn{3}{c}{No correction} & \multicolumn{3}{c}{+ D3(BJ)} & \multicolumn{3}{c}{+ D4} \\
\cmidrule(lr){2-4}\cmidrule(lr){5-7}\cmidrule(lr){8-10}
 & Chem. & Phys. & All & Chem. & Phys. & All & Chem. & Phys. & All \\
\midrule
PBE & 0.29 & 0.48 & 0.36 & 0.45 & 0.18 & 0.35 & 0.44 & 0.16 & 0.34 \\
MACE-MatPES-PBE & 0.35 & 0.52 & 0.41 & 0.42 & 0.26 & 0.36 & 0.41 & 0.23 & 0.35 \\
TensorNet-MatPES-PBE & 0.34 & 0.57 & 0.42 & 0.48 & 0.48 & 0.48 & 0.48 & 0.44 & 0.46 \\
QET-MatPES-PBE & 0.45 & 0.56 & 0.49 & 0.65 & 0.49 & 0.59 & 0.65 & 0.47 & 0.58 \\
\addlinespace
r$^2$SCAN & 0.36 & 0.23 & 0.32 & 0.48 & 0.24 & 0.39 & 0.46 & 0.16 & 0.35 \\
MACE-MatPES-r2SCAN & 0.66 & 0.87 & 0.73 & 0.72 & 0.55 & 0.66 & 0.72 & 0.61 & 0.68 \\
TensorNet-MatPES-r2SCAN & 0.43 & 0.51 & 0.46 & 0.51 & 0.61 & 0.55 & 0.49 & 0.58 & 0.52 \\
QET-MatPES-r2SCAN & 0.49 & 0.76 & 0.59 & 0.53 & 0.68 & 0.59 & 0.53 & 0.69 & 0.59 \\
\bottomrule
\end{tabular}
\end{table}

\subsection{Relaxation of adsorbates on Ni}
\label{si:ads-relax}

\begin{figure}[!htbp]
\centering
\includegraphics[width=\textwidth]{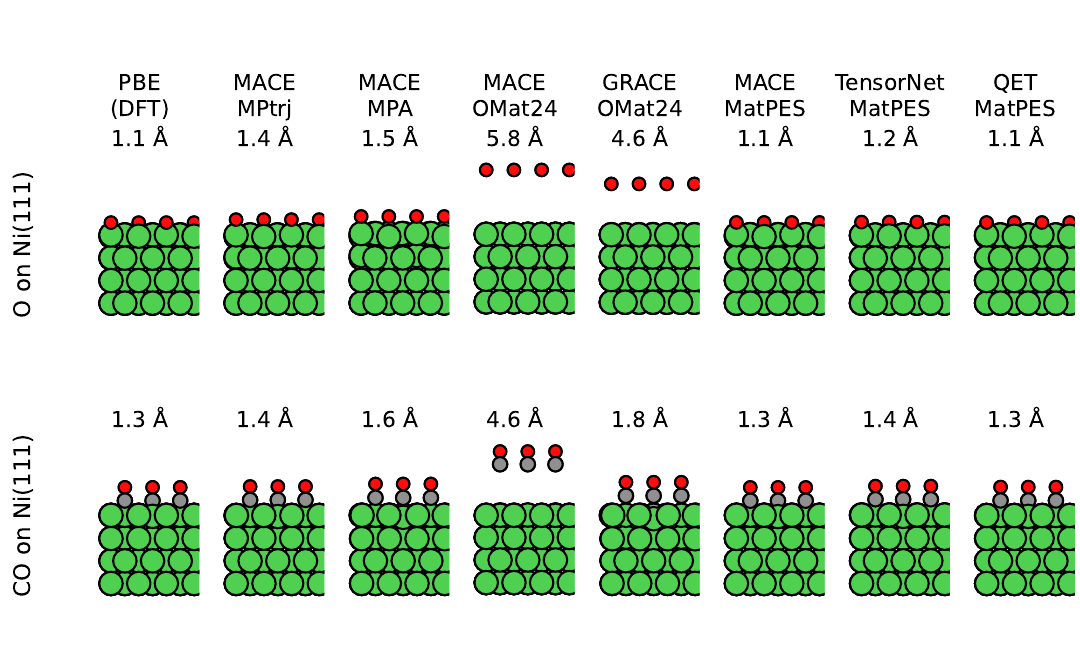}
\caption{\textbf{O and CO on Ni(111) after relaxation with the FPs.} Side views of the ADS41 structures of O (top) and CO (bottom) on Ni(111), as relaxed with PBE and after relaxation with the seven PBE-trained FPs starting from the PBE geometry, with the bottom two layers fixed. The number above each panel is the height of the lowest adsorbate atom above the top Ni layer. Ni, green; O, red; C, gray.}
\label{si:ads-desorption}
\end{figure}

\begin{figure}[!htbp]
\centering
\includegraphics[width=\textwidth]{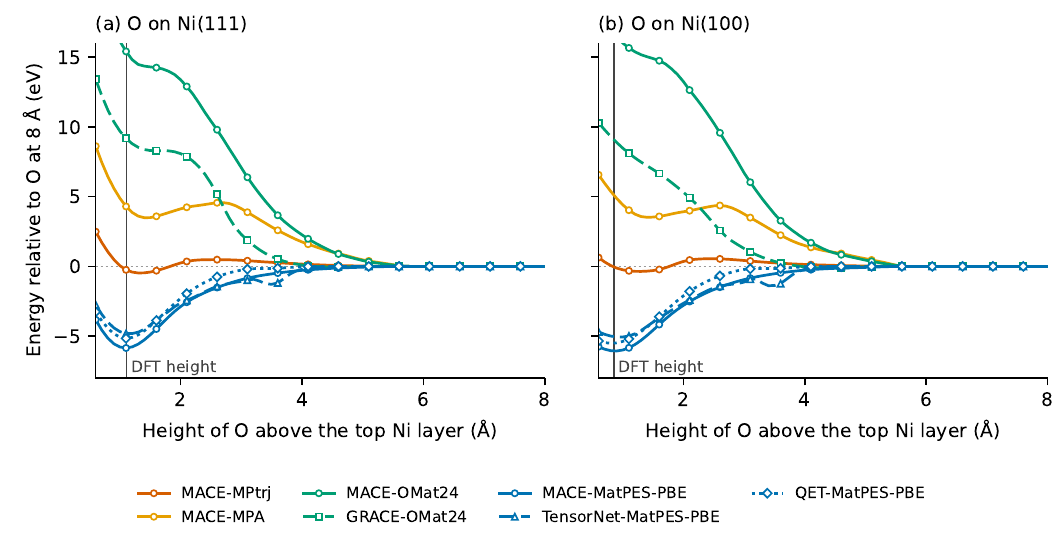}
\caption{\textbf{Energy of O on Ni as a function of its height.} Energy of the ADS41 structures of O on (a) Ni(111) and (b) Ni(100) as the O atom is moved along the surface normal, with all other atoms fixed at the DFT positions, relative to the energy with O at 8~\AA{}. The vertical line marks the height of O in the DFT structure.}
\label{si:ads-o-scan}
\end{figure}

To test whether the adsorption energies change when the FPs relax the adsorption structures, the ADS41 structures\cite{trepteDatadrivenConstrainedOptimization2022} were relaxed with each FP from the DFT geometries to 0.05~eV~\AA$^{-1}$ using FIRE, with the bottom two layers fixed as in the reference calculations. Figure~\ref{si:ads-desorption} shows O and CO on Ni(111) after the relaxation. MACE-OMat24 and GRACE-OMat24 moved O on Ni(111) and Ni(100) from 1.1 and 0.85~\AA{} to 4.4--5.8~\AA{} above the top Ni layer, and MACE-OMat24 also moved CO on Ni(111) to 4.6~\AA{}; the adsorbates desorbed, and the total energies of these structures decreased by 9--17~eV. MACE-MPtrj and MACE-MPA moved O outward by 0.3--0.7~\AA{}, whereas the three MatPES-trained FPs kept it within 0.1~\AA{} of the DFT position. The lower MAEs of MACE-OMat24 and GRACE-OMat24 after relaxation (0.86 and 0.66~eV, against 2.13 and 1.36~eV at fixed geometry) therefore result from desorption rather than from a better description of the adsorbed state. Rigid scans of the height of O above Ni(111) and Ni(100), with the other atoms fixed at the DFT positions (Figure~\ref{si:ads-o-scan}), show that MACE-MPA has a local minimum near the surface 3.6~eV above the desorbed state, separated from it by a barrier of 0.8--1.0~eV, and that MACE-MPtrj binds O by 0.36--0.47~eV, compared with 4.8--6.1~eV for the MatPES-trained FPs; MACE-OMat24 and GRACE-OMat24 have no minimum near the surface. Similar behavior, the absence of binding for MACE-OMat24 and a barrier with underbinding for MACE-MPA, was reported by Warford et al.\ for an O atom on top sites of 54 elemental slabs, together with the separation of the metal and oxide slabs during the relaxation of Fe/FeO and Ni/NiO interfaces.\cite{warfordBetterWithoutImpact2026} The repulsion between O and Ni is consistent with the Materials Project scheme, in which the Hubbard $U$ correction is applied to Ni and several other transition metals only in the presence of O or F,\cite{jainFormationEnthalpiesMixing2011} so that the training data of these FPs contain energies of these metals with and without O computed at different levels of theory, as discussed by Warford et al.\cite{warfordBetterWithoutImpact2026}

\subsection{Adsorption energies on transition metal compounds from Catalysis-Hub}
\label{si:ads-cathub}

To test adsorption on transition metal compounds, the PBE-trained FPs were evaluated on the full set of 12,935 adsorption reactions of Yohannes et al.\cite{yohannesCombinedHighThroughputDFT2023} in Catalysis-Hub,\cite{wintherCatalysisHubOpenElectronic2019} which comprises H, OH, CO, and CHO on the (100) and (110) surfaces of binary transition metal compounds. The reference calculations used VASP with the PBE functional and the D3 dispersion correction with Becke--Johnson damping.\cite{grimmeConsistentAccurateInitio2010,grimmeEffectDampingFunction2011,yohannesCombinedHighThroughputDFT2023} The stored structures were evaluated with each FP at fixed geometry. Since the FPs were trained on data without a dispersion correction, the D3 energy of each slab, slab with adsorbate, and gas-phase molecule was computed with the parameters of the reference calculations (PBE, Becke--Johnson damping, two-body terms) and added to its FP energy before the reaction energies were calculated.

Table~\ref{si:ads-cathub-mae} gives the MAEs without and, in parentheses, with the D3 correction. As for the CMR set, the errors of MACE-MPtrj, MACE-MPA, MACE-OMat24, and GRACE-OMat24 are concentrated on compounds of V, Cr, Mn, Fe, Co, Ni, Mo, and W and on the adsorbates that contain O, whereas the MatPES-trained FPs have similar errors on both groups of substrates. The D3 correction changes the reaction energies by $-0.18$~eV on average. Adding it changes the MAEs over all reactions by less than 0.1~eV.

\begin{table}[!htbp]
\centering\small
\caption{\textbf{MAEs of the adsorption energies on transition metal compounds from Catalysis-Hub (eV).} Rows give all reactions, each adsorbate, and the reactions on compounds that contain or do not contain V, Cr, Mn, Fe, Co, Ni, Mo, or W ($U$ metals). Values in parentheses include the D3 correction added to the FP energies.}
\label{si:ads-cathub-mae}
\begin{tabular}{l ccccccc}
\toprule
 & \multicolumn{1}{c}{MPtrj} & \multicolumn{1}{c}{MPA} & \multicolumn{2}{c}{OMat24} & \multicolumn{3}{c}{MatPES-PBE} \\
\cmidrule(lr){2-2}\cmidrule(lr){3-3}\cmidrule(lr){4-5}\cmidrule(lr){6-8}
 & MACE & MACE & MACE & GRACE & MACE & TensorNet & QET \\
\midrule
All & 1.11 (1.02) & 1.37 (1.28) & 3.07 (3.03) & 1.20 (1.17) & 0.38 (0.41) & 0.42 (0.40) & 0.40 (0.40) \\
H & 0.38 (0.37) & 0.32 (0.29) & 0.26 (0.25) & 0.20 (0.20) & 0.36 (0.32) & 0.44 (0.40) & 0.43 (0.40) \\
OH & 2.03 (1.92) & 2.44 (2.34) & 5.59 (5.49) & 2.49 (2.44) & 0.37 (0.29) & 0.35 (0.33) & 0.30 (0.34) \\
CO & 0.79 (0.66) & 1.16 (1.06) & 2.73 (2.72) & 0.84 (0.80) & 0.28 (0.41) & 0.32 (0.33) & 0.31 (0.30) \\
CHO & 1.22 (1.13) & 1.54 (1.44) & 3.68 (3.65) & 1.28 (1.24) & 0.50 (0.62) & 0.58 (0.54) & 0.56 (0.56) \\
$U$ metals & 1.71 (1.57) & 2.33 (2.17) & 5.61 (5.45) & 2.12 (1.99) & 0.39 (0.41) & 0.43 (0.42) & 0.42 (0.42) \\
Others & 0.49 (0.45) & 0.36 (0.35) & 0.41 (0.50) & 0.24 (0.31) & 0.37 (0.41) & 0.41 (0.38) & 0.38 (0.39) \\
\bottomrule
\end{tabular}
\end{table}

\section{Interfacial energies and works of adhesion}
\label{si:interface-metal}

\begin{figure}[!htbp]
\centering
\includegraphics[width=\textwidth]{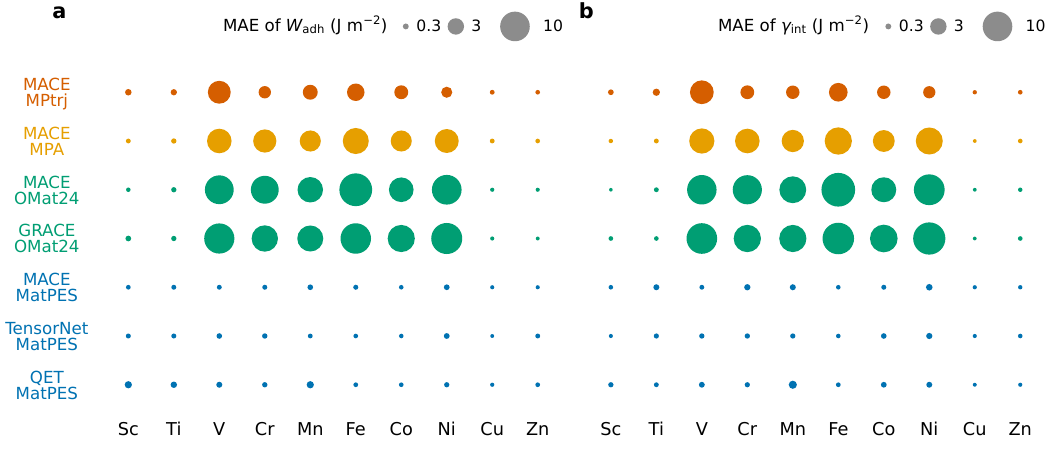}
\caption{\textbf{Interfacial energies and works of adhesion by metal.} MAE of (a)~the work of adhesion $W_\mathrm{adh}$ and (b)~the interfacial energy $\gamma_\mathrm{int}$ for each metal, obtained over all interfaces of the metal with its oxides. The area of each circle is proportional to the MAE.}
\label{si:interface-bubbles}
\end{figure}

Figure~\ref{si:interface-bubbles} shows the MAEs of the work of adhesion $W_\mathrm{adh}$ and the interfacial energy $\gamma_\mathrm{int}$ for each metal.

\section{Comparison with experiment}
\label{si:expt}

\begin{figure}[!htbp]
\centering
\includegraphics[width=0.6\textwidth]{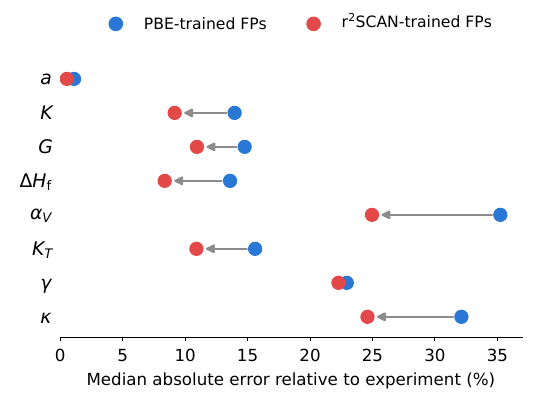}
\caption{\textbf{Absolute deviations of the FPs from experiment.} Median absolute relative error against experiment of the eight properties of the main text. The median is taken over the materials for each FP and then averaged over the PBE-trained (blue) and r$^2$SCAN-trained (red) FPs. Arrows point from the PBE-trained to the r$^2$SCAN-trained FPs.}
\label{si:expt-unsigned}
\end{figure}

Figure~\ref{si:expt-unsigned} shows the median absolute relative errors of the FPs against experiment, the unsigned counterpart of the median relative errors in the main text.

\section{Computational cost}
\label{si:cost}

This section gives the timings summarized in the main text, measured as described in Methods on one NVIDIA A100 GPU and on 32 cores of a CPU node.

\subsection{Molecular dynamics}
\label{si:cost-basic}

Table~\ref{si:speed} gives the time of an MD step for the seven PBE-trained FPs. For MACE, TensorNet, and QET, two implementations are compared on the GPU. The energies, forces, and stresses of the implementations of each FP, and those on the GPU and the CPU, agree within $6\times10^{-6}$\,eV\,atom$^{-1}$, $7\times10^{-5}$\,eV\,\AA$^{-1}$, and $5\times10^{-6}$\,eV\,\AA$^{-3}$. Molecular dynamics of 10,648 atoms exceeds the 40\,GB of GPU memory with the e3nn implementation of MACE and the PyG implementation of TensorNet and QET.

For 1,000 atoms on the GPU, a step takes 31 and 35\,ms with TensorNet-MatPES-PBE and QET-MatPES-PBE, 57--64\,ms with the MACE models, and 89\,ms with GRACE-OMat24. These timings use the Warp implementation of TensorNet and QET and the cuEquivariance implementation of MACE, which are 1.6--1.9 times faster than the default implementations for 1,000 atoms and are the only ones that fit 10,648 atoms into 40\,GB of GPU memory. In double precision, a step is typically about 30\% slower. On the CPU node, a step for 1,000 atoms takes 0.65--0.77\,s with the MACE, TensorNet, and QET models and 4.2\,s with GRACE-OMat24.

\begin{table}[!htbp]
\centering\small
\caption{\textbf{Time of an MD step for the PBE-trained FPs.} MD of diamond Si in the NVT ensemble. $^{*}$Configuration used for the benchmark results. OOM: out of memory.}
\label{si:speed}
\setlength{\tabcolsep}{4pt}
\resizebox{\textwidth}{!}{%
\begin{tabular}{lllccc}
\toprule
 & & & \multicolumn{3}{c}{MD (ms per step)} \\
\cmidrule(lr){4-6}
FP & Implementation & Precision & 1,000 atoms & 4,096 atoms & 10,648 atoms \\
\midrule
\multicolumn{6}{l}{\textit{GPU, NVIDIA A100}} \\
MACE-MPtrj & e3nn & single & 109 & 410 & OOM \\
 & cuEquivariance$^{*}$ & single & 56.8 & 131 & 352 \\
MACE-MPA & e3nn & single & 114 & 426 & OOM \\
 & cuEquivariance$^{*}$ & single & 62.4 & 144 & 384 \\
MACE-OMat24 & e3nn & single & 114 & 426 & OOM \\
 & cuEquivariance$^{*}$ & single & 63.5 & 146 & 387 \\
 & e3nn & double & 148 & -- & -- \\
 & cuEquivariance & double & 60.5 & -- & -- \\
MACE-MatPES-PBE & e3nn & single & 114 & 429 & OOM \\
 & cuEquivariance$^{*}$ & single & 62.5 & 147 & 392 \\
\addlinespace
GRACE-OMat24 & TensorFlow$^{*}$ & single & 88.5 & 339 & 896 \\
 & TensorFlow & double & 102 & 418 & 1,160 \\
\addlinespace
TensorNet-MatPES-PBE & PyG$^{*}$ & single & 49.0 & 183 & OOM \\
 & Warp & single & 30.6 & 73.7 & 190 \\
 & PyG & double & 65.6 & -- & -- \\
QET-MatPES-PBE & PyG$^{*}$ & single & 55.3 & 190 & OOM \\
 & Warp & single & 35.1 & 80.0 & 198 \\
 & PyG & double & 71.3 & -- & -- \\
\midrule
\multicolumn{6}{l}{\textit{CPU, 32 cores}} \\
MACE-MPtrj & e3nn & single & 767 & -- & -- \\
MACE-MPA & e3nn & single & 769 & -- & -- \\
MACE-OMat24 & e3nn & single & 765 & -- & -- \\
MACE-MatPES-PBE & e3nn & single & 774 & -- & -- \\
GRACE-OMat24 & TensorFlow & single & 4,203 & -- & -- \\
TensorNet-MatPES-PBE & PyG & single & 648 & -- & -- \\
QET-MatPES-PBE & PyG & single & 652 & -- & -- \\
\bottomrule
\end{tabular}}
\end{table}

\subsection{Cost of property workflows}
\label{si:cost-workflows}

Supplementary Table~\ref{si:speed-workflows} gives the wall time per material of six property workflows for the seven PBE-trained FPs, measured for 8 materials each for the elastic constants, $C_V$, and the EOS, 4 materials for the QHA, 2 migration paths for NEB, and one compound (BN) for $\kappa$. Each workflow was run once in a new process, so that compilation is included. For GRACE-OMat24, the compilation is repeated whenever a structure has more atoms or neighbor pairs than any structure evaluated before in the process, which takes about 8--13\,s and makes up most of its time for the larger materials of the elastic-constant and EOS workflows and for most materials of the phonon workflow. The FP accounts for 82--99\% of the time of the elastic constants, the QHA, and NEB, and for 25--69\% of that of $\kappa$.

\begin{table}[!htbp]
\centering\small
\caption{\textbf{Time per material of the property workflows on one NVIDIA A100 GPU.} Median wall time in seconds, with the range over the materials in parentheses, in single precision and with the implementation of each FP used for the benchmarks.}
\label{si:speed-workflows}
\setlength{\tabcolsep}{3.5pt}
\resizebox{\textwidth}{!}{%
\begin{tabular}{lcccccc}
\toprule
FP & Elastic constants & Phonon $C_V$ & EOS & QHA & NEB & $\kappa$ (Phonon3) \\
\midrule
MACE-MPtrj & 1.4 (0.8--4.0) & 2.2 (1.3--7.9) & 3.8 (0.7--19.4) & 50.0 (35.3--227) & 26.7 (24.9--28.5) & 29.5 \\
MACE-MPA & 1.8 (1.1--10.0) & 2.2 (1.6--8.8) & 5.4 (0.7--21.9) & 77.6 (42.6--262) & 34.6 (31.7--37.5) & 30.0 \\
MACE-OMat24 & 2.3 (1.2--9.1) & 2.3 (1.9--8.8) & 5.8 (0.9--21.7) & 80.6 (44.1--261) & 33.0 (31.9--34.0) & 29.5 \\
MACE-MatPES-PBE & 1.8 (1.1--8.9) & 2.6 (2.0--8.9) & 4.8 (0.7--21.1) & 72.0 (42.6--261) & 30.5 (30.5--30.5) & 31.4 \\
GRACE-OMat24 & 0.3 (0.2--9.0) & 13.6 (0.5--21.2) & 1.5 (0.2--18.7) & 47.2 (33.6--137) & 24.3 (19.3--29.3) & 53.7 \\
TensorNet-MatPES-PBE & 0.8 (0.6--1.3) & 1.1 (0.8--5.8) & 1.6 (0.4--14.9) & 45.4 (24.8--166) & 14.7 (11.1--18.2) & 22.2 \\
QET-MatPES-PBE & 1.1 (0.8--1.5) & 1.2 (0.9--7.0) & 1.7 (0.3--17.2) & 52.8 (29.6--190) & 17.3 (13.3--21.2) & 22.2 \\
\bottomrule
\end{tabular}}
\end{table}

\subsection{Dependence on the number of atoms}
\label{si:cost-natoms}

\begin{figure}[!htbp]
\centering
\includegraphics[width=0.8\textwidth]{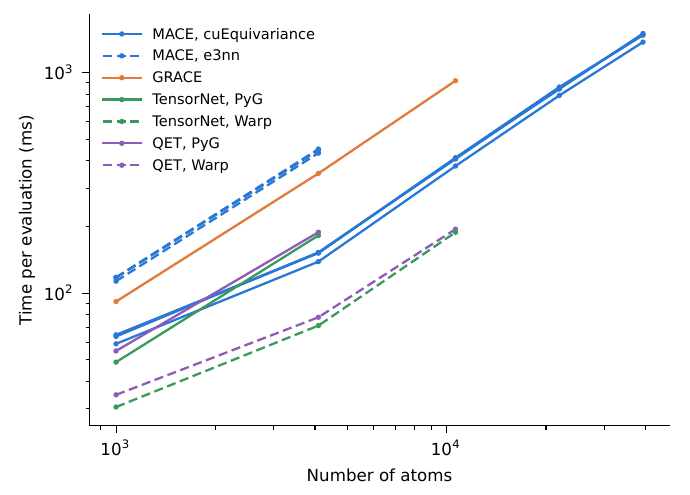}
\caption{\textbf{Evaluation time as a function of the number of atoms.} Median time of one evaluation of the energy, forces, and stress of diamond Si supercells on one NVIDIA A100 GPU in single precision. Each line ends at the largest supercell that fits into 40\,GB of GPU memory. The four MACE models are shown with the same color.}
\label{si:speed-natoms}
\end{figure}

Supplementary Fig.~\ref{si:speed-natoms} and Table~\ref{si:speed-size} give the time of a single evaluation of the energy, forces, and stress for diamond Si supercells of 1,000 to 85,184 atoms. The time increases with the number of atoms, in proportion above 4,096 atoms for all FPs. The cuEquivariance implementation of MACE is 2.9--3.1 times faster than the e3nn implementation for 4,096 atoms, and the Warp implementation of TensorNet and QET is 2.4--2.6 times faster than the PyG implementation. The largest supercell that fits into 40\,GB of GPU memory has 4,096 atoms with e3nn and PyG, 10,648 atoms with Warp and for GRACE-OMat24, and 39,304 atoms for MACE with cuEquivariance.

\begin{table}[!htbp]
\centering\small
\caption{\textbf{Evaluation time (ms) as a function of the number of atoms.} Median time of one evaluation of the energy, forces, and stress of diamond Si supercells on one NVIDIA A100 GPU in single precision. $^{\dagger}$Implementation used for the benchmarks. OOM: out of GPU memory.}
\label{si:speed-size}
\setlength{\tabcolsep}{3.5pt}
\resizebox{\textwidth}{!}{%
\begin{tabular}{llrrrrrr}
\toprule
 & & \multicolumn{6}{c}{Number of atoms} \\
\cmidrule(lr){3-8}
Model & Implementation & 1{,}000 & 4{,}096 & 10{,}648 & 21{,}952 & 39{,}304 & 85{,}184 \\
\midrule
MACE-MPtrj & cuEquivariance$^{\dagger}$ & 59.0 & 139 & 378 & 789 & 1{,}376 & OOM \\
MACE-MPtrj & e3nn & 114 & 432 & OOM & -- & -- & -- \\
MACE-MPA & cuEquivariance$^{\dagger}$ & 63.8 & 153 & 406 & 856 & 1{,}484 & OOM \\
MACE-MPA & e3nn & 118 & 450 & OOM & -- & -- & -- \\
MACE-OMat24 & cuEquivariance$^{\dagger}$ & 64.6 & 153 & 409 & 843 & 1{,}505 & OOM \\
MACE-OMat24 & e3nn & 118 & 449 & OOM & -- & -- & -- \\
MACE-MatPES-PBE & cuEquivariance$^{\dagger}$ & 64.8 & 152 & 412 & 861 & 1{,}480 & OOM \\
MACE-MatPES-PBE & e3nn & 118 & 445 & OOM & -- & -- & -- \\
GRACE-OMat24 & TensorFlow$^{\dagger}$ & 91.8 & 349 & 920 & OOM & -- & -- \\
TensorNet-MatPES-PBE & PyG$^{\dagger}$ & 48.8 & 183 & OOM & -- & -- & -- \\
TensorNet-MatPES-PBE & Warp & 30.5 & 71.4 & 189 & OOM & -- & -- \\
QET-MatPES-PBE & PyG$^{\dagger}$ & 54.8 & 189 & OOM & -- & -- & -- \\
QET-MatPES-PBE & Warp & 34.7 & 77.9 & 195 & OOM & -- & -- \\
\bottomrule
\end{tabular}}
\end{table}

\subsection{ASE and LAMMPS backends}
\label{si:cost-backends}

The two simulation backends were compared for TensorNet-MatPES-PBE with MatCalc v0.8.0 on one NVIDIA A100 GPU (80\,GB) and on 32 CPU cores; on the GPU, LAMMPS used the Kokkos variant of the pair style. For single points on 17 structures of 5 to 1,000 atoms, the energies, forces, and stresses of the two backends agree within $2\times10^{-6}$\,eV\,atom$^{-1}$, $3\times10^{-5}$\,eV\,\AA$^{-1}$, and $2\times10^{-7}$\,eV\,\AA$^{-3}$. Table~\ref{si:speed-backends} gives the timings of MD in the NVT ensemble. On the GPU, a step takes 17\% less time with LAMMPS for 1,000 atoms and 26\% less for 10,648 atoms, and the peak GPU memory is half of that with ASE.

\begin{table}[!htbp]
\centering\small
\caption{\textbf{Timings of MD with the ASE and LAMMPS backends for TensorNet-MatPES-PBE.} MD of diamond Si in the NVT ensemble at 300\,K: time per step and peak GPU memory. One run of each test.}
\label{si:speed-backends}
\begin{tabular}{llcc}
\toprule
 & & ASE & LAMMPS \\
\midrule
\multicolumn{4}{l}{\textit{GPU}} \\
 & Time per step, 1,000 atoms (ms) & 47 & 39 \\
 & Time per step, 10,648 atoms (ms) & 506 & 373 \\
 & Peak GPU memory, 1,000 atoms (GB) & 6.0 & 3.1 \\
 & Peak GPU memory, 10,648 atoms (GB) & 63 & 32 \\
\multicolumn{4}{l}{\textit{CPU}} \\
 & Time per step, 1,000 atoms (ms) & 826 & 797 \\
\bottomrule
\end{tabular}
\end{table}

\section{Conservative and direct-force models}
\label{si:nonconservative}

Table~\ref{si:nc-results} compares two direct-force models, eqV2-OMat24\cite{Liao2024EquiformerV2,barroso-luqueOpenMaterials20242024} and ORB-OMat24 with direct forces,\cite{rhodesOrbv3AtomisticSimulation2025} with three conservative models, ORB-OMat24 with conservative forces, MACE-OMat24, and GRACE-OMat24, all trained on OMat24. The two ORB-OMat24 models differ only in how the forces are obtained. The benchmarks follow the protocols of the main text and use 300 randomly selected WBM structures with one perturbation, 500 randomly selected structures of the elasticity benchmark, the compounds of the $C_V$ and $\kappa$ comparisons of Section~\ref{si:cv-r2scan}, and 300 randomly selected migration paths. The relaxed structures and formation energies of the direct-force models are as accurate as those of the conservative models, whereas their $K_\mathrm{VRH}$ MAEs are two to three times as large. For $C_V$, the direct-force and conservative ORB-OMat24 models have the same MAE. For $\kappa$, the MAE of $\log_{10}\kappa$ of the direct-force ORB-OMat24 model is six times that of the conservative one, and both direct-force models underestimate $\kappa$, with median ratios to the reference of 0.60 (ORB-OMat24) and 0.11 (eqV2-OMat24). For $E_\mathrm{m}$, the MAE of the direct-force ORB-OMat24 model is 0.03\,eV higher than that of the conservative one.

\begin{table}[!htbp]
\centering\small
\caption{\textbf{Direct-force and conservative FPs trained on OMat24.} $\bar{d}_\mathrm{fp}$ and MAEs against PBE reference values of $E_\mathrm{f}^{0}$ (meV\,atom$^{-1}$), $K_\mathrm{VRH}$ and $G_\mathrm{VRH}$ (GPa), $C_V$ at 300\,K (J\,mol-atom$^{-1}$\,K$^{-1}$), $\log_{10}\kappa$ at 300\,K, and $E_\mathrm{m}$ (eV). The MAEs of $K_\mathrm{VRH}$ and $G_\mathrm{VRH}$ are computed over the structures for which all five FPs give a result, and those of $E_\mathrm{m}$ over the paths that converged. eqV2-OMat24: EquiformerV2 with 153M parameters (\texttt{eqV2\_153M\_omat}); ORB-OMat24: \texttt{ORB-OMat24-v3-direct-inf} and \texttt{ORB-OMat24-v3-conservative-inf}; MACE-OMat24: \texttt{MACE-OMat24-0-medium}; GRACE-OMat24: \texttt{GRACE-OMat24-2L-large-base}.}
\label{si:nc-results}
\setlength{\tabcolsep}{4pt}
\begin{tabular}{llccccccc}
\toprule
FP & Forces & $\bar{d}_\mathrm{fp}$ & $E_\mathrm{f}^{0}$ & $K_\mathrm{VRH}$ & $G_\mathrm{VRH}$ & $C_V$ & $\log_{10}\kappa$ & $E_\mathrm{m}$ \\
\midrule
eqV2-OMat24 & direct & 0.15 & 25 & 33 & 21 & 0.10 & 0.90 & 0.29 \\
ORB-OMat24 & direct & 0.22 & 34 & 48 & 18 & 0.04 & 0.36 & 0.18 \\
ORB-OMat24 & conservative & 0.22 & 36 & 15 & 13 & 0.04 & 0.06 & 0.15 \\
MACE-OMat24 & conservative & 0.23 & 38 & 15 & 18 & 0.07 & 0.05 & 0.15 \\
GRACE-OMat24 & conservative & 0.21 & 32 & 15 & 14 & 0.03 & 0.03 & 0.16 \\
\bottomrule
\end{tabular}
\end{table}

Table~\ref{si:nc-conservation-tab} gives three tests of the consistency between forces and energies. The forces were compared with central finite differences of the energy for 100 force components of ten rattled structures, the net force was evaluated for the same structures, and the drift of the total energy was measured in NVE simulations of 10\,ps with a time step of 1\,fs for Si and MgO (64 atoms) and Cu (108 atoms), started at 600\,K. MACE-OMat24 and GRACE-OMat24 were run in double precision, and eqV2-OMat24 and the ORB-OMat24 models in single precision; the ORB-OMat24 models were also run in double precision for the finite differences, the net force, and Si. In double precision, the forces of the conservative ORB-OMat24 model agree with the finite differences to $10^{-3}$\,eV\,\AA$^{-1}$ and its energy drift is below $10^{-4}$\,meV\,atom$^{-1}$\,ps$^{-1}$, whereas the forces of the direct-force model deviate by 0.06\,eV\,\AA$^{-1}$ and its total energy in Si rises by 4.5\,meV\,atom$^{-1}$\,ps$^{-1}$. In single precision, the finite differences of both ORB-OMat24 models are limited by rounding errors of the energy, and the conservative model drifts by 0.13\,meV\,atom$^{-1}$\,ps$^{-1}$ in Si. The net force of eqV2-OMat24 reaches $2\times10^{-3}$\,eV\,\AA$^{-1}$ per atom, and its total energy decreases by 0.01--0.03\,meV\,atom$^{-1}$\,ps$^{-1}$. In relaxations of 19 WBM structures with at most 2,000 steps, both ORB-OMat24 models in single precision converge for all structures to $f_\mathrm{max}=0.05$\,eV\,\AA$^{-1}$, for 74\% (direct) and 84\% (conservative) to 0.01\,eV\,\AA$^{-1}$, and for none to $10^{-3}$\,eV\,\AA$^{-1}$, whereas GRACE-OMat24 in double precision converges for all structures to $10^{-4}$\,eV\,\AA$^{-1}$.

\begin{table}[H]
\centering\small
\caption{\textbf{Consistency of forces and energies.} $\Delta F$: root-mean-square difference between the forces and central finite differences of the energy with a displacement $h$ (eV\,\AA$^{-1}$). $|\sum\mathbf{F}|/N$: largest net force per atom (eV\,\AA$^{-1}$). Drift: slope of the total energy in NVE simulations (meV\,atom$^{-1}$\,ps$^{-1}$).}
\label{si:nc-conservation-tab}
\setlength{\tabcolsep}{4pt}
\resizebox{\textwidth}{!}{%
\begin{tabular}{lllcccccc}
\toprule
 & & & \multicolumn{2}{c}{$\Delta F$} & & \multicolumn{3}{c}{Drift} \\
\cmidrule(lr){4-5}\cmidrule(lr){7-9}
FP & Forces & Precision & $h=0.005$\,\AA & $h=0.02$\,\AA & $|\sum\mathbf{F}|/N$ & Si & MgO & Cu \\
\midrule
eqV2-OMat24 & direct & single & $1.7\times10^{-1}$ & $6.5\times10^{-2}$ & $2\times10^{-3}$ & $-0.013$ & $-0.027$ & $-0.018$ \\
ORB-OMat24 & direct & single & $1.1\times10^{-1}$ & $6.5\times10^{-2}$ & $1\times10^{-7}$ & $4.3$ & $0.17$ & $0.20$ \\
 & & double & $6.8\times10^{-2}$ & $6.2\times10^{-2}$ & $8\times10^{-17}$ & $4.5$ & -- & -- \\
ORB-OMat24 & conservative & single & $1.1\times10^{-1}$ & $3.1\times10^{-2}$ & $1\times10^{-7}$ & $0.13$ & $0.019$ & $-0.002$ \\
 & & double & $1.2\times10^{-3}$ & $8.8\times10^{-4}$ & $1\times10^{-16}$ & $8\times10^{-5}$ & -- & -- \\
MACE-OMat24 & conservative & double & $2.6\times10^{-5}$ & $4.2\times10^{-4}$ & $2\times10^{-16}$ & $-3\times10^{-5}$ & $2\times10^{-5}$ & $-4\times10^{-6}$ \\
GRACE-OMat24 & conservative & double & $2.6\times10^{-5}$ & $4.1\times10^{-4}$ & $2\times10^{-16}$ & $-3\times10^{-5}$ & $-2\times10^{-5}$ & $6\times10^{-6}$ \\
\bottomrule
\end{tabular}}
\end{table}

\clearpage
\bibliography{references,new_refs}